\documentclass[prd,nofootinbib,preprintnumbers,floats,aps]{revtex4}
\usepackage{graphicx}
\usepackage{graphics}
\usepackage{amssymb}
\usepackage{amsmath}
\usepackage{amsfonts}
\usepackage{subfigure}
\usepackage{wrapfig}
\usepackage[usenames]{color}
\usepackage{bm}
\usepackage{amsbsy}
\usepackage{epsfig}
\newcommand{\nc}{\newcommand}

\nc\beq{\begin{equation}}
\nc\eeq{\end{equation}}
\nc\beqa{\begin{eqnarray}}
\nc\eeqa{\end{eqnarray}}
\nc\bi{\begin{itemize}}
\nc\ei{\end{itemize}}

\long\def\teq#1{\hbox{$#1$}}				% for longer in-text equations

\def\gsim{\mathrel{\rlap{\lower4pt\hbox{$\sim$}}
    \raise1pt\hbox{$>$}}}					%greater than or approx. symbol
\def\lsim{\mathrel{\rlap{\lower4pt\hbox{$\sim$}}
    \raise1pt\hbox{$<$}}}					%less than or approx, symbol

\def\Dsl{\,\raise.15ex \hbox{/}\mkern-12.8mu D}

\def\OMIT#1{}

\newcommand{\ba}{\begin{array}}
\newcommand{\ea}{\end{array}}

\def\gsim{\mathrel{\rlap{\lower4pt\hbox{\small \hskip1pt$\sim$}}
\raise1pt\hbox{$>$}}}

\def\lsim{\mathrel{\rlap{\lower4pt\hbox{\small \hskip1pt$\sim$}}
\raise1pt\hbox{$<$}}}
\newcommand{\nucl}[4]{\ensuremath{_{#2}^{#4}{\rm #1}^{\phantom{1}}_{#3}}}

\newcommand{\fract}[2]{{\textstyle\frac{#1}{#2}}}

\newcommand{\half}{\frac{1}{2}} \newcommand{\p}{\partial}
\newcommand{\be}{\begin{equation}}
\newcommand{\ee}{\end{equation}}
\newcommand{\bea}{\begin{eqnarray}}
\newcommand{\eea}{\end{eqnarray}}
\newcommand{\ben}{\begin{enumerate}}
\newcommand{\een}{\end{enumerate}}
\newcommand{\bit}{\begin{itemize}}
\newcommand{\eit}{\end{itemize}}
\newcommand{\eq}[1]{eq.~(\ref{#1})}

\newcommand{\Eq}[1]{Eq.~(\ref{#1})}
\newcommand{\Eqs}[1]{Eqs.~(\ref{#1})}
\numberwithin{equation}{section}

\definecolor{BrickRed}{cmyk}{0,0.89,0.94,0.28}%%%PANTONE 1805
\definecolor{MidnightBlue}{cmyk}{0.98,0.13,0,0.43}%%%PANTONE 302
\definecolor{DarkGreen}{rgb}{0.100806,0.495968,0.209979}
\definecolor{orange}{rgb}{0.587167,0.354498,0.146197}
\newcommand{\braket}[3]{\left\langle #1\right| #2 \left| #3 \right\rangle }
\newcommand{\braketnoline}[1]{\left\langle #1 \right\rangle }

\begin{document}

\preprint{MIT-CTP-3968}
\preprint{0809.1647 [hep-ph]}

\title{Quark Masses:  An Environmental Impact Statement}
 
\author{Robert\ L.\ Jaffe}
\author{Alejandro\ Jenkins}
\author{Itamar\ Kimchi}

\affiliation{Center for Theoretical Physics,
Laboratory for Nuclear Science and Department of Physics,
Massachusetts Institute of Technology,
Cambridge, MA 02139, USA}

\published{in Phys. Rev. D {\bf 79}, 065014 (2009)}

%%%%%%%%%%
%%% ABSTRACT
%%%%%%%%%%
\begin{abstract}

We investigate worlds that lie on a slice through the parameter space of the Standard Model over which quark masses vary.  We allow as many as three quarks to participate in nuclei, while fixing the mass of the electron and the average mass of the lightest baryon flavor multiplet.  We classify as {\it congenial\/} worlds that satisfy the environmental constraint that the quark masses allow for stable nuclei with charge one, six, and eight, making organic chemistry possible.  Whether a congenial world actually produces observers capable of measuring those quark masses depends on a multitude of historical contingencies, beginning with primordial nucleosynthesis and including other astrophysical processes, which we do not explore.  Such constraints may be independently superimposed on our results.  Environmental constraints such as the ones we study may be combined with information about the {\it a priori\/} distribution of quark masses over the landscape of possible universes to determine whether the measured values of the quark masses are determined environmentally, but our analysis is independent of such an anthropic approach.

We estimate baryon masses as functions of quark masses via first-order perturbation theory in flavor $SU(3)$ breaking.  We estimate nuclear masses as functions of the baryon masses using two separate tools: for a nucleus made of two baryon species, when possible we consider its {\it analog} in our world: a nucleus with a similar binding energy, up to Coulomb contributions.  For heavy nuclei or nuclei made of more than two baryons, we develop a generalized Weizs\"acker semi-empirical mass formula, in which strong kinematic flavor symmetry violation is modeled by a degenerate Fermi gas.

We check for the stability of nuclei against fission, strong particle emission (analogous to $\alpha$ decay), and weak nucleon emission.  For two light quarks with charges 2/3 and $-1/3$, we find a band of congeniality roughly 29 MeV wide in their mass difference, with our own world lying comfortably away from the edges.  We also find another, less robust region of congeniality with one light, charge $-1/3$ quark, and two heavier, approximately degenerate quarks with charges $-1/3$ and 2/3.  No other assignment of light quark charges yields congenial worlds with two baryons participating in nuclei.   We identify the region in quark-mass space where nuclei would be made from three or more baryon species.  We discuss issues relevant to the congeniality of such worlds, but a full characterization of them is left for future investigation.
 
\end{abstract}


\maketitle

%%%%%%%%%%
%%% OVERVIEW
%%%%%%%%%%
\section{Overview}
\label{overview}

\subsection{Introduction}
\label{introduction}

It is logically possible that certain physical quantities might be environmentally selected, rather than uniquely determined by fundamental principles.  For instance, although early scientists such as Johannes Kepler sought an explanation for the sizes of the orbits of the planets in terms of fundamental mathematical laws, we now recognize that these result from complicated dynamics that involve significant environmental contingencies.  Many solar systems are possible (indeed, many are now known to exist) with very different configurations.  Today we understand that the size and shape of the Earth's orbit is environmentally selected:  we happen to live on a planet whose orbit allows intelligent observers to evolve.  If the orbit were much different, we would not find ourselves inhabiting this particular planet. The extent to which the size of other planetary orbits in our solar system are environmentally constrained is a subject of active inquiry.  For instance, Jupiter may play a role in stabilizing the inner solar system over large time scales.  Other orbits, such as that of the planet Mercury, might have little to do with the emergence of life on Earth and consequently might not be explained by environmental selection.

From time to time modern scientists have suggested that the values of certain physical constants, and perhaps even the form of some of the laws of nature, might similarly be environmentally determined \cite{carter,barrowtipler}.  This idea, known loosely as the \emph{anthropic principle} has received support recently on both the theoretical and phenomenological fronts.  Inflationary cosmology suggests that our universe may be one of a vast, perhaps infinite, number of causally disconnected universes, and studies of flux compactification in string theory suggest that---at least in that version of quantum gravity---the laws of nature and physical constants could be different in each of those universes \cite{douglaskachru}.  If that were the case, then each universe would evolve according to its own laws and only some would lead to the evolution of physicists who would try to make sense of what they observe.  In those universes some of the laws of nature and physical constants may be explained not by some deep principle, but rather by the requirement that they be consistent with the evolution of intelligent observers \cite{weinberg-review,hogan-review}.

Whatever one may think of these theoretical arguments, the \emph{prediction} of the value of the cosmological constant $\Lambda$ on essentially these grounds deserves to be taken seriously.  At a time when the experimentally measured value of $\Lambda$ was consistent with zero, Weinberg \cite{weinberg1} pointed out that values outside of a very small range (on the scale of fundamental physics) would prevent galactic structure formation and that probabilistic considerations suggested a value close to the environmental bound imposed by the structure we see around us (suggestions along similar lines had been made earlier by Linde \cite{linde}, Banks \cite{banks}, and by Barrow and Tipler in Sec. 6.9 of \cite{barrowtipler}).  Shortly after Martel, Shapiro, and Weinberg refined this anthropic argument in \cite{weinberg2}, a positive value of $\Lambda$ was measured that was consistent with their prediction \cite{darkenergy}.  Such an environmental explanation is all the more persuasive in the absence of any attractive alternative derivation of $\Lambda$ from a dynamical principle.

Anthropic reasoning raises some deep questions that will be difficult (perhaps even impossible) to answer scientifically:  Is there a multiverse to begin with?  Are string theory and eternal inflation the correct tools with which to study the space of possible universes?  What parameters of the Standard Model (SM) are environmentally selected, and what is the {\it a priori\/} distribution over which environmentally selected parameters range?  These questions would have to be addressed before one could claim that some parameter, like the up-down quark mass difference, is environmentally \emph{selected}.  However, there is another, less daunting question to be asked about the parameters of the SM:  To what extent is a particular parameter environmentally \emph{constrained}?  Over what range of a parameter do the laws of physics allow for the existence of an observer?  This is the question we attempt to formulate and, at least in part, answer in this paper.  The answer to this question might be of little scientific significance outside of an anthropic framework; nevertheless, the question is well formulated and can, at least in principle, be answered if we understand the relevant physics well enough.

In order to determine the environmental constraints on the interactions and physical constants in our universe, we must be able to deduce the physical consequences of \emph{varying the laws and parameters} that are familiar to us.  If we do not understand the theoretical landscape well enough to declare certain regions environmentally barren of observers or others fruitful, it will not be possible to place environmental constraints on the various ingredients of the SM.  The generally-accepted environmental explanation for the radius of the earth's orbit provides a good example of how this process might work:  There are environmental \emph{constraints} on carbon-based life forms.  Biochemical structures cannot survive if it is too hot and biochemical reactions grind to a halt if it is too cold.  One could, in principle, make these limits precise by studying the dependence of carbon chemistry and biology on temperature.  At the end of a long and careful study one could imagine arriving at an understanding of the temperature constraints on the emergence of carbon based life forms.    Combined with a similar understanding of planetary science, these would translate into environmental \emph{constraints} on the orbital radius of a planet inhabited by carbon-based life.  These questions are clearly within the purview of science, and this is the type of analysis we would like to carry out within the SM.  The question of whether the earth's radius was environmentally \emph{selected} depends on whether the universe contains many planetary systems and on the distributions of orbital radii, compositions, and so forth.  Likewise, the question of whether some aspect of the SM is environmentally \emph{selected} depends on whether there is a multiverse and on the {\it a priori\/} distribution of SM properties and parameters over that multiverse.  This second question involves totally different physics---cosmology, grand unification, {\it etc.\/}---and is much harder to answer than whether some parameter of the SM is environmentally \emph{constrained}.

In order to carry out an environmental analysis of the SM we must first clearly specify the criterion for an anthropically acceptable universe, {\it i.e.\/} one in which we believe observers may exist.  Then we must adopt and justify a specific scheme for varying (``scanning'') some parameters over the landscape of possible worlds, while holding other parameters fixed.  This procedure defines the \emph{slice} through the space of SM parameters which is to be investigated.  Then we must use our knowledge of the physics of the SM to predict the consequences of scanning those parameters and, in particular, to determine which choices of parameters would correspond to worlds that meet our criterion of anthropic acceptability.\footnote{Note that the approach we are taking allows only for \emph{local} variations about the laws of physics that we see in our universe.  We have nothing to say about universes with entirely different interactions and particle content.}

In this paper we attempt to carry out such an environmental analysis for the masses of the three lightest quarks.  We have chosen to study quark masses because the dependence of the strong interaction on the masses of the quarks is non-trivial and yet is understood well enough to offer some hope that we can make definitive statements about the probability of the evolution of observers as a function of them.

We propose to vary the masses of quarks \emph{holding as much as possible of the rest of SM phenomenology constant}.  In particular we leave the mass and charge of the lightest lepton, the electron, and the mass of the lightest baryon flavor multiplet unchanged.  In pursuit of our objective we are forced to vary the masses of the muon and tauon (and other mass scales) relative to the QCD mass scale, but this has no effect on the questions of interest.  We also leave the weak interactions of quarks and leptons unchanged.  It is \emph{not} sufficient to simply hold all the other parameters of the SM fixed and vary only the quark masses.

If one insisted on varying quark masses while keeping all the other parameters of the SM fixed, for example, at some unification scale or at the scale of electroweak symmetry breaking, then the resulting changes to the low energy physics would be too extreme for us to analyze.  In particular, nuclear binding in the SM depends strongly both on the quark masses and on $\Lambda_{\rm QCD}$, the scale below which QCD becomes strongly coupled.  If we imagined that $\Lambda_{\rm QCD}$ were set by the renormalization group running of a unified gauge coupling held fixed at some high mass scale $M_{0}$, then $\Lambda_{\rm QCD}$ would change with the masses of the quarks and leptons.

Holding $\Lambda_{\rm QCD}$ fixed as the light quark masses are varied is not only inconvenient for calculation, but also as theoretically unmotivated as adjusting it in a mild way, as we prefer to do here.  In Sec. \ref{rules} we describe in detail which parameters of the SM must vary along with the quark masses in order that  we have enough control of the strong interactions to make sharp predictions about nuclear masses.
 
The worlds we will study will look much like our own.  They will have some stable baryons, some of them charged.  The charged stable baryons, whether positively or negatively charged, will capture electrons or positrons to form neutral atoms\footnote{If the stable hadrons have negative charge, neutrinos would offset the negative lepton number of the positrons.} with chemistry that will be essentially identical to ours.  This leads us to adopt the existence of stable nuclei with charge one (some isotope of hydrogen) and charge six (some isotope of carbon) to be the criterion for an acceptable universe.  One could argue that other elements, especially oxygen (charge eight) are necessary for carbon-based life.  We find that worlds with stable carbon also have stable oxygen and a variety of other elements, so it is not necessary to refine our criteria in this regard.  Note that carbon might have very different \emph{nuclear physics} in worlds with different quark masses.  If, for example, the charges of the stable baryons were different from our world, then the baryon number of carbon would not be 12, but the \emph{chemistry} of the element with charge six would be nearly the same as the chemistry of carbon in our world.

For convenience, from now on we will call a universe \emph{congenial} if it seems to support the evolution of an observer.  An \emph{uncongenial} universe is the opposite.  Our criterion for congeniality is, as already stated, the existence of stable nuclei with $Z=1$ and $Z=6$.  We realize that these may not be necessary if some exotic form of life not based on organic molecules were possible.  Likewise they may not be sufficient if there were some other obstruction to the development of carbon based life.  The former is interesting but very difficult to study.  The latter is very important and requires further investigation. 

For instance, Hoyle famously predicted---with better than 1\% accuracy---the location of a 7.65 MeV excited state of $^{12}$C, based on the fact that, without such a resonance, stars would produce only a negligible amount of carbon, since the fusion of three $\alpha$ particles would proceed very slowly \cite{3alpha}.  It is difficult, however, to translate the requirement of efficient carbon synthesis into a statement about the fine tuning of the fundamental SM parameters, since the rate of carbon production depends on the {\it spacing} between two different nuclear energy levels (which are very difficult to compute from first principles) and also on stellar temperatures (which could conceivably be different in other universes with different interactions).  Furthermore, Hoyle's argument presupposes that most of the carbon available for the evolution of organic life has to be produced in stars by the fusion of the three $\alpha$'s, which might or might not be true of all universes in the landscape that have stable isotopes of carbon.

Another example of an argument for an astrophysical obstruction to the development of carbon-based life comes from the work of Hogan, who has argued that the existence of a stable, neutral baryon would short circuit big-bang nucleosynthesis and/or leave behind deadly clouds of neutral baryons~\cite{Hogan:2006xa}.  However, some of the worlds we study have stable neutral baryons but may have different early histories than Hogan considers: for example, in some of our worlds the deuteron is not stable while $^{3}$H (the tritium nucleus, or ``triton'') is, a possibility not contemplated by Hogan.  The lack of a stable deuteron could interrupt the reaction path that Hogan identifies as leading to an explosive Big-Bang $r$-process in worlds with stable neutrons.

Furthermore, the availability of stable $^3$H would change stellar dynamics significantly, since two tritium nuclei can make $^{4}$He plus two free neutrons by a \emph{strong interaction}, in contrast to the weak interaction, $pp \to ^2$H$\, e^+ \, \nu_e$ in our world.  Finally, whether the presence, in some given universe, of roaming clouds of stable neutral baryons left over from Big-Bang nucleosynthesis would prevent the evolution of life {\it everywhere} presumably depends on very complicated astrophysical considerations. 

This highlights a distinction between two types of criteria for congeniality.  The first, which we adopt, is that the \emph{laws of nature} should be suitable for the evolution of life (as we know it). The second, which we put aside for later work, is that the history of a universe evolving according to those laws should lead to intelligent life.  Clearly the latter is a much more difficult problem, involving aspects of astrophysics, celestial dynamics, planetary physics, {\it etc.\/}, each of which has its own complex rules.  Returning to the planetary analogy, it is akin to showing that planets in the habitable zone would have any number of other attributes, like atmospheres, water, plate tectonics, {\it etc.\/}, that might be necessary for life.  For example, none of the astrophysics or celestial mechanics leading to earth-like planets seems to rely on the existence of unstable, but long-lived $\alpha$-emitting nuclei.  However, the earth would be a featureless ocean with very different life forms if it were not for plate tectonics powered by uranium and thorium decays in the earth's core. Such considerations are clearly important---Weinberg's constraints on the cosmological constant follow from consideration of the history of the universe, not the form of the SM---but they are beyond the scope of this paper.  Constraints on the formation of life from such cosmological or astrophysical considerations can always be cleanly superimposed upon the constraints from nuclear physics studied in this paper.

\subsection{Variation of quark masses}
\label{variation}

Quark masses in the SM, $m_{a}$, are determined by Yukawa couplings, $g_{a}$, by $m_{a}=g_{a}v$, where $v$ is the Higgs  vacuum expectation value (vev).  In our world, the quark Yukawa couplings range from $\sim 10^{-5}$ for the $u$ quark up to $\sim 1$ for the $t$ quark.  The {\it a priori\/} connection, if any, between the quark masses and the scale of the strong interactions, $\Lambda_{\rm QCD}$, is unknown.  We would like to consider the widest possible range of variation of quark masses, limited only by our ability to study their consequences with the tools of the SM.  Quarks fall into two classes:  light quarks, with masses less than or of the order of $\Lambda_{\rm QCD}$, and heavy quarks, with masses above $\Lambda_{\rm QCD}$.  Heavy quarks do not participate in the physics of nuclei and atoms.  If Yukawa couplings are varied widely, the number of light quarks can range from zero to six, and their charges can be $+2/3$ or $-1/3$, constrained only by the fact that there are at most three of either charge.

In our world, with light $u$ and $d$ quarks and a somewhat heavier $s$ quark, the methods of flavor $SU(3)$ perturbation theory have been moderately successful in baryon spectroscopy.  Using flavor $SU(3)$ perturbation theory we can extract enough information from our world to predict the masses of the lightest baryons in worlds with \emph{up to three light quarks}, provided their mass differences are small enough to justify the use of first order perturbation theory.  In practice this is not an important limitation because we find that a quark species does not participate in nuclear physics unless its mass is considerably less than the mass of the $s$ quark in our world.  So first order perturbation theory in quark masses should be even better in the other worlds we explore than it is in our world.  In fact, the principal limitation to our use of perturbation theory in the quark masses comes from possible higher order corrections to the extraction, from the experimentally measured baryon masses, of the reduced matrix elements of $SU(3)$ tensor operators.   For a similar reason, we are limited to at most three species of light quarks by the fact that we cannot extract flavor $SU(4)$ reduced matrix elements from the measured baryon spectrum, given how badly that symmetry is broken in our world.

To summarize, we consider worlds where baryons made of up to three species of light quarks participate in nuclear physics.  We circumvent ambiguities associated with the extraction and renormalization scale dependence of quark masses by presenting results as functions of the ratio of quark masses to the sum of the three light quark masses, $m_T$, in our world. 

\subsection{Our slice through the Standard Model}
\label{introslicing}

Previous research has considered the environmental constraint on the value of the Higgs vev $v$, with the Yukawa couplings fixed to the values they have in our world \cite{vev-selection}.  Along this particular slice through the SM parameter space, all quark and lepton masses scale together.  The requirement that stable atoms exist is found to roughly impose the constraint
\beq
0 < \frac{v}{v^\oplus} \lesssim 5~,
\label{vevconstraint}
\eeq
where $v^\oplus$ is the Higgs vev in our universe.\footnote{Throughout this paper we will use the super- or subscript ``$\oplus$'' to label the value of a parameter in our own world.}  A general argument has been proposed for why this particular choice of a slice might correspond approximately to the landscape resulting from string theory and eternal inflation \cite{friendly-landscape}.

\begin{wrapfigure}{r}{0.45\textwidth}
\begin{center}
\includegraphics[width=0.25\textwidth]{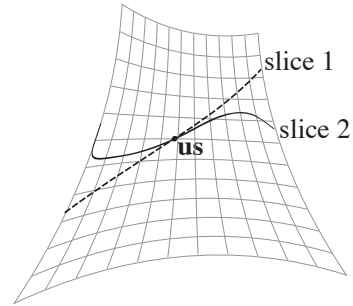}
\end{center}
\caption{\footnotesize Two slices through the parameter space of the SM are shown in this cartoon representation of that space.  For simplicity the space is drawn as two-dimensional and the slices as one-dimensional.  The two slices meet at the point the corresponds to the parameters of our own world (labelled ``us'' in the figure).  We may imagine slice 1 to correspond to varying quark masses at fixed $\alpha_s (M_Z)$, while slice 2 would correspond to the subspace we investigate in this paper.}
\label{slice}
\end{wrapfigure}

In this paper we will consider a very different slicing (see Fig. \ref{slice}), in which the three light quark masses are allowed to vary independently, with minimal variation on the other phenomenological aspects of the SM.  First we must fix a scale with respect to which all other relevant masses will be defined.  We do this by fixing the electron mass equal to its value in our world, $m_{e}=m_{e}^\oplus=0.511$ MeV.  Next, we choose a set of quark mass values.  Our world, for example, is specified by $m_{u}=m_{u}^\oplus=1.5-3$ MeV, $m_{d}=m_{d}^\oplus= 2.5-5.5$ MeV, and $m_{s}=m_{s}^\oplus=95\pm 25$ MeV \cite{PDG}.  Finally, we adjust $\Lambda_{\rm QCD}$ so that the average mass of the lightest baryon flavor multiplet (usually an $SU(2)$ doublet) equals the nucleon mass in our world, $M_{N}^\oplus=940$ MeV.     

This procedure defines our ``slice'' through the parameter space of the SM.  The other parameters of the SM including the masses of heavy leptons and neutrinos, the parameters of the Cabibbo-Kobayashi-Maskawa (CKM) and Pontecorvo-Maki-Nakagawa-Sakata (PMNS) matrices, the scale of weak symmetry breaking, {\it etc.\/}, do not impact nuclear physics significantly, so our analysis applies to wide families of worlds where those parameters vary from their values in our world.  

Because of this relatively clean separation into one set of mass scales that influence nuclear physics significantly $(\{m_{q}\},m_{e},\Lambda_{\rm QCD})$ and all the rest that do not, there is an additional mass rescaling that enlarges the set of worlds to which our analysis applies.  Consider a point on our slice through the SM, labeled by quark masses, $\Lambda_{\rm QCD}$, and $m_{e}$.  Now consider a world where all these mass scales are multiplied by some factor, $\alpha$.  Nuclear physics in this  world can be mapped back to ours by redefining 1 MeV $\to 1/\alpha$ MeV.  All the other mass parameters of the SM, including for example $m_{\mu}$, $m_{\tau}$, $M_{W,Z}$, and so forth, are scaled by $1/\alpha$, but these have negligible effect on the question of congeniality that interests us.  In effect, by fixing $m_{e}$ we are ``choosing a gauge'' with respect to this additional invariance.

We use $\Lambda_{\rm QCD}$ to rescale the lightest baryon's mass to 940 MeV.  The masses of other hadrons, in particular of the pion, the $\rho(770)$ and the $f_{0}(600)$, (or the equivalent mesons in worlds with different light quarks), which determine the properties of nuclei, are affected by the choice of quark masses and of $\Lambda_{\rm QCD}$.  Changing these mesons' masses affects the binding of nuclei and introduces uncertainty into our results.  As a result there are some regions of quark mass parameter space where we can make reliable quantitative statements and others where only qualitative statements are possible.  This question is discussed in considerably more detail in Sec.~\ref{range}.  

We have chosen a slice through the parameter space of the SM along which it is relatively simple to examine congeniality as the quark masses vary.  This is only one of many possible slices, and might not be the slice dictated by a deeper understanding of the landscape of possible universes.  We choose our approach simply because \emph{the resulting universes can be studied with some certainty}. So doing, we accomplish two things:  first, we give an example of what is required to analyze the environmental impact of non-trivial variations in SM parameters, and second, we find some of the limits of the domain of congeniality in which we live, and also discover some different (disconnected) domains of congeniality elsewhere in light quark mass parameter space.  As our knowledge of QCD improves, in particular as lattice QCD matures, it may become possible to explore much larger domains of parameter space by simulating hadrons and nuclei on the lattice.

\subsection{Plan of the paper and survey of results}
\label{survey}

In Sec. \ref{rules} we confront some of the (meta)physical issues that arise when attempting an environmental analysis of quark masses.  First we define more carefully the difference between \emph{environmental constraint} and \emph{environmental selection}.  We discuss what features the {\it a priori\/} distribution of quark masses might have.  Next we explain why we ignore quarks with masses greater than the confinement scale of QCD.  This leads us to a parameterization of the space of light quark masses and finally to some discussion of baryon masses, nuclear forces, and stability.

In general, only members of the familiar flavor $SU(3)$ octet of spin-1/2 baryons can play a role in nuclear physics.\footnote{Two exceptions are mentioned:  when all quark masses are small one must consider the possibility that the dihyperon is the most stable baryon \cite{dihyperon} (see Sec.~\ref{dihyp}), and when only one quark is light, the spin-3/2 decuplet must be considered (see Sec. \ref{onebaryon}).}   In Sec. \ref{baryonmasses} we express the masses of the octet baryons as functions of the three light quark masses.  Most of the information necessary to relate octet baryon masses to quark masses can be obtained from a study of octet baryon mass differences in our world, using first order perturbation theory in the quark masses.  However, baryon mass differences cannot tell us about the dependence of baryon masses on the \emph{average quark mass}.  We take this information from the $\sigma$-term in $\pi N$ scattering (see \cite{quarkmasses} and references therein).  To get as accurate an estimate as possible of the connection between baryon masses and quark masses, we include estimates of the electromagnetic contributions to baryon masses.

Except in the case where all three light quark masses are nearly equal, only the two lightest baryons participate in nuclear dynamics.  The others, like hyperons in our world, make only exotic, very short lived nuclei (usually called ``hypernuclei'') \cite{hypernuclei-review}.  When all three quark masses are nearly equal, then all eight members of the baryon octet may be involved in nuclear dynamics.  In Sec. \ref{nuclearmasses}, we estimate the masses of nuclei in our slice through the landscape as functions of the baryon masses.  For some universes a given nuclear mass can be estimated very accurately by considering its \emph{analog}: a nucleus in our world whose binding energy should be the same except for easily calculable Coulomb corrections that occur if the participating baryon charges differ from those in our own world \cite{analog}.  For other universes where direct analogs are not available, we build a generalized Weizs\"acker semi-empirical mass formula (SEMF), with parameters more or less closely related to those in our world.  Kinematic flavor-breaking contributions are obtained from a degenerate Fermi gas model \cite{walecka}, supplemented by a term proportional to the quadratic Casimir of flavor $SU(2)$ or $SU(3)$, depending on whether there are two or three very light quarks.

With these tools in hand, in Sec. \ref{congenial} we explore the congeniality of worlds along our slice through the SM.  Since nuclear masses can be estimated, it is possible to check for the stability of various nuclei against fission, $\alpha$ decay, and ``weak nucleon emission'' ({\it i.e.\/}, the emission of a nucleon and a lepton-antilepton pair, mediated by a weak interaction).  For universes with only two light quarks and charges $+2/3$ and $-1/3$, $u$ and $d$, we find a band of congeniality roughly 29 MeV wide in $m_{u}-m_{d}$, with our own world lying comfortably away from the edges.  We also find another region of congeniality with small $m_d$ and $m_u \approx m_s$ (or equivalently, small $m_{s}$ and $m_{u}\approx m_{d}$), whose exact width in $m_{u}-m_{s}$ is difficult to estimate because nuclear forces in these worlds differ qualitatively from ours.  Worlds with two light quarks \emph{of the same charge}, either $+2/3$ or $-1/3$, are not congenial since, regardless of the quark masses, carbon is always unstable.

Finally, we identify the region in the space of quark masses where nuclei would be made from three or more baryon species.  We discuss some issues relevant to nuclear stability in such worlds, but a full characterization of them as congenial or uncongenial is left for future study.  We point out several possible sources of uncongeniality for these worlds: it is possible that in these worlds octet baryons would decay into dihyperons, which are bosons and would therefore behave very differently from nucleons in our universe.  In worlds with one positively-charged and two negatively-charged light quarks, nuclei with net electric charge would be very heavy, and might well be unstable to decay by emission of light, uncharged nuclei composed of a pair of each baryon species.

%%%%%%%%%%
%%% SETTING UP THE PROBLEM
%%%%%%%%%%
\section{Setting up the problem}
\label{rules}

In this section we delve more deeply into some of the issues raised in Sec. \ref{overview}.  We argue that a study of the environmental constraints on quark masses is a well-posed problem, and explain its relation to more ambitious anthropic inquiry.  Also the slice through the SM parameter space introduced qualitatively in Section \ref{overview} must be delineated more quantitatively.

\subsection{{\it A priori\/} distribution of quark masses}
\label{apriori}

As explained in Sec. \ref{overview}, we focus on the relatively well-defined problem of environmental \emph{constraints} on quark masses.  Nevertheless we recognize that there is much interest in the theoretical physics community in whether physical parameters might be environmentally {\it selected}.  The jump from constraint to selection requires some theoretical input (or prejudice) about the {\it a priori\/} distribution of quark masses.  Also, a prejudice about the {\it a priori\/} quark mass distribution can suggest where to focus our analysis of environmental constraints.  Therefore, we review here the relation between environmental constraint and environmental selection, and then summarize our thoughts about possible {\it a priori\/} quark mass distributions.
 
Anthropic reasoning can be presented in the language of contingent probabilities~\cite{carter2}.  According to Bayes's theorem, the probability distribution for measuring some set of values for the quark masses can be expressed as 
\beq
p( \{ m_i \} | {\rm observer} ) \propto p({\rm observer} | \{m_i \}) \times p(\{ m_i \})~,
\label{bayes}
\eeq
where $p(\{ m_i \})$ is the {\it a priori\/} distribution of quark masses over the landscape.  The quantity $p( \{ m_i \} | {\rm observer} )$ is the contingent probability distribution for the quark masses over the subset of the landscape that contains intelligent beings likes us, capable of measuring quark masses.   In simple terms, it is the probability that some set of quark masses would be measured.   The quantity $p({\rm observer} | \{m_i \})$, the contingent probability for an observer existing, given a choice of quark masses, is the stuff of \emph{environmental constraint} and the primary subject of this paper. 

The connection between environmental constraint and environmental selection is provided by the {\it a priori\/} distribution, $p(\{m_{i}\})$.  Environmental selection requires the existence of an ensemble of universes over which $p(\{m_{i}\})$ is defined.  Anthopic reasoning is based on some prejudice about the form of $p(\{m_{i}\})$ as (presumably) determined by some fundamental theory, such as string theory in an eternally-inflating scenario.  This should be contrasted with a situation where one or more quark mass is \emph{dynamically} determined.  Then $p(m_{a})$ (for some $a$) would be a delta-function and $m_{a}$ would not be environmentally selected.

In his analysis of the cosmological constant $\Lambda$, Weinberg argued that the {\it a priori\/} probability distribution $p(\Lambda)$ should not vary significantly over the range where the contingent probability $p(\rm{observer}| \Lambda)$ is non-zero, in which case the {\it a priori\/} distribution factors out of the expression for $p(\Lambda | {\rm observer})$.  Similarly, if $p(\{m_{i}\})$ does not vary significantly over the region where $p({\rm observer}|\{m_{i}\})$ is non-zero, then the probability that we observe some set of quark masses, which is the object of fundamental interest in an anthropic context, is simply proportional to $p({\rm observer} | \{m_i \})$, which we analyze here.

We believe that if anthropic considerations have any relevance at all, then there is reason to assume that the \emph{logarithms} of the quark masses are smoothly distributed over a range of masses small compared to the Planck scale.  Note, first, that an environmental analysis is predicated on the absence of any other underlying explanation for quark masses.  Therefore we assume that the probability of a pattern of masses is the product of independent probabilities,
\begin{equation}
p(\{m_{i}\})=\prod_{i=1}^{6}p(m_{i})\,.
\label{indep}
\end{equation}
In this paper we will usually ({\it i.e.\/}, unless otherwise specified) label the positively-charged quarks as $u$, $c$, and $t$, and the negatively-charged quarks as $d$, $s$, and $b$, in ascending order of mass.  (Therefore the identification of the label $i=1, ..., 6$ in \eq{indep} with $u,d,s,c,b,t$ varies according to the values of the $\{m_{i}\}$.)  In general, we shall care little about the details of the flavor structure of the weak interactions, but in Secs. \ref{nuclearmasses} and \ref{congenial} we will assume that no element of the CKM matrix is accidentally zero, so that no baryon that would normally decay by weak interactions is accidentally stable.

\begin{figure}[t]
\centering
\includegraphics[width=0.45\textwidth]{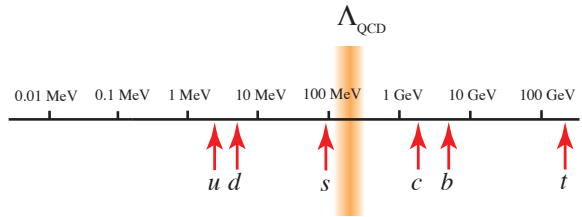}
\caption{\footnotesize The quark masses and $\Lambda_{\rm QCD}$, shown on a logarithmic scale.}
\label{logscale}
\end{figure}

Taking an anthropic point of view, there is a hint that, at least at masses below the scale of electroweak symmetry violation, the {\it a priori\/} distribution of the quark masses might be logarithmic.  This hint comes from the observed distribution of quark masses in our world, which is shown in Fig.~\ref{logscale}:  The $s$, $c$, $b$, and $t$ quarks are so heavy (on the scale set by $\Lambda_{\rm QCD}$) and short lived that they seem to play no role in nuclear and atomic physics (and consequently not in chemistry or biology either).  Therefore, referring to \Eq{bayes}, it is reasonable to take $p({\rm observer}|\{m_{i}\})$ to be essentially independent of $m_s, m_c,m_b$, and $m_t$.  Then the pattern of heavy quark masses that we see could well be a measure of the {\it a priori\/} probability distribution $p(m)$.  

Note, though, that the authors of \cite{topquark} proposed that $m_t$ might be anthropically selected to be near its measured value, $\sim 170$ GeV.  This proposal was based on the assumption (unmotivated by any dynamical consideration about the landscape) that the {\it a priori\/} distribution for the Higgs quartic self-coupling, $\lambda$, strongly favors small values.  The environmental requirement that the SM have a stable or metastable vacuum would then select $m_t$ through its effect on the renormalization group running of $\lambda$.  If we relax the assumption about the {\it a priori\/} value for $\lambda$, however, there would not seem to be a reason to expect strong anthropic constraints on $m_t$.  In any event, such a constraint could only apply to the heaviest quark, leaving the masses of $s,c,$ and $b$ unconstrained.

Although we have only four (or three) data points, the measured values of heavy quark masses range over more than two orders of magnitude, which suggests that the {\it a priori\/} cumulative distribution
\beq
P (m) \equiv \int^m dm' \, p(m')~,
\label{CDF}
\eeq
might be a smooth function in the logarithm of the mass, over some range that contains the six quark masses measured in our world \cite{logmasses}.  The {\it a priori\/} distribution of quark masses is determined by the landscape, which presumably is generated by physics at the Planck scale.  Therefore the scale for variation in the quark mass probability distribution should be $M_{\rm Planck}$ and we expect
\beq
P (m) = a \log \frac{m}{M_{\rm Planck}} + \ldots
\label{logdistrib2}
\eeq
where the omitted terms vary more slowly with $m$ than the logarithm.

Keeping only the leading term in \Eq{logdistrib2} implies that quark masses in that range are {\it a priori\/} equally likely to fall between any two consecutive orders of magnitude, and corresponds to a {\it scale-invariant} distribution
\beq
p (m) \propto \frac{1}{m}~.
\label{scaleinvariant}
\eeq
Approximately scale invariant distributions are observed in a variety of phenomena associated with complex underlying dynamics:  {\it e.g.\/}, the distribution by area of lakes in Minnesota has such a form, with a lower cutoff given by the conventional definition of a lake and the upper cutoff given by the area of Minnesota.\footnote{Scale invariance also explains the well-known phenomenon---known as Benford's law---that the first digits of numbers from many different kinds of real-life data are logarithmically distributed \cite{benford}.}  In the context of a possible landscape of quark and lepton masses, a scale-independent {\it a priori\/} distribution was investigated in \cite{donoghue-yukawas}.

Note that an {\it a priori\/} logarithmic distribution of quark masses puts a much greater weight on masses that are very small compared to $\Lambda_{\rm QCD}$ than would a linear distribution.  Therefore it is interesting to explore the congeniality of worlds where all light quark masses are very small compared to $\Lambda_{\rm QCD}$.  We have been able to make some progress in formulating the tools needed to explore this region of parameter space, and we can make some qualitative statements about its congeniality.  Nevertheless we have not been able to explore this region as fully as we would like.

Finally, we emphasize again that this excursion into a discussion of {\it a priori\/} probabilities is independent of the analysis of environmental constraints on quark masses which will concern us in the rest of this paper.  For a more explicit model of a possible landscape of Yukawa couplings, coming from the overlap of localized wavefunctions randomly distributed over compact extra dimensions, see \cite{mike}.

\subsection{Light and heavy quarks}
\label{lightandheavy}
The energy scale at which QCD becomes strongly coupled, $\Lambda_{\rm QCD}$, provides us with a distinction between light and heavy quarks.  Heavy quarks are those with $m_{j}>\Lambda_{\rm QCD}$ and light quarks have $m_{j}\lesssim\Lambda_{\rm QCD}$.  Even though the strange quark mass in our world is heavy enough that it decouples from nuclear physics and chemistry, it is nevertheless light compared with $\Lambda_{\rm QCD}^{\oplus}$, so it falls among the ``light'' quarks in this discussion.  If any quarks are lighter than $\Lambda_{\rm QCD}$, then assuming no accidental zeros in the CKM matrix, the heavy quarks will decay rapidly (on the time scales of atomic physics or chemistry) into light quarks by  means of weak interactions.  In this case only light quark masses would be environmentally constrained by nuclear physics, atomic physics, and chemistry---the constraints explored in this paper.

If all quarks were heavy, then the emergence of an observer would require a \emph{dynamical explanation}, which would be an interesting counterexample to anthropic reasoning: an observer who found him/her self in a world where all quark masses are large compared to $\Lambda_{\rm QCD}$ would have to dismiss an anthropic origin and instead seek a dynamical explanation for the pattern of quark masses.

To see why, suppose $c$ were the lightest quark in a world with all quark masses greater than $\Lambda_{\rm QCD}$.  If no other quark had mass close to $m_{c}$ then only the $ccc$ baryon (analogous to the $\Delta^{++}$ for $Q_{c}=2/3$ or to the $\Omega^{-}$ for $Q_{c}=-1/3$) would participate in building nuclei.  It is easy to see that Coulomb and symmetry effects would destabilize all systems with baryon number greater than one (remember, no single species nucleus, starting with the diproton or dineutron, is stable in our world), and there would be no congenial worlds.  

Suppose, on the other hand, that two heavy quarks had masses close enough to one another that both could participate in nucleus building.  The most favorable situation, from the point of view of Coulomb interactions, occurs when the quarks have different charges, so to make the discussion concrete we can call these heavy, nearly degenerate quarks the $c$ and $b$ quarks.  The two lightest baryons, $ccb$ and $bbc$, analogs of the proton and neutron, would experience attractive forces with a range $\sim 1/\Lambda_{\rm QCD}.$\footnote{Although there are no mesons with masses of order $\Lambda_{\rm QCD}$ in such a world---the lightest meson would be $b\bar c$, the analog of the $B_{c}$---nevertheless the nucleons would have radii of order $\Lambda_{\rm QCD}$ (up to logarithms), and when they overlap, they would attract by quark exchange processes. This apparent violation of the rules of effective field theory (the longest range force is not associated with the nearest threshold in the complex plane) was previously encountered in heavy meson effective field theory and is explained in Ref.~\cite{ineffective}.}. Attractive forces between massive particles bind when the range of the force is much greater than the particle's mass\footnote{The criterion is $mV_{0}R^{2}\gtrsim\,\mbox{constant}$, where $V_{0}$ and $R$ are characteristic of QCD and independent of $m$ (up to logarithms), while $m$ is assumed to be large.}.  So it is reasonable to expect stable nuclei when $m_{b}\approx m_{c} > \Lambda_{\rm QCD}$.  However the two heavy quarks' masses would have to be within a few tens of MeV of one another, otherwise the nuclei would rapidly decay away by weak nucleon emission.  This process is discussed in Sec. \ref{weakemission} and the corresponding constraints on the baryon masses are covered in Sec. \ref{udworld}.  For now, it is sufficient  to know that the quark mass difference cannot be greater than of order tens of MeV.  

Thus there is a possible window of congeniality if two heavy quarks are nearly degenerate.  However, if the {\it a priori\/} quark mass distributions, $p(m_{j})$ are independent, then this is a subspace of very small measure in the ensemble of universes.  If $p(m)$ is a smooth function of $\log m$ up to the Planck scale, then this congenial domain has essentially measure zero.  Physicists who lived in that universe would be confronted by an ``accidental'' symmetry, similar to (but even more exaggerated than) the one that puzzled particle physicists in our universe before the discovery of QCD.  Before {\it circa} 1970, it was thought that isospin was a good symmetry because the $u$ and $d$ quarks, both believed to have masses of order 300 MeV, had a mass \emph{difference} of only a few MeV.  This approximate degeneracy appeared mysterious, and physicists sought a dynamical explanation.  When QCD explained that the mass scale of hadronic bound states was set by $\Lambda_{\rm QCD}$, not by $u$ and $d$ quark masses, then the mysterious approximate equality was replaced by the less mysterious inequality, $m_{u},m_{d}\ll \Lambda_{\rm QCD}$.

Analogously to the pre-1970 situation in our universe, physicists in the heavy quark world would seek a \emph{dynamical} reason for the approximate equality of heavy quark masses.  In environmental terms, they would have to explain why the {\it a priori\/} quark mass distribution should have a (approximate) $\delta$ function, $\delta(m_{c}-m_{b})$, since a world with only heavy quarks would necessarily be uncongenial, unless there were a nearly-perfect degeneracy between the masses of the quarks, and such a degeneracy would not admit a compelling environmental explanation.\footnote{Of course, just as we do not know how small quark masses can be in the landscape, it is possible that physicists in a universe with only heavy quarks would not realize that quark masses {\it are} dynamically allowed to fall below $\Lambda_{\rm QCD}$.  Until they had learned, as we already know, that the {\it a priori\/} distribution allows quarks to be lighter than $\Lambda_{\rm QCD}$, they might accept the anthropic argument that only nearly degenerate (heavy) quark masses lead to a congenial world.}

\subsection{Parameterizing the masses of light quarks}
\label{parameterization}

\begin{figure*}[t]
\centering 
\subfigure[]{\includegraphics[width=0.22\textwidth]{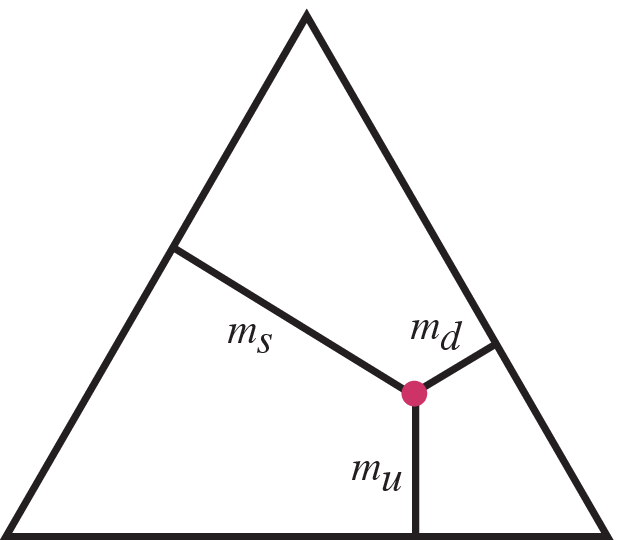}}\hspace*{2cm}
\subfigure[]{\includegraphics[width=0.17\textwidth]{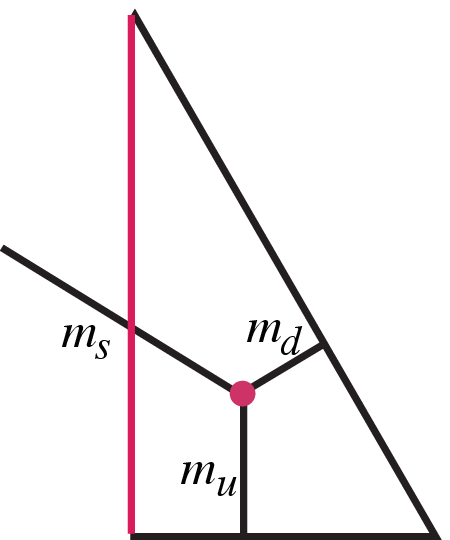}}
\caption{\footnotesize (a).  Graphical representation of the landscape of light quark masses for fixed \teq{m_T=m_u + m_d + m_s};  (b). Reduced landscape, assuming $m_s > m_d$.}
\label{triangles}
\end{figure*}

As mentioned in Subsection \ref{introslicing}, we will consider the variation of three light quark masses over a slice through the space of parameters of the SM in which the average mass of the two lightest octet baryons is fixed with respect to the mass of the electron.

To be definite we assume  the charge assignments $(2/3,-1/3,-1/3)$.  The extension to the other mixed charge possibility, $(2/3,2/3,-1/3)$, is mentioned as needed.  We denote the quarks $(m_{u},m_{d},m_{s})$ in the former case and $(m_{u},m_{c},m_{d})$ in the latter case.  The case where all three light quarks have the same charge, either $+2/3$ or $-1/3$ does not yield a congenial world with one possible exception:  If all three quarks have charge $-1/3$ and all quark masses are much lighter than $\Lambda_{\rm QCD}$, then it is possible that such a world would be congenial.  We mention this regime in Section~\ref{congenial}.

\begin{wrapfigure}{r}{0.45\textwidth}
\centering
\includegraphics[width=0.3\textwidth]{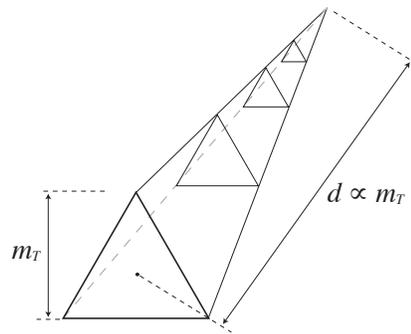}
\caption{\footnotesize Representation of the full three-dimensional space of light quark masses as a triangular prism, with four different slices of constant $m_T=m_u + m_d + m_s = \sqrt{6} \, m_{0}$ shown.}
\label{prism}
\end{wrapfigure}

For a fixed value of $m_T = m_u + m_d + m_s$, we may represent the landscape of light quark masses by the points in the interior of an equilateral triangle with altitude of $m_T$, as shown in Fig. \ref{triangles}(a).  The value of each quark mass is given by the perpendicular distance from the point to the corresponding side of the triangle. (Recall that the sum of those perpendicular distances from any point inside an equilateral triangle is always equal to the altitude of the triangle.)

As we first discussed in Sec. \ref{apriori}, we shall not need to make any assumptions about the structure of the CKM matrix beyond it not having accidental zeros.  This removes any distinction between the worlds described by points in the right and left halves (related by $d\leftrightarrow s$) of the triangle of Fig.~\ref{triangles}(a), so we will \emph{define} the $s$ quark to be the more massive of the two charge -1/3 quarks and (when appropriate) restrict our attention to the reduced landscape shown in Fig.~\ref{triangles}(b).  The triangular landscapes relevant for quark charges $(2/3,2/3,-1/3)$ can be obtained by the replacements:  $m_{s}\to m_{c}, m_{d}\to m_{u}, m_{u}\to m_{d}$ with the convention that $m_{c}>m_{u}$. 

The full space of light quarks is, of course, three-dimensional, and may be represented as a triangular prism, as in Fig. \ref{prism}.  The altitude of any triangular slice is $m_T$, the sum of the three light quark masses.  Remember that the $\Lambda_{\rm QCD}$ has been adjusted so that the average mass of the lightest flavor multiplet of baryons is held fixed at 940 MeV. In practice this multiplet is always a {flavor} doublet. The prism shown in Fig.~\ref{prism} summarizes the range of universes explored in this paper.

The variation of baryon masses and nuclear forces over the prism is influenced by the adjustments of $\Lambda_{\rm QCD}$.  Those adjustments are described in the following subsection, and the resulting variation of baryon masses is discussed in Sec. \ref{baryonmasses}.

\subsection{Strength of the nuclear interaction}
\label{range}

As described in Sec. \ref{overview}, our aim is to explore universes with at least one and at most three light quarks, in which nuclear physics resembles ours as closely as possible.  In this section we discuss more carefully the extent to which this can be accomplished.

\begin{figure*}[t]
\centering
\subfigure[]{\includegraphics[width=0.2\textwidth]{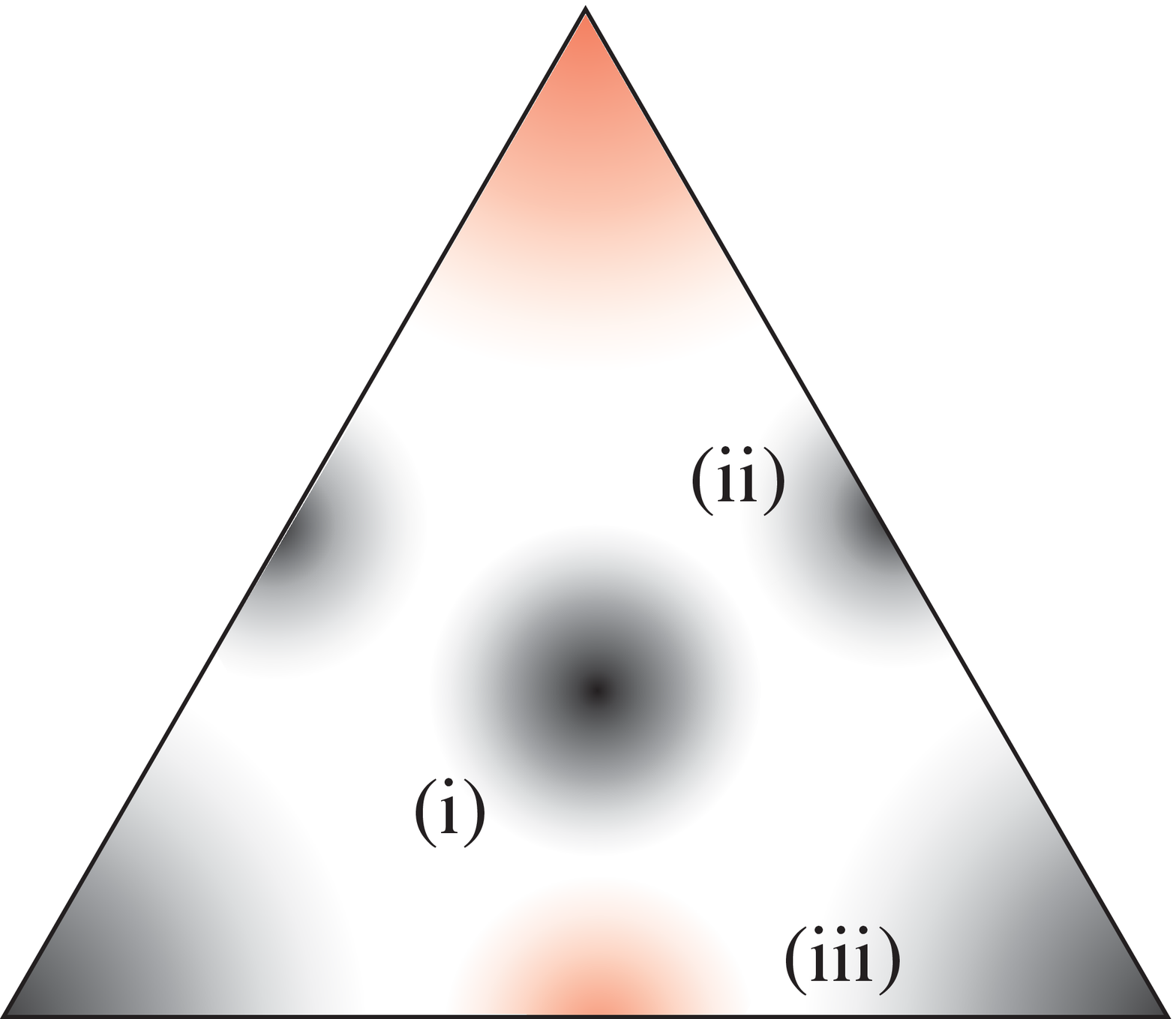}}\hspace*{2cm}
\subfigure[]{\includegraphics[width=0.1\textwidth]{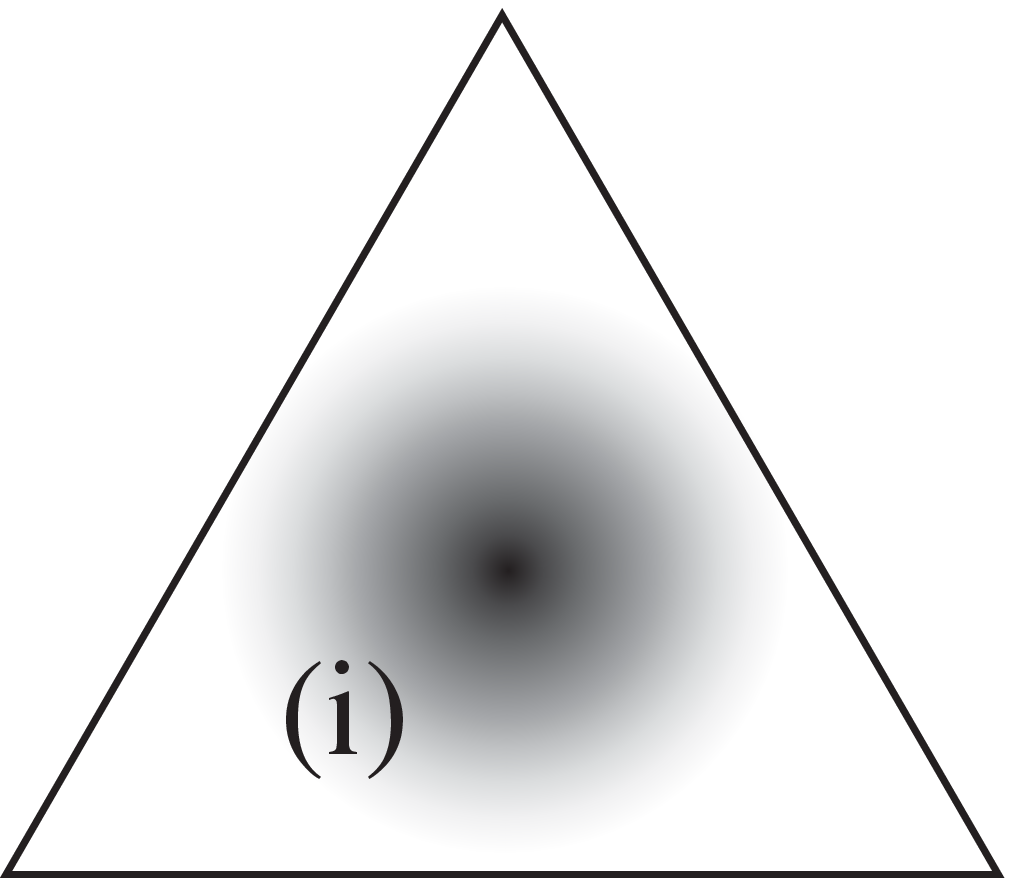}}
\caption{\footnotesize (a). Three regions in quark mass parameter space in which complex nuclei form.  The sum of quark masses here is large compared to the energy scale of nuclear binding.  The red shaded regions are not congenial because both light baryon species have the same charge (see Sec.~\ref{congenial}).  (b). In a slice where the sum of quark masses is smaller, the peripheral regions disappear, leaving only the central region where all quark masses are small compared to nuclear binding energies.}
\label{regions}
\end{figure*}

We find three regimes where long lived nuclei can form: 
\begin{enumerate}
\item [(i)] Universes where all three quarks are very light compared to $\Lambda_{\rm QCD}^\oplus$; 
\item [(ii)] universes where one quark is much lighter than the other two, which are approximately degenerate; and  
\item [(iii)] universes similar to ours where two quarks are considerably lighter than the third.  
\end{enumerate}
The regions are sketched on the quark mass triangle in Fig.~\ref{regions}.  These regions are only well separated when quark masses are large.  When all quark masses are small compared to the $\sim 10$ MeV scale of nuclear binding, then only case (i) can occur.  This is shown in Fig.~\ref{regions}(b).  Other universes outside of regions (i), (ii), or (iii), for example those with a hierarchy like\, $m_{1}\ll m_{2} (\sim 10-20\,\text{MeV}) \ll m_{3} < \Lambda_{\rm QCD}^\oplus$ have no stable nuclei.
  
In universes of type (i) all eight pseudoscalar Goldstone bosons ($\pi$, $K$, and $\eta$) participate in nuclear interactions.  This makes it impossible to describe nuclei quantitatively in those worlds based on information from ours.  In universes of type (ii) there is an approximate flavor $SU(2)$ symmetry, but unlike our world, the lightest pseudoscalar Goldstone boson is a flavor singlet and there is a complex doublet analogous to the $K/\bar K$ system that is not much heavier.  Again, nuclear physics in these worlds will differ quantitatively from ours.  It is still possible to make qualitative statements about nuclear physics in these regions because the qualitative features of baryon-baryon forces should not change:  they are short range, attractive at long distances and strongly repulsive at short distances.

In the last case, (iii), the lightest pseudoscalar bosons form a triplet under the $SU(2)$ symmetry associated with the two lightest quarks.  In this very important case, nuclear physics is quantitatively similar to our world over a significant range of quark masses.  Without loss of generality, we can call the two lightest quarks $u$ and $d$ and the heavier one $s$.  The rest of this subsection is given over to a discussion of case (iii).

Nuclear forces depend on the masses and coupling constants of the mesons that couple strongly to the light baryons.  In case (iii) this includes the pion, which gives the long range tail of the attractive nucleon-nucleon force, the $f_{0}(600)$, also known as the $\sigma$, which is a short hand for the correlated two pion exchange forces that generate the intermediate range attraction in the nucleon-nucleon force \cite{brown}, and the $\rho(770)$ and $\omega(783)$, which generate the short range repulsion.  The same changes of parameters that keep the mass of the nucleon fixed over regions (iii) of the quark mass prism, tend also to keep the masses of hadrons that scale linearly with quark masses fixed as well, thereby reducing the effect of quark mass variations on the nature of the nuclear force.

If it were not for the pion, we would argue that non-strange meson masses and coupling constants are so insensitive to the small variations in the $u$ and $d$ quark masses, that the variation of the nuclear force could be ignored over the entirety of type (iii) regions.  The pion mass, however, varies like  the \emph{square root} of the average $u$ and $d$ quark masses, $\hat m = (m_{u}+m_{d})/2$, and therefore changes more significantly over the prism.  For example, as $\hat m \to 0$  the pion masses get very small, although the $\rho$, $\omega$, and $f_{0}(600)$ masses hardly change at all (recall that we are adjusting $\Lambda_{\rm QCD}$ to keep $M_{N}$ fixed as we send $\hat m\to 0$.)  

Let us make this more precise:  We are only interested in quark masses that are small enough compared to $\Lambda_{\rm QCD}$ so that hadron masses, other than those of the pseudoscalar Goldstone bosons, can be accurately estimated by first order perturbation theory,
\begin{equation}
\label{hadmass}
M_\alpha \approx a_\alpha \Lambda_{\rm QCD} + b_\alpha \hat m + c_\alpha m_{s}
\end{equation}
for hadron $\alpha$.  Here $a_\alpha$, $b_\alpha$, and $c_\alpha$ are dynamically determined constants that are independent of quark masses. The triplet of pseudoscalar bosons, which we will call pions, are different.  Their masses vary dramatically for small quark masses, as dictated by the Gell-Mann--Oakes--Renner relation \cite{GOR},\footnote{In this discussion we ignore electromagnetic mass differences.}
\beq
m_\pi = \frac{\sqrt{ \hat m\left| \langle \bar q q \rangle \right| }}{f_\pi}~,
\label{mpi}
\eeq 
where the pion decay constant  $f_\pi$ and the chiral-symmetry breaking vev of QCD $\left\langle \bar q q \right\rangle$ are believed to be nearly independent of light quark masses.  So schematically,
\begin{equation}
\label{mpi2}
m_\pi = d_\pi \sqrt{\Lambda_{\rm QCD} \, \hat m}~.
\end{equation}

It is convenient to express the hadron and quark masses in terms of their values in our world,
\beqa
&&y_{u,d,s}\equiv m_{u,d,s}/m^\oplus_{u,d,s},\quad\hat y \equiv \hat m/\hat m^\oplus,\quad y_{T} \equiv m_{T}/m_{T}^\oplus\nonumber\\
&& y_{N} \equiv M_{N}/M_{N}^\oplus, \quad y_{\Lambda}\equiv \Lambda_{\rm QCD}/\Lambda_{\rm QCD}^\oplus, \quad y_\pi \equiv m_\pi / m_\pi^\oplus~.
\label{e.an.1}
\eeqa
Then
\begin{eqnarray}
\label{e.an.1.0}
m_{\pi} &=& \sqrt{y_\Lambda \hat y}\,\,m_\pi^\oplus ~, \\
\label{e.an.1.1}
M_{N}&=  & (783 \mbox{~MeV})\,y_\Lambda + (35 \mbox{~MeV})\,\hat y + (124 \mbox{~MeV})\,y_{s} ~, \\
m_\rho&=& m^\Lambda_\rho y_\Lambda + \hat m_\rho \hat y + m^s_\rho y_s ~.
\label{e.an.2}
\end{eqnarray}
The mass of the $\rho(770)$, a representative light meson, is decomposed in a similar way, with unknown coefficients.  We have borrowed the numerical coefficients in \Eq{e.an.1.1} from the fit performed in Sec.~\ref{baryonmasses}.

The numerical coefficients in \Eq{e.an.1.1} reflect the small contribution to the nucleon mass from $u$ and $d$ quarks in our world (35 MeV), the rather surprisingly large---though very uncertain---contribution from the $s$ quark (124 MeV), and the dominant QCD contribution (783 MeV).  We are interested in how meson masses vary for values of $\hat m$ and $m_{s}$ that remain in region (iii), where $\hat m\ll m_{s}$.  As $\hat m$ moves away from its value in our world, $\Lambda_{\rm QCD}$ must be adjusted to keep the nucleon's mass constant at 940 MeV.  The masses of non-chiral mesons like the $\rho(770)$ and $f_{0}(600)$ can be expected to vary rather little as $\hat m$ changes because their proportional dependence on $\Lambda_{\rm QCD}$ and $\hat m$ should be similar to the nucleons'.

The pion mass scales differently with $\hat m$ and $\Lambda_{\rm QCD}$, and it cannot be expected to remain nearly constant as $\hat m$ varies. 
So the extent to which nuclear physics changes over the quark mass parameter space of interest (type (iii) regions in the prism) can be examined in two stages:  (a) how much does the pion mass change? and (b) how important is the mass of the pion in determining the parameters of nuclear binding?  The first question can be answered quantitatively, but the second remains a matter of speculation.

To answer the first question it is helpful to combine the \Eqs{e.an.1.0} and (\ref{e.an.1.1}),
\begin{equation}
M_{N} =  (783\,\mbox{MeV})\frac{y_{\pi}^{2}}{\hat y}+(35\,\mbox{MeV})\,\hat y + (124\,\mbox{MeV})\,y_{s} ~.
\label{e.an.2.1}
\end{equation}
It is clear from this equation that there is a subspace of the quark mass parameter space on which both the nucleon and the pion mass remain fixed as $\hat m$ and $m_{T}$ vary.  To display this constraint more usefully, we set $M_{N}=M_{N}^\oplus= 940$ MeV,  and substitute values of quark masses from our world, where $\hat m^\oplus \approx 3.6$ MeV and $m_{s}^\oplus \approx 93$ MeV\footnote{The errors on the quark masses are very large, but since this analysis is only intended to yield qualitative conclusions we will ignore them in this section.} into \Eq{e.an.2.1}, and switch variables to $\hat m$ and $m_{T}$,
\begin{equation}
940\,\mbox{MeV} = \frac{(2820\,\mbox{MeV}^{2})y_{\pi}^{2}}{\hat m}+ 7.0 \,\hat m + 1.3\, m_{T}~,
\label{e.an.3}
\end{equation}
where $\hat m$ and $m_{T}$ are in MeV.  We leave the $y_{\pi}$ dependence in \Eq{e.an.3} explicit so we can study the constraint between $\hat m$ and $m_{T}$ as $m_{\pi}$ varies.
       
\begin{figure*}
\centering 
\subfigure[]{\includegraphics[width=0.3\textwidth]{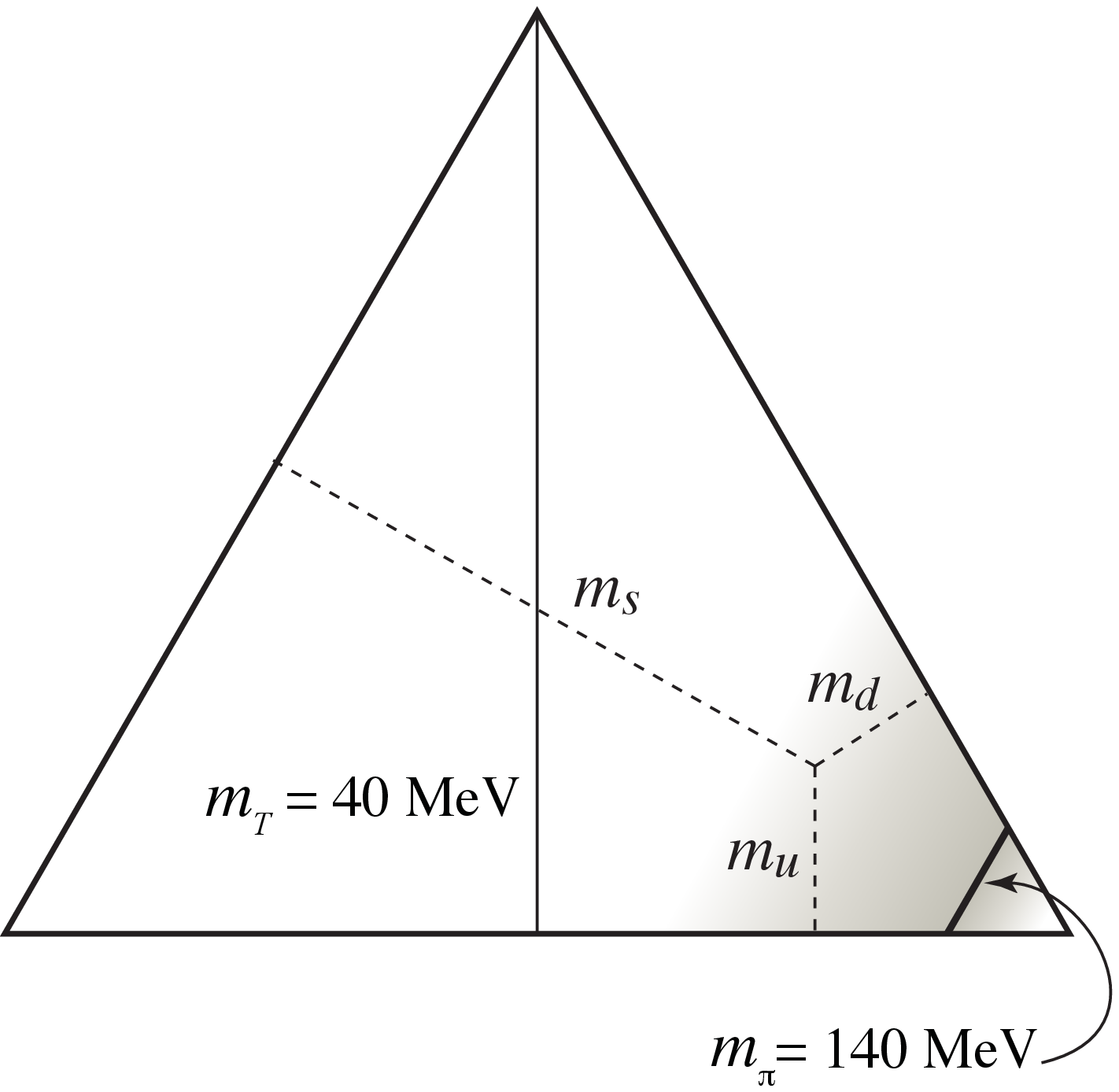}}\hspace*{2cm}
\subfigure[]{\includegraphics[width=0.38\textwidth]{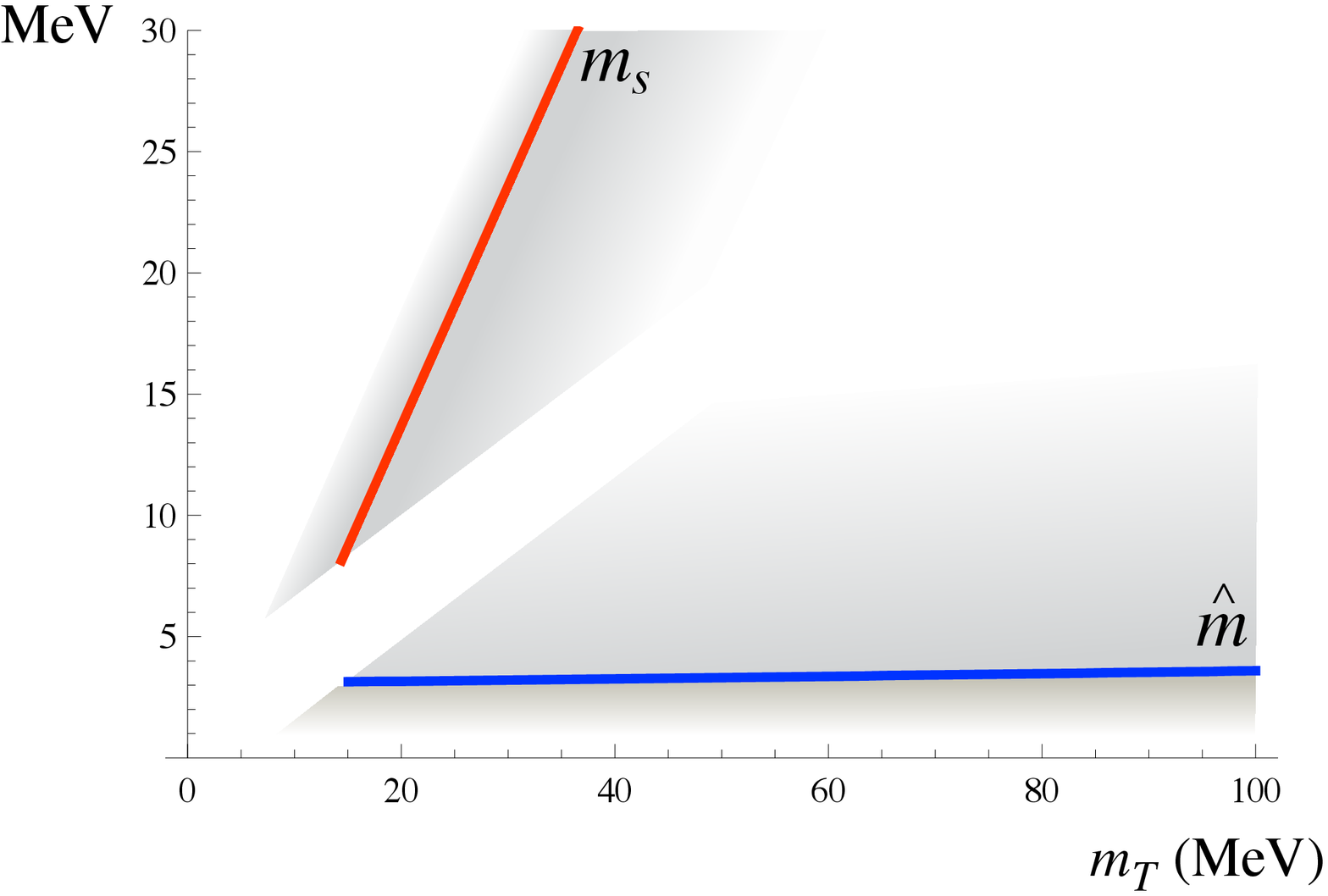}}
\caption{\footnotesize \small (a) ~The quark mass triangle for $m_{T}=40$ MeV showing the the line along which $m_{\pi}=m_{\pi}^\oplus=140$ MeV with shading corresponding to pion mass variation by a factor of two above and below its value in our world.  Note that the entire region satisfies $m_{s} \gg \hat m$.  (b)  The constraint on $\hat m$ and $m_s$ as functions of $m_T$, generated by fixing $m_{\pi}$ and $M_{N}$ to their values in our world.  As before, the shading correspond to varying $m_\pi$ by a factor of 2.  The shaded regions are truncated when $m_s - \hat m < 5$ MeV, because the corresponding worlds would have nuclei containing strange baryons.}
\label{pionmassrange}
\end{figure*}

Equation (\ref{e.an.3}) constrains the quark masses to a two dimensional subspace of the prism.\footnote{There are two solutions of \Eq{e.an.3} for $\hat m$ as a function of $y_\pi$ and $m_T$.  We can identify the relevant one by requiring $\hat m^\oplus = 3.6$ MeV for $m_T^\oplus = 100$ MeV.}   Since the pion mass depends only on $\hat m$, the constraint is independent of the \emph{difference} of the light quark masses, and defines a line parallel to one side in a fixed $m_{T}$ triangle, as shown in Fig.~\ref{pionmassrange}(a).  The subspace is also constrained by the demand that the third (in this case, the $s$) quark must be heavy enough so that strange baryons decouple from nuclear physics.  
In  Fig.~\ref{pionmassrange}(b) we plot the average light quark mass, $\hat m$, that satisfies  \Eq{e.an.3} as a function of $m_{T}$.    The value of $\hat m$ does not depart far from its value in our world because a small variation in $\hat m$ compensates for the variation in $m_{T}$ in \Eq{e.an.3}.  The shaded band in the figure shows the range of $\hat m$ over which $m_{\pi}$ varies by a factor of two.  The plot is truncated on the left by the constraint that the mass of the strange quark must be greater than $\hat m$ by $\sim 5$ MeV, to insure that the domain remains a type (iii) region.  (The choice of a 5 MeV splitting is justified by \Eq{2baryonlimit}, in Sec. \ref{3baryons}.)

If nuclear binding were dominated by single-pion exchange, \Eq{mpi} would imply quite a strong dependence of the nuclear interaction on the masses of the light quarks.  In particular, \emph{increasing} the masses of the $u$ and $d$ quarks in our universe would be expected to significantly decrease and eventually turn off nuclear binding \cite{vev-selection,hogan-review}.  It is believed, however, that the attractive forces responsible for nuclear binding receive significant contributions from correlated two-pion exchange in the region of 400 -- 600 MeV invariant mass, an effect that is usually associated with the $f_{0}(600)$ meson \cite{brown}.  The mass of the $f_{0}(600)$ is not singular in the chiral limit, nor does it vary dramatically when the pion mass is increased well above its value in our world \cite{pelaez}, so there is reason to believe that nuclear binding depends less dramatically on light quark masses than the single pion exchange mechanism would lead one to believe. 

Most studies of the effect on the nuclear interaction of changing the pion mass have focused on the $S$-wave nucleon-nucleon scattering lengths and on the related problem of deuteron binding.  Various attempts have been made to compute these numerically, either by solving the Schr\"odinger equation with approximate potentials \cite{martin, epelbaum} or from lattice QCD calculations \cite{aoki}.  The exact dependence of deuteron binding on the quark masses, however, still remains uncertain.\footnote{Another approach to studying the dependence of nuclear binding on the pion mass is to consider the analytic structure of nucleon-nucleon scattering amplitudes and to study the effects---on the corresponding dispersion integrals---of varying the pion mass \cite{donoghue, damour}.  This approach might be subtle because, as was explored in \cite{ineffective} in a different context, the expectation from effective field theory---that low-energy scattering is controlled by the nearest singularity in the amplitude (and therefore, in this case, by the pion mass)---can fail.  For the $S$-wave channels, the effective nucleon-nucleon contact interactions at low energies are {\it not} of order $1 / m_\pi^2$, as naive effective field theory might suggest, but instead are significantly larger (as can be seen from the size of the absolute values of the corresponding scattering lengths).  The presence (or absence) of a spin-1 nucleon-nucleon bound state (the deuteron) could, through Regge pole theory, alter the asymptotic behavior of the dispersion integrals that connect the low-energy scattering amplitude to the poles and branch points set by the pion mass \cite{Adams:2008hp}.}

In our world, the deuteron plays an important role in primordial nucleosynthesis, but it is unclear whether a deuteron bound state is necessary in all worlds to ensure that heavy elements eventually become abundant enough to support intelligent life (as we have noted in Sec. \ref{introduction}, tritium might serve as an acceptable form hydrogen and its binding is more robust than the deuteron's).  For this reason, an in keeping with the philosophy of this investigation as outlined in Sec. \ref{introduction}, we will {\it not} focus specifically on the deuteron binding, but rather with the stability of any isotopes of hydrogen, carbon, and oxygen.

Based on general considerations of dimensional analysis, we expect that all of the mass scales $M_i$ relevant to nuclear physics---including the masses of the baryons as well as the masses of the $f_0(600)$, $\rho(770)$, and $\omega(783)$ resonances---should vary roughly as
\beq
\frac{\Delta M_i}{M_i^\oplus} \sim \frac{\Delta m_q}{\Lambda_{\rm QCD}^\oplus}~,
\label{deltaM}
\eeq
where $\Delta m_q$ is the quark mass variation being considered.\footnote{In fact, investigations by \cite{bruns} and \cite{gockeler} suggest that, for the mass of the $\rho(770)$, non-linearities in $\Delta m_{u,d}$ are relevant near the $m_u = m_d = 0$ chiral symmetry point, but these non-linearities actually make the quark mass dependence of $M_\rho$ {\it milder} than \Eq{deltaM} would predict.}  The impact of the baryon mass variations on nuclear structure, however, will be much enhanced compared to the effect of the variation of other mass scales.  In fact, if it were not for the sensitivity of nuclear structure to the baryon masses, the breaking of $SU(2)$ and $SU(3)$ flavor symmetries by the strong interaction would be largely irrelevant to a nuclear physicist.

As an illustration of the exquisite sensitivity of nuclear structure to the baryon masses, consider the fact that (as we shall work out in detail in Sec. \ref{3baryons}) the  $\Lambda$ baryon could be as little as 20 MeV heavier than the proton and the neutron and still not participate in stable nuclei.  Thus, a breaking of flavor $SU(3)$ by a mere 2\% in the baryon masses would leave almost no trace of that symmetry in the composition of stable nuclei.  It is therefore lucky that baryon masses can be reliably estimated as functions of the light quark masses using first-order perturbation theory in flavor $SU(3)$ breaking (as we shall work out in Sec. \ref{baryonmasses}) and that the sensitivity of nuclear structure to baryon masses can be qualitatively understood by modeling the nucleus as a free Fermi gas subject to a confining pressure (as we shall explain in Sec. \ref{s4.22}).  Thus, in this paper we shall retain the effect of quark mass variations on the baryon masses, but we shall mostly assume that we can ignore the variation in the strength of the nuclear interactions due to the relatively small changes to the quark masses that we consider.

%%%%%%%%%%
%%% OCTET BARYON MASSES
%%%%%%%%%%
\section{Octet baryon masses}
\label{baryonmasses} 

In this section we use flavor and chiral symmetry violation among the baryons in our world to find the masses of the light baryons as functions of the quark masses.  These are  used as inputs in Sec. \ref{nuclearmasses} to determine the spectrum of nuclear masses, which in turn allow us to study the congeniality of worlds with different quark masses.

In our world the lightest baryons are in the octet representation of the flavor $SU(3)$ symmetry group associated with the $u$, $d$, and $s$ quarks.  These are the nucleons (proton and neutron), the $\Sigma$'s, the $\Xi$'s, and the $\Lambda$.  Quark mass effects (which break flavor $SU(3)$) and the electromagnetic interaction give rise to mass differences within the octet.  Electromagnetic contributions to the masses are small and calculable \cite{CG_EM}.  They can be subtracted from the total baryon mass to leave only the contributions from QCD and from quark mass differences.

To first order, the quark mass contributions are proportional to the reduced matrix elements of $SU(3)$ tensor operators.  These matrix elements can be extracted from the measured mass differences in our world.  These can then be used to compute the masses of the octet baryons in worlds with quark masses that differ from those in our world.  One complication is that the analysis of baryon mass \emph{differences} does not enable us to estimate the matrix element of the flavor \emph{singlet} operator.  This, however, can be extracted from the analysis of chiral symmetry violation in the baryon sector, through the well-known $\sigma$-term analysis.

As we pointed out in Sec. \ref{variation}, $SU(3)$ flavor symmetry will be only very mildly broken in most of the worlds studied in this paper, and first-order perturbation theory will therefore be an excellent approximation.  Higher-order corrections principally affect the extraction of the $SU(3)$ reduced matrix elements from the measured baryon masses.  However, we shall neglect such corrections in light of the level of accuracy that is called for in this kind of investigation.

\subsection{Flavor $SU(3)$ and baryon masses}
\label{flavorSU3}

\begin{wrapfigure}{r}{0.45\textwidth}
\centering
\includegraphics[width=0.3\textwidth]{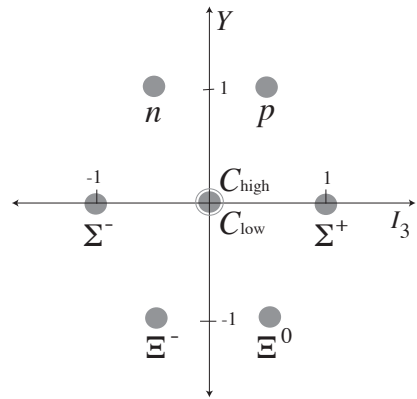}
\caption{\footnotesize The $SU(3)$ weight diagram for the octet baryons showing their conventional names and their assignment of hypercharge $Y$ and the third component of isospin $I_{3}$.  The two central states, denoted $C_{\rm high}$ and $C_{\rm low}$, mix at lowest order in $SU(3)$ perturbation theory.}
\label{octet}
\end{wrapfigure}
 
The mass of a baryon is given by the eigenvalue of a Hamiltonian $H$, which we may decompose as
\beq
H = H_{\rm QCD}  + H_{\rm flavor} + H_{\rm EM}~,
\label{baryonH}
\eeq
where $H_{\rm QCD}$ corresponds to QCD interactions and is independent of quark masses and electric charges, $H_{\rm flavor}$ is the quark mass dependent piece
\begin{equation}
H_{\rm flavor} = \sum_{i}m_{i}\bar q_{i}q_{i}\, ,
\label{e.an.4}
\end{equation}
and $H_{\rm EM}$ gives the electromagnetic contribution to mass of the baryon. $H_{\rm EM}$ is the only part of $H$ that depends on the electric charges of the component quarks.  Using the renormalized Cottingham formula \cite{cottingham,gunion}, it is possible to relate baryon matrix elements of $H_{\rm EM}$  to experimental data for electron-baryon scattering.  Data is only available for electron-nucleon scattering, but there is evidence that $\langle H_{\rm EM}\rangle$ is dominated by the Born terms which in turn are dominated by the charge and magnetic moment contributions, which are known.  Here we use the values  as quoted in Gasser and Leutwyler in \cite{quarkmasses}.\footnote{Ref.~\cite{quarkmasses} does not list an electromagnetic contribution to the $\Lambda$ mass.  We have computed it using their methods.}  They are tabulated in Table~\ref{t.an.1}.
 
The quark mass dependent term in $H$ can be decomposed in terms of irreducible tensor operators, 
\beq
H_{\rm flavor} =  m_{0}\Theta_{0}^{1} + m_{3}\Theta_{3}^{8} + m_{8}\Theta_{8}^{8}\,,
\label{Hoctet}
\eeq
with
\begin{align}
&\Theta_{0}^{1} = \sqrt{\frac{2}{3}} \left( \bar u u + \bar d d + \bar s s  \right), ~~
& &\Theta_{3}^{8} = \bar u u - \bar d d, ~~
& &\Theta_{8}^{8} = \sqrt{\frac{1}{3}} \left( \bar u u + \bar d d - 2 \bar s s \right) \nonumber\\
&m_{0} = \frac{1}{\sqrt{6}} (m_{u} + m_{d} + m_{s}), ~~
& &m_{3} =  \frac{1}{2}(m_{u} - m_{d}), ~~
& &m_{8} = \frac{1}{2\sqrt{3}} (m_{u} + m_{d} -2 m_{s})~. 
\label{thetas}
\end{align}

The Hamiltonian, \Eq{Hoctet} violates $SU(3)$ symmetry but commutes with the generators $Q_{3}$ and $Q_{8}$.\footnote{The subscripts for the generators follow the usual convention for the Gell-Mann matrices.}  Therefore, no matter how light quark masses vary, we may label octet baryons by the eigenvalues of the third component of isospin, $I_{3}$, and of the hypercharge, $Y$ (which is the sum of baryon number plus strangeness).   Total isospin is approximately conserved when $m_{u}=m_{d}$ (and electromagnetic effect are ignored) and can be used to label baryon states in our world. However, when light quark masses differ arbitrarily, there is no conserved $SU(2)$ subgroup of $SU(3)$, and total isospin loses its utility.  All of the baryons at the periphery of the $SU(3)$ weight diagram in Fig.~\ref{octet} have different $I_{3}$ and $Y$ and therefore do not mix with one another for any quark masses.  However, the two states at the center of the octet weight diagram have the same $I_{3}$ and $Y$, and can mix.  In our world, the resulting eigenstates---the $\Sigma^{0}$ and the $\Lambda$---are approximate eigenstates of total isospin.  In other worlds they will not be close to eigenstates of any $SU(2)$ subgroup of $SU(3)$ unless two light quarks have approximately equal masses.  In general we label the mass eigenstates at the center of the weight diagram $C_{\rm high}$ and $C_{\rm low}$.

It is a straightforward exercise in first order perturbation theory to compute the effect of the Hamiltonian, \Eq{Hoctet} on the spectrum of octet baryons.  For completeness, the analysis is presented in Appendix~\ref{a.an.1} and its results are summarized in Table~\ref{t.an.1}.  The key point is that the Wigner reduced matrix elements (WRME's), conventionally labeled $F$ and $D$, can be extracted from octet baryon mass differences in our world and provide the necessary dynamical information for computing octet baryon mass differences in any of the other worlds we study.
  
\subsection{Fitting parameters to the baryon mass spectrum}

\begin{table}[bt]
\begin{center}
\begin{tabular}{|l|c|c|c|c|c|c|}
\hline
Baryon  & Flavor-dependent & Electromagnetic & Experimental & Fitted & Residual \\ [-1.5ex] 
species & contribution & correction (MeV) & (MeV) & (MeV) & (MeV) \\ \hline
\,\,$p$ & $\left(\frac{3F-D}{\sqrt{3}}\right)m_{8} + \left(F+D\right)m_{3}$& $ +0.63$ &\,\,938.27 &  \,\,939.87& \,\,$1.60$\\
\,\,$n$  & $ \left(\frac{3F-D}{\sqrt{3}}\right)m_{8} - \left(F+D\right)m_{3}$ &$-0.13$  &\,\,939.57 &  \,\,942.02 &\,\,$2.45$ \\
\,\,${\Xi^{0}}$  & $- \left(\frac{3F+D}{\sqrt{3}}\right)m_{8} + \left(F-D\right)m_{3}$ &$-0.07$ & \,\,1314.83 &  \,\,1316.81 &\,\,$1.98$   \\
\,\,${\Xi^{-}}$ &$ - \left(\frac{3F+D}{\sqrt{3}}\right)m_{8} - \left(F-D\right)m_{3} $ &$+0.79 $ &\,\,1321.31 & \,\,1323.38 &\,\,$2.07$   \\
\,\,${\Sigma^{+}}$ & $\left(\frac{2D}{\sqrt{3}}\right)m_{8} + 2F \, m_{3}$ &$ +0.70$  &\,\,1189.37 & \,\,1188.42&\,\,$-0.96$ \\
\,\,${\Sigma^{-}}$ & $\left(\frac{2D}{\sqrt{3}}\right)m_{8} - 2F \, m_{3}$ & $+0.87$ & \,\,1197.45 & \,\,1197.21 &\,\,$-0.24$ \\
\,\,${C_{\rm high}}$ & $\left(\frac{2D}{\sqrt{3}}\right)\sqrt{m_{8}^{2} + m_{3}^{2}} $ &$-0.21$     & \,\,1192.64  & \,\,1191.82 &\,\,$-0.82$ \\
\,\,${C_{\rm low}}$ &$ - \left(\frac{2D}{\sqrt{3}}\right)\sqrt{m_{8}^{2} + m_{3}^{2}}$ &$ +0.21$  & \,\,1115.68  & \,\,1109.61 & \,\,$-6.07$\\
\hline
\end{tabular}
\end{center}
\caption{\footnotesize \small Contributions to the octet baryon masses in our world.  The flavor-independent contribution is $A_0 \equiv \langle H_{\rm QCD} \rangle + m_{0}\braketnoline{\Theta^1_0}$.  The quark mass parameters $m_{0,3,8}$ are defined in Eq. (\ref{thetas}).}
\label{t.an.1}
\end{table}
 
In order to extract the values of the parameters $F$ and $D$, we fit the matrix elements of $H_{\rm flavor}$ to the measured octet baryon masses after subtracting the electromagnetic mass contributions. In our universe $\sqrt{m_{8}^{2} + m_{3}^{2}} = 1.0002 \ m_{8} \approx m_{8}$, so the quark masses $m_{3}$ and $m_{8}$ always appear in combination with the WRME's $F$ and $D$. We choose  
\begin{equation}
\label{e.an.5}
 F \, m_{8}\,,\quad D \, m_{8}\,,\quad  \frac{m_3}{m_8} \,,\quad \mbox{and}\quad A_{0} \equiv \langle H_{\rm QCD} \rangle + m_{0}\braketnoline{\Theta^1_0}
\end{equation}
as our fit parameters.

There are eight baryon masses to be fit in terms of our four parameters.  One linear combination of baryon masses is the Gell-Mann--Okubo relation and is independent of our parameters.  $A_{0}$ is simply the average octet baryon mass.  This leaves six mass differences to be fit by three parameters, $m_{8}F$, $m_{8}D$, and $m_{3}/m_{8}$.  To keep a strong connection to chiral dynamics we take the ratio $m_{3}/m_{8}$ from the analysis of pseudoscalar meson masses rather than from our fit to baryon mass differences.  According to the Particle Data Group (PDG),
\beq
\frac{m_u}{m_d} = 0.56 \pm 0.15;  ~ \frac{m_s}{m_d} = 20.1 \pm 2.5~,
\label{PDGratios}
\eeq
from which the PDG extracts $m_{3}/m_{8} =  0.0197 \pm 0.0071$ MeV \cite{PDG}.   If we fit $m_{3} / m_{8}$, we obtain $m_{3}/m_{8} =  0.0181$ MeV, which is quite close to the value quoted by the PDG.  Note that we have not attempted to do a careful error analysis since the uncertainties in our theoretical modeling assumptions exceed the uncertainties in the parameters that we have extracted from data in our world.  The values of the parameters of \Eq{e.an.5} that we use in our analysis are given in Table~\ref{params}. 

\begin{table}[tb]
\centering
\begin{tabular}{|c|c|} \hline
Parameter &  Fitted value
\\ \hline 
$A_{0}$ &    $1150.8$  MeV  \\ \hline
$F  m_{8}$&$ -109.4$  MeV \\ \hline
$D m_{8}$ & $35.7 $ MeV\\ \hline
${m_{3}/m_{8}}$ & 0.0197	\\
\hline		
\end{tabular}
\caption{\footnotesize $SU(3)$ linear perturbation theory parameters.}
\label{params}
\end{table}

\subsection{The matrix element of $m_{0}\langle \Theta^{1}_{0}\rangle$ and the pion-nucleon $\sigma$ term}
\label{sigmaterm}

Analysis of flavor $SU(3)$ symmetry breaking cannot give us information about either the average quark mass or the matrix element of the $SU(3)$ invariant operator $\Theta^{1}_{0}$.  However, $SU(2)\times SU(2)$ chiral symmetry violation can.  It is well known that a measure of chiral symmetry violation in the baryon sector known as the pion-nucleon sigma term, $\sigma_{\pi N}$, can be extracted (with considerable theoretical input) from experiment, and can be related to the matrix element of $m_{0}\Theta^{1}_{0}$.  This is not the place for a critical review of the (still somewhat controversial) analysis of the chiral dynamics and pion nucleon scattering that leads to a value for $\sigma_{\pi N}$ \cite{Gasser:1990ce}.  Instead we briefly quote the results of the standard analysis of Gasser and Leutwyler in \cite{quarkmasses} and refer the reader to their review for further details.  The extraction of the $\sigma$ term from low energy $\pi N$ scattering remains controversial.  For more recent evaluations, see \cite{Borosoy} and \cite{GWU}.

The $\sigma_{\pi N}$ term is defined as the matrix element of the $SU(2)\times SU(2)$ violating scalar density,
\begin{eqnarray}
\sigma_{\pi N} &=& \frac{m_{u}+m_{d}}{2} \braket{N}{\bar u u + \bar d d}{N}\nonumber\\
&=&  \braket{N}{m_{u}\bar u u + m_{d}\bar d d}{N}-\frac{1}{2}(m_{u}-m_{d})
\braket{N}{\bar u u - \bar d d}{N}\nonumber\\
&\approx&\braket{N}{m_{u}\bar u u + m_{d}\bar d d}{N}
\label{e.an.6}
\end{eqnarray} 
where $|N\rangle$ is shorthand for the average of proton and neutron matrix elements.  In going from the second to the third line, we have dropped a term that is second order in $SU(2)$ violation, and therefore is negligible in our world.  The combination $m_{u}\bar u u + m_{d}\bar d d$ can be written as a linear combination of the operators $m_{0}\Theta ^{1}_{0}$ and $m_{8}\Theta^{8}_{8}$.  Since the matrix elements of the latter are determined by baryon mass differences, knowledge of the $\sigma$-term fixes the matrix element of $m_{0}\Theta^{1}_{0}$.
  
Rewriting $\sigma_{\pi N}$ in terms of irreducible tensor operators, we have,
\beq
\sigma_{\pi N} = \frac{\sqrt{2}m_{0} + m_{8}}{3} \left \langle N \left|  \sqrt{2} \Theta^1_0 + \Theta^8_8 \right| N \right \rangle ~.
\label{e.an.7}
\eeq
Using the known matrix elements of $\Theta^8_8$ we obtain
\beq
\sigma_{\pi N} = \left( 1 + \sqrt{2}\frac{m_{0}}{m_{8}} \right)   \left[ \frac{\sqrt{2}}{3}\frac{m_{8}}{m_{0}}  \braket{N}{ m_{0}\Theta^1_0}{N}  + \frac{m_{8}}{\sqrt{3}}\left(F   - \frac{D}{3}\right) \right] ~.
\label{e.an.8}
\end{equation}

Given a value for $\sigma_{\pi N}$, we can use the $F m_{8}$ and $ D m_{8}$ fit values and $m_{0}/m_{8}$ derived from the PDG quark mass ratios to solve for the singlet flavor mass term $\braket{N}{ m_{0}\Theta^1_0}{N}$, thereby providing the needed separation of the average mass of the   octet baryons, $A_{0}$, into a $QCD$ component that does not scale with quark masses and this flavor singlet component that scales with the sum of quark masses $m_{0}$. 

Gasser and Leutwyler conclude that the value of $\sigma_{\pi N}$ to be used in \Eq{e.an.8} is $\sigma_{\pi N}= 35\pm 5$ MeV.  Substituting into \Eq{e.an.8} and keeping track of the experimental uncertainties, we obtain
\begin{equation}
m_{0}^\oplus\braketnoline{\Theta^1_0} = 368 \pm 101  \text{~MeV} ~.
\end{equation}

Using the values of the matrix elements of the other scalar densities, we can separate the contributions of the average $u$ and $d$ quark masses and the $s$ quark mass,
\begin{eqnarray}
\label{e.an.8.1}
\langle N|\hat m^\oplus(\bar uu+\bar d d)|N\rangle &=& 35\pm 5\,\,\mbox{MeV}\nonumber\\
\langle N| m_{s}^\oplus\bar ss|N\rangle &=& 124\pm 94 \text{~MeV}~,
\end{eqnarray}
which were the values quoted in \Eq{e.an.2.1}.

Having determined $m_{0}\langle\Theta^{1}_{0}\rangle$, albeit with large errors, we can separate the QCD dynamical mass out of the the flavor independent term in $\langle H\rangle$,
\begin{equation}
\langle H_{\rm QCD}\rangle \,\,\equiv C_{0} = A_{0}-m_{0}\langle \Theta^{1}_{0}\rangle = 783 \pm 101\,\,\mbox{MeV}
\label{e.an.8.2}
\end{equation}
as quoted in \Eq{e.an.2.1}.  Although the uncertainties in $ \langle N|m_{0}\Theta^{1}_{0}|N\rangle$, and especially in $\langle N| m_{s}^\oplus\bar ss|N\rangle$ are large, the fact that $\langle H_{\rm QCD}\rangle$ and $\langle m_{0}\Theta^{1}_{0}\rangle$ always appear in the combination $A_{0}$ makes the analysis relatively insensitive to these uncertainties.

It is well known that higher-order corrections in $m_s$ are important to the baryon mass spectrum in our world.  Such non-linearities mostly affect the flavor-singlet contribution to the baryon masses---as evidenced by the fact that the Gell-Mann--Okubo formula is accurate to within 1\%---and are therefore primarily relevant to the issue of extracting of the $\sigma_{\pi N}$-term \cite{higher-ms,Gasser:1990ce}.  Since we have set up the problem for this investigation in such a way that we are largely insensitive to the uncertaintly in the value of $\sigma_{\pi N}$, nonlinearities in $m_s$ will not affect our results significantly.  Notice also that the central values for the quantities in Eqs. (\ref{e.an.8.1}) and (\ref{e.an.8.2}) agree with those given by Borosoy and Meissner \cite{Borosoy}, who take non-linearities in $m_s$ into account.

\subsection{Coordinates in the space of light quark masses}
\label{coordinates}

Let us introduce dimensionless coordinates $x_T$, $x_3$ and $x_8$ to describe the space of quark masses as pictured in Fig. \ref{prism}.  To sidestep the problem that quark masses are not very accurately measured (and that they are renormalization-scale dependent), we will always work with quark masses and quark mass combinations as ratios to $m_T^\oplus$ (the sum of the values of the light quark masses in our universe).  With respect to the $m_{0}$, $m_{3}$, and $m_{8}$ defined in Eq. (\ref{thetas}):
\beqa 
x_T &\equiv&  m_T \times \frac{100}{m_T^\oplus} = \sqrt 6 \, m_0 \times \frac{100}{m_T^\oplus}
= \frac{100 (m_u + m_d + m_s)}{m_T^\oplus} ~, \nonumber \\
x_3 &\equiv& \frac{2 m_3} {\sqrt 3} \times \frac{100}{m_T^\oplus}
= \frac{100 (m_u - m_d)}{\sqrt 3 \, m_T^\oplus} ~, \nonumber \\
x_8 &\equiv&  \frac{2 m_8} {\sqrt 3} \times \frac{100}{m_T^\oplus}
= \frac{100 (m_u + m_d - 2 m_s)}{3 m_T^\oplus} ~.
\label{axes}
\eeqa

\begin{figure}[tb]
\centering
\includegraphics[width=0.27\textwidth]{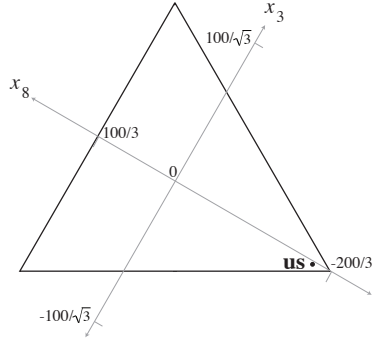}
\caption{\footnotesize A slice through the quark mass prism at fixed $x_T = x_T^\oplus \equiv 100$.  The axes $x_{3}$ and $x_{8}$, as well as $x_T$ are defined in Eq. \ref{axes}.  Our universe corresponds to the point marked ``us,'' near the lower right-hand corner.}
\label{x3x8}
\end{figure}

For a given $x_T$ ({\it i.e.\/}, for a fixed sum of the light quark masses) the resulting two-dimensional space of quark masses may be pictured as the interior of an equilateral triangle with altitude $x_{T}$.  Choices of quark masses may be described by the orthogonal coordinates $x_{3,8}$ as shown in Fig. \ref{x3x8}, with ranges
\beq
-\frac{2x_T}{3} \leq x_8 \leq \frac{x_T}{3} ~; ~~ - \frac{2 x_{T} / 3 + x_{8}}{\sqrt 3} \leq x_3 \leq \frac{ 2 x_{T} / 3 +x_{8}}{\sqrt 3} ~.
\label{x3x8range}
\eeq
For our world, $x_T^\oplus \equiv 100$, $x_3^\oplus = -1.17 \pm 0.43$, and $x_8^\oplus = -59.5 \pm 1.1$.

In order to express the baryon masses as functions of the scaled variables, $x_{T}$, $x_{3}$, and $x_{8}$, we must separate the flavor $SU(3)$ invariant contribution to the baryon masses, $\overline M(x_{T})$, into the piece independent of quark masses, $C_{0}$, and the term proportional to $\langle m_{0}\Theta^{1}_{0}\rangle$, as defined in \Eq{e.an.8.2}:
\begin{equation}
\label{e.an.11}
\overline M(x_{T}) = A_{0} = C_{0}+ \frac{x_{T}}{100}\langle N|m_{0}^\oplus\Theta^{1}_{0}|N\rangle
\equiv C_{0}+c_{T}x_{T}
\end{equation}
We can now express the mass of an octet baryon $B$ as the sum of five terms:
\begin{equation}
M_B = C_{0}+ c_{T}x_{T}+c_{8}x_{8}+c_{3}x_{3}+ \langle B|H_{\rm EM}|B\rangle~.
\label{e.an.12}
\end{equation}
The coefficients $c_{T}, c_{3}$, and $c_{8}$ and $\langle H_{\rm EM}\rangle$ are given in Table~\ref{t.an.2}.  Although both $C_{0}$ and $c_{T}x_{T}$ have large uncertainties, the fact that $C_0$ (via its dependence on $\Lambda_{\rm QCD}$) is adjusted throughout the parameter space to keep the average mass of the lightest flavor multiplet fixed removes the uncertainty from the parameterization of hadron masses.  Therefore the errors for $C_{0}$ and $\langle  m_{0} \Theta^{1}_{0} \rangle$ are not quoted in the table.
 
\begin{table}[bt]
\begin{center}
\begin{tabular}{|l|c|c|c|c|c|}
\hline
Baryon &$c_T$ &$c_8$ &$c_3$ &$\langle H_{\rm EM}\rangle$ \\ [-1ex] 
 & (MeV) & (MeV) & (MeV) & (MeV) \\ \hline
\,\,$p$ & $3.68$ &  $3.53 $  & $  1.24 $ &$0.63$  \\
\,\,$n$ &$ 3.68 $&  $3.53 $  &  $-1.24 $ &$-0.13$\\
\,\,${\Xi^{0}}$  & $3.68$&     $-2.84 $  & $ 2.44 $  &$ -0.07$ \\
\,\,${\Xi^{-}}$ & $ 3.68$ & $-2.84 $   &$ -2.44$    & $0.79 $\\
\,\,${\Sigma^{+}}$ &$3.68$ &  $-0.69 $ & $ 3.68 $ &  $0.70$\\
\,\,${\Sigma^{-}}$ &$ 3.68$ &$ - 0.69$ &  $ -3.68 $   & $0.87 $\\
\hline
 \,\,${C_{\rm high}}$ & 3.68 & \multicolumn{2}{|c|}{ \quad $ 0.69  \sqrt{x_{8}^{2} + x_{3}^{2} } \quad$}& -0.21\\
 \,\,${C_{\rm low}}$ & 3.68 & \multicolumn{2}{|c|}{ \quad $-0.69  \sqrt{x_{8}^{2} + x_{3}^{2} } \quad$}& 0.21\\
\hline
\end{tabular}
\end{center}
\caption{\footnotesize \small Values of the parameters in the equation baryon mass equation \teq{M_B = C_0 + c_T x_T + c_8 x_8 + c_3 x_3 + \langle B |H_{\rm EM}|B\rangle}, for the quark mass coordinates $x_{T}$, $x_{3}$ and $x_{8}$ are defined in Eq. (\ref{axes}).  As $x_{T,3,8}$ vary, the QCD contribution, $C_0$, is adjusted to keep the mass of the lightest baryon flavor multiplet fixed  at 940 MeV.  Note that $C_0^\oplus = 783 \pm 101$ MeV.} 
\label{t.an.2}
\end{table}

\subsection{Baryon and meson spectra for a fixed sum of quark masses}
\label{perimeterspectrum}

\begin{wrapfigure}{r}{0.45\textwidth}  
\centering
\includegraphics[width=0.3\textwidth]{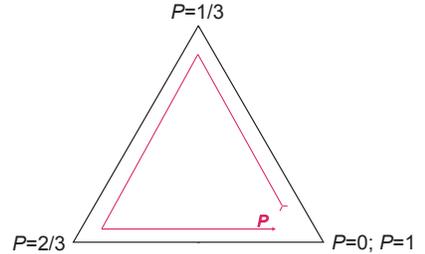}
\caption{\footnotesize The definition of the perimeter coordinate used in Figs.~\ref{b_perimeter} through \ref{spectra} is shown graphically.  Regardless of the value of $x_{T}$, the coordinate $P$ is defined to run from zero to unity around the triangle.}
\label{perimeter}
\end{wrapfigure}

As we discussed in Sec. \ref{range}, a triangular slice through the $m_{u,d,s}$ mass prism corresponding to a fixed value of $x_T$ will have a central region of type (i), where flavor $SU(3)$ is a good symmetry and all of the octet baryons can appear in nuclei.  If $x_T$ is large enough, then along the perimeter of the triangle ({\it i.e.\/}, where the quark mass differences are greatest) there will be regions of type (ii) and (iii), where only the lightest pair of baryons can participate in nuclear physics.

\begin{figure}[tb]
\centering
\includegraphics[width=0.75\textwidth]{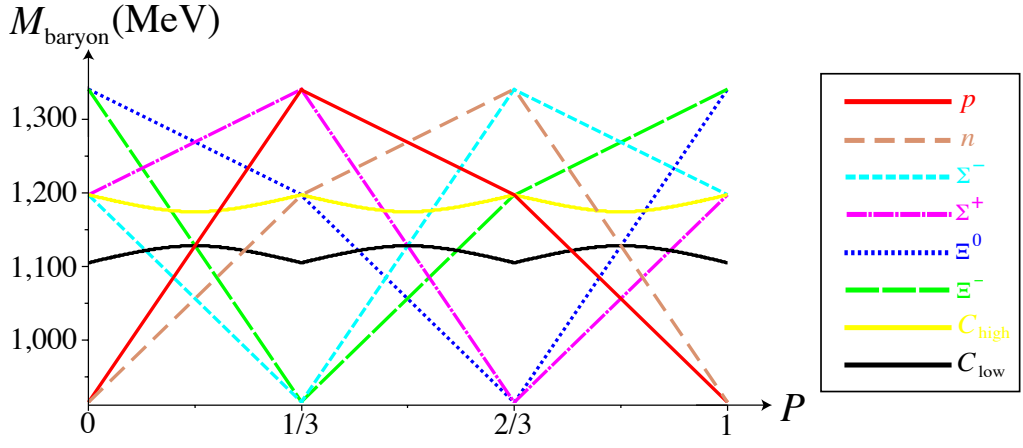}
\caption{\footnotesize Baryon masses (without electromagnetic corrections) at points along the perimeter of the quark mass triangle with $x_{T}=x_{T}^\oplus \equiv 100$ ({\it i.e.\/}, for the sum of the three light quark masses equal to what it is in our world) and with fixed $\Lambda_{\rm QCD} = \Lambda_{\rm QCD}^\oplus$. See Figure~\ref{perimeter} for a definition of the $P$ axis.}
\label{b_perimeter}
\end{figure}

\begin{figure}[tbh]
\centering
\includegraphics[width=0.6\textwidth]{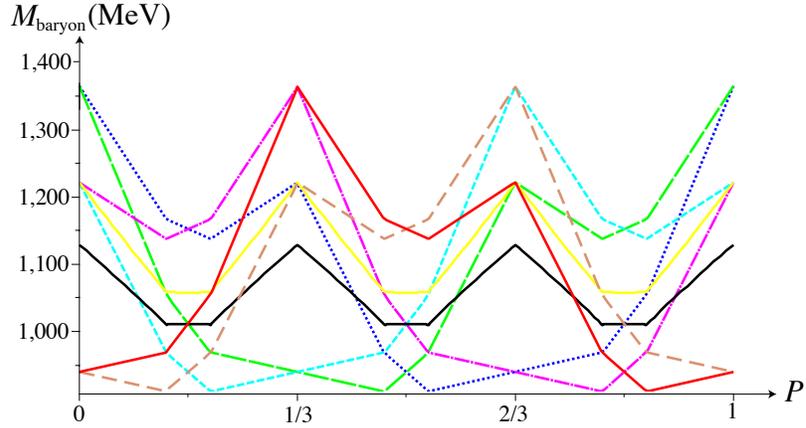}
\caption{\footnotesize The same baryon masses plotted in Fig. \ref{b_perimeter} are shown again, but now adjusted by an appropriate rescaling of $\Lambda_{\rm QCD}$ so that the average mass of the lightest pair of baryons is held fixed at 940 MeV.  For the baryon masses shown in this plot, $\Lambda_{\rm QCD}$ varies between 1.03$\Lambda_{\rm QCD}^\oplus$ and 0.85$\Lambda_{\rm QCD}^\oplus$.}
\label{b_perimeter_adjusted}
\end{figure}

To understand the rescaling that defines our slice through the SM and some of the physical phenomena we have identified, it is useful to examine the way that baryon masses vary over a triangular slice.  First we track the baryon spectrum as we move along the perimeter of the triangle corresponding to $x_T = x_T^\oplus = 100$.  Figure \ref{perimeter} shows how we define the coordinate $P$ to indicate where along the perimeter we are.  Figures~\ref{b_perimeter} and~\ref{b_perimeter_adjusted} show the octet baryon masses computed from Eq. (\ref{e.an.12}) (without the electromagnetic corrections), as functions of $P$.

The baryon masses in Fig.~\ref{b_perimeter} are computed with a fixed value of $\Lambda_{\rm QCD}=\Lambda_{\rm QCD}^\oplus$.  However, as explained in Sec. \ref{introslicing}, we adjust $\Lambda_{\rm QCD}$ to keep the average mass of the lightest pair of baryons fixed at 940 MeV everywhere along our slice through the parameter space of the SM.  Figure~\ref{b_perimeter_adjusted} shows the mass distribution after the rescaling of $\Lambda_{\rm QCD}$.  Candidates for congenial worlds occur when the two lightest baryons have nearly equal mass.  As the figure shows, this occurs in two characteristic circumstances, first at $P=0,1/3$ and $2/3$ (type (iii) regions in Fig.~\ref{regions}), and second at $P=1/6, 1/2$, and $5/6$ (type (ii) regions in Fig.~\ref{regions}).  We will discuss these areas in detail in Section~\ref{congenial}.

\begin{wrapfigure}{r}{0.45\textwidth}
\centering
\includegraphics[width=0.17\textwidth]{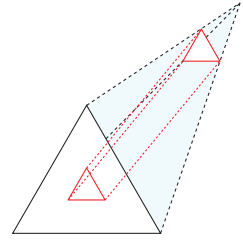}
\caption{\footnotesize The perimeter of a triangle of small constant $x_T$ is shown projected onto a central region within a triangle corresponding to a larger value of $x_T$.  The two triangular trajectories, drawn in red, give the same spectrum of octet baryon masses, but different meson spectra.}
\label{rescaling}
\end{wrapfigure}

The formulas for the baryon masses are linear in $x_3$ and $x_8$ (or in the radius $\sqrt{x_8^2 + x_3^2}$), so considering a smaller triangle would yield similar but rescaled plots, with the curves in Figs. \ref{b_perimeter} and \ref{b_perimeter_adjusted} squeezed together.  This is true of both the perimeter of a triangle that corresponds to a smaller value of $x_T$, as well as of a triangular trajectory interior to the original triangle.  Two such triangles with the same baryon spectrum are shown in red in Fig. \ref{rescaling}.

The adjustment of $\Lambda_{\rm QCD}$ to keep the average mass of the two lightest baryons fixed at 940 MeV eliminates the dependence of the baryon spectrum on $x_T$, but even though the two triangles in red in Fig. \ref{rescaling} give the same baryon masses as one goes around their perimeter, their physics is not equivalent because the meson spectrum is not the same {in the two cases}.
To see this, consider the masses of the light (flavor $SU(3)$ octet) pseudoscalar mesons.  From chiral perturbation theory:
\beq
m^2_\pi = B_0 (m_u + m_d ); ~~ m^2_{K^\pm} = B_0 (m_u + m_s); ~~
m^2_{K^0} = B_0 (m_d + m_s)~,
\label{mesonmasses}
\eeq
where  $B_0$ is proportional to $\Lambda_{\rm QCD}$.  These vary as functions of $x_{T,3,8}$ because of the change in the quark masses and also because of the change in $\Lambda_{\rm QCD}$ required to keep the lightest pair of octet baryons at an average mass of 940 MeV.  Notice, though, that the two central states of the meson octet, $\pi^0$ and $\eta$, mix.  We label the corresponding mass eigenstates as $\xi_{\rm high}$ and $\xi_{\rm low}$, with masses:
\beq
m^2_{\xi_{\rm high/low}} = \frac{2 B_0}{3} \left( m_T \pm \sqrt{m_u^2 + m_d^2 + m_s^2 - m_u m_d - m_u m_s - m_d m_s} \right) ~.
\label{chis}
\eeq

\begin{figure}[tbh]
\centering
\subfigure[]{\includegraphics[width=0.52\textwidth]{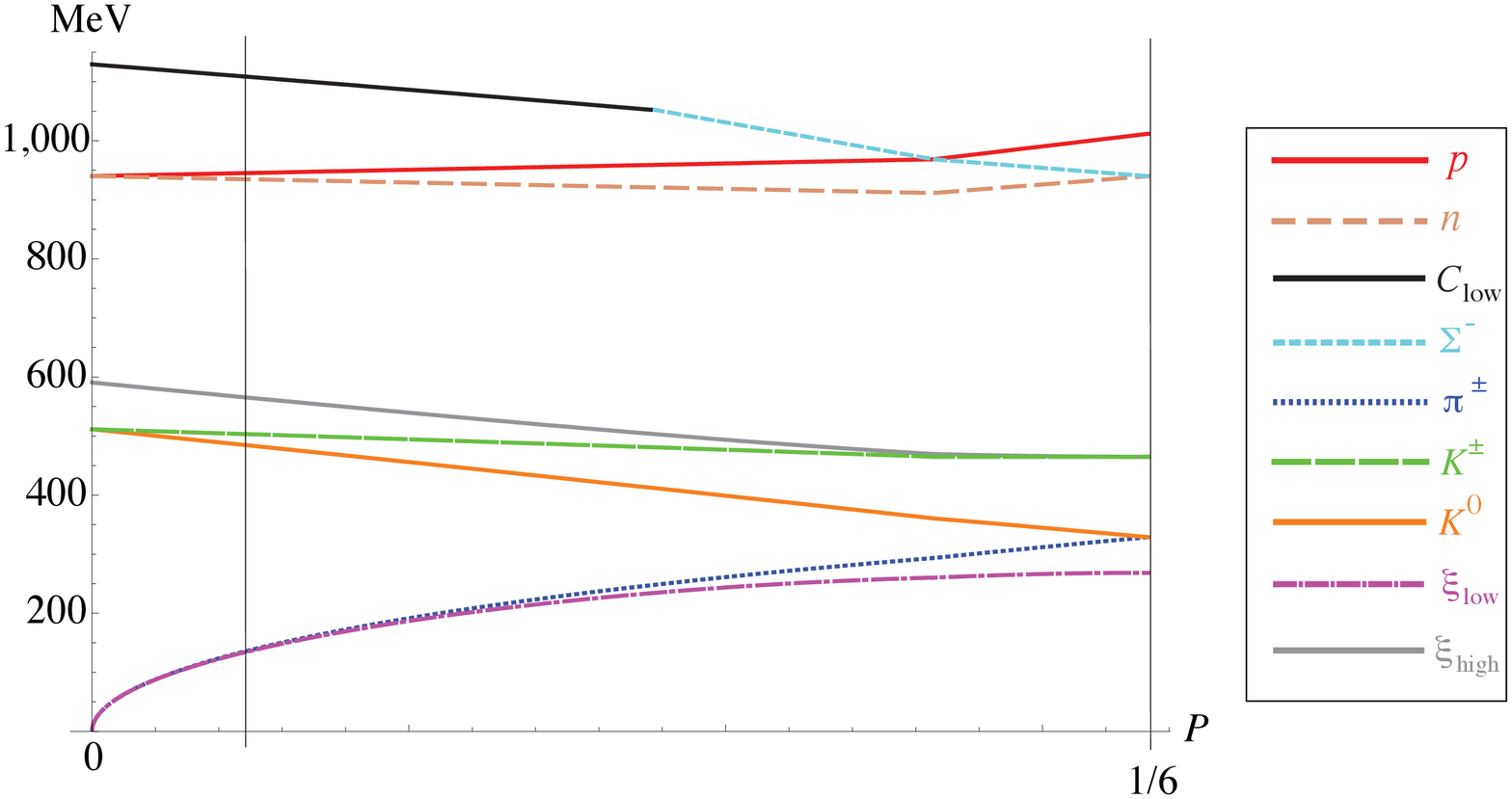}} \\
\subfigure[]{\includegraphics[width=0.4\textwidth]{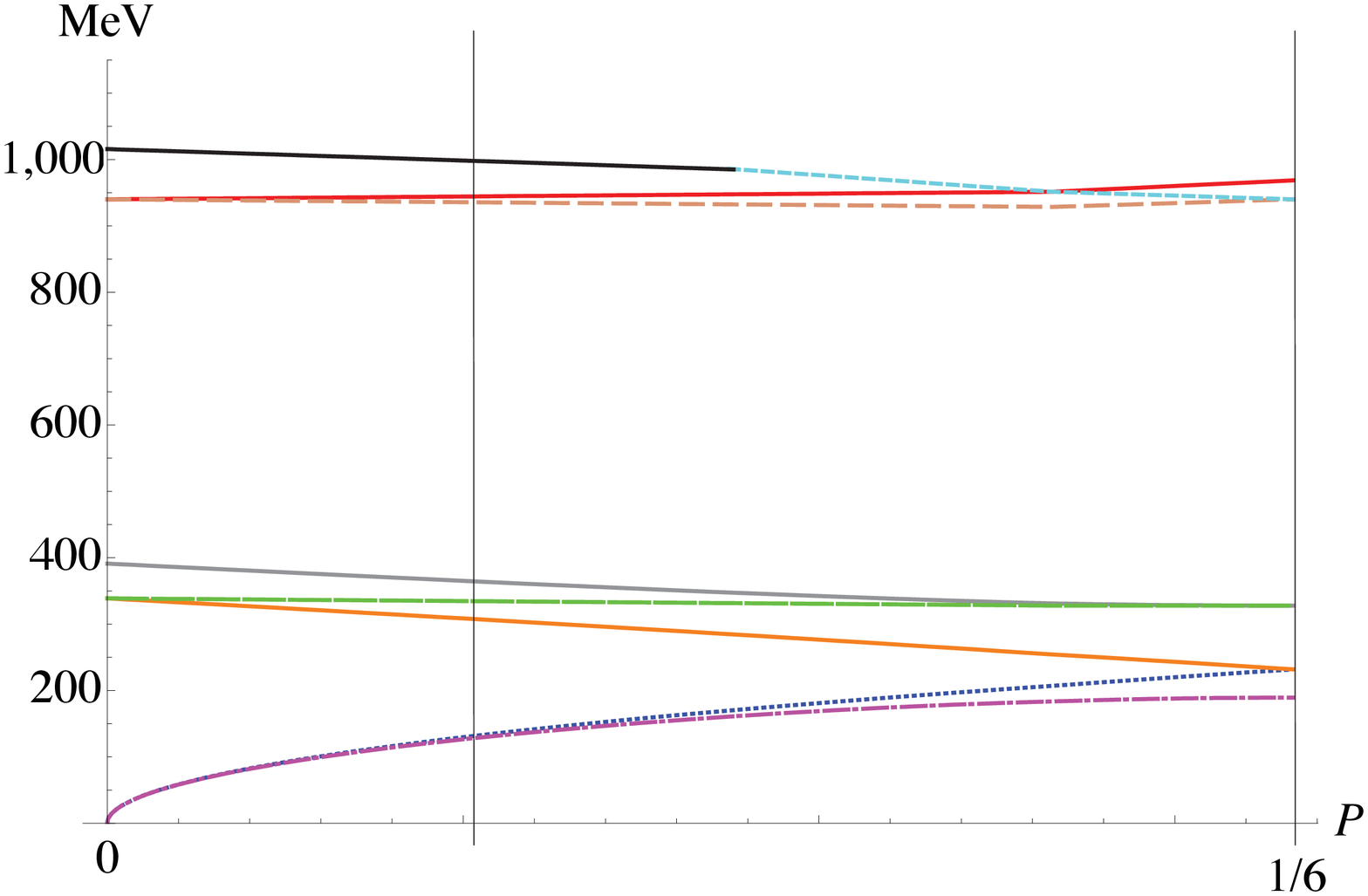}} \hspace*{2cm}
\subfigure[]{\includegraphics[width=0.4\textwidth]{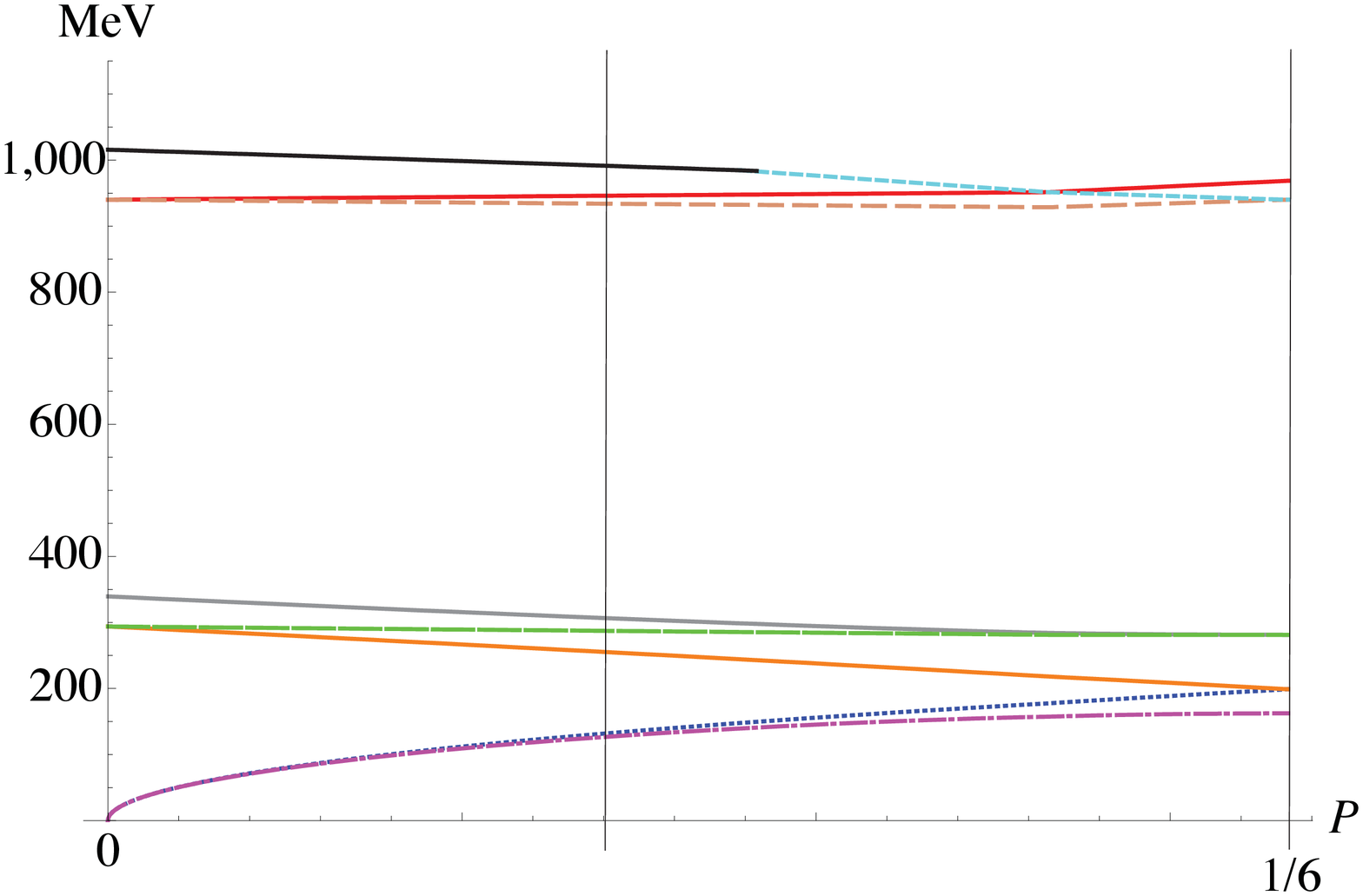}}
\caption{\footnotesize The masses of the three lightest octet baryons and of the the light pseudoscalar mesons are shown as functions of the perimeter coordinate $P$, for $0 \leq P \leq 1/6$.  The vertical line identifies the point where the pion mass is the same as in our world.  The spectra are shown for: (a) the perimeter of the $x_T = 100$ triangle, (b) the perimeter of the $x_T = 40$ triangle, and (c) a trajectory within the $x_T = 100$ triangle that corresponds to the projection of the perimeter of the $x_T = 40$ triangle.}
\label{spectra}
\end{figure}

Figure \ref{spectra} (a) shows the spectra of baryons and mesons, as a function of $0 \leq P \leq 1/6$, for the perimeter of the $x_T = 100$ triangle.  The other sectors in $P$ are obtained by symmetries.  For clarity only the three lightest baryons are plotted.  Figures \ref{spectra} (a) and (b) show, respectively, these spectra for the perimeter of the $x_T = 40$ triangle, and for a trajectory inside the $x_T = 100$ triangle that corresponds to the projection of the $x_T = 40$ triangle.  The baryon spectra are identical in these last two cases (because of the way we have defined our slice through the parameter space of the SM), but the meson spectra are somewhat different.  As previously mentioned nuclear binding may depend more strongly on the mass of the $f_{0}(600)$, which is likely to be less affected by changing quark masses (especially when $\Lambda_{\rm QCD}$ is readjusted as we have described), than on the masses of pseudoscalar bosons.  So we are optimistic that our analysis can be applied over a relatively wide range in $x_{T}$ as long as low energy nuclear physics does not change qualitatively, as it does when more than two baryons appear in stable nuclei.

In Fig. \ref{spectra}, a vertical line identifies the value of $P$ for which the mass of the pion is the same as in our world.  This corresponds to the position where the $m_\pi = 140$ MeV line shown in Fig. \ref{pionmassrange}(a) cuts the perimeter of the corresponding triangle.  Around that value of $P$, the nuclear interaction between the two lightest baryons (the proton and the neutron) would be {very} similar to what it is in our world.  For $P = 1/6$ the meson spectrum is qualitatively quite different.  The lightest meson is a mixture of an isosinglet ($\eta$) and the $I_3 = 0$ component of an isovector ($\pi^0$).  Furthermore, the neutral kaons and the charged pions have become degenerate in mass and are only slightly heavier than the lightest meson.  It would be very difficult to describe precisely the features of the nuclear interaction between the lightest baryons (the neutron and the $\Sigma^-$) in such a world.

%%%%%%%%%%
%%% MODELING NUCLEAR INTERACTIONS IN OTHER WORLDS
%%%%%%%%%%
\section{Modeling nuclear interactions in other worlds}
\label{nuclearmasses}

In order to explore the congeniality of worlds with different quark masses we need to at best compute or at least estimate the binding energy of nuclei in those alternative worlds.  In worlds where nuclear forces are the same as ours but the masses and charges of participating nucleons are different, we can compute nuclear binding energies of low atomic mass nuclei accurately by comparing with \emph{analog nuclei} in our world \cite{analog}.  This method enables us to judge the congeniality of all the worlds where two light quarks participate in building nuclei.  To study higher atomic mass nuclei in these worlds and to estimate the properties of nuclei in worlds with more than two participating baryons, we turn to a generalization of the Weizs\"acker semi-empirical mass formula (SEMF).  As noted in Sec. \ref{introduction}, we have only succeeded in making qualitative statements about worlds with three very light quarks.  Nevertheless the generalization of the SEMF to these worlds is interesting in itself and appears to be new, so we include it here.

\subsection{Analog nuclei}
\label{s4.1}

As explained in Section II, we can explore a wide class of interesting worlds while keeping the dynamics that determine nuclear interactions approximately fixed.  Nuclei in these worlds differ from ours in two important ways:  First, the charges of the two participating baryons may differ from ours.  If the $s$ and $d$ quarks are lightest then the $\Sigma^{-}$ and $\Xi^{-}$ replace the proton and neutron.  If the $c$ and $u$ are lightest, then the $\Sigma_{c}^{++}$ ($uuc$) and the $\Xi_{cc}^{++}$ ($ccu$) are the constituents of nuclei.  Second, the masses may differ.  However, the average mass of the nuclear constituents has been kept at 940 MeV and the mass difference ranges only over $\pm\, 20 - 30$ MeV.  Altering the charges affects only the nuclear Coulomb energy, a change that is easily made in the framework of the semi-classical estimate of the Coulomb energy, $E_{\rm c} = (0.7\,\mbox{MeV})\, Z(Z-1)/A^{1/3}$.  Altering the masses does not affect the nuclear \emph{binding energy}.  It does, however, affect nuclear decay patterns since it determines the thresholds for $\beta$-decays and for weak nucleon emission.

It remains only to determine the appropriate analog nucleus for the case of interest and make the necessary mass and charge adjustments.  As a simple (but important) example, consider the stability of ``oxygen'' in a world where the $s$ and $d$ quarks are lightest.  In that world the nucleus with the chemistry of oxygen has the baryon content $(\Sigma^{-})^{4}(\Xi^{-})^{4}$.  It is helpful to introduce a new notation to describe the nuclear and chemical properties of atoms in worlds where the constituents of nuclei have different charges than in our world.  If we limit ourselves to worlds with two participating baryons, then an atom can be labeled by the number of nucleons of each type, $N_{1}$ and $N_{2}$, the baryon number, $A=N_{1}+N_{2}$, and the electric charge, $Z$, which we denote by the chemical element symbol independent of whether the charge is positive or negative ({\it e.g.\/} $Z=\pm 4 \Rightarrow$ Be).  For convenience, we will represent the form of element El with $N_1$ baryons of type 1 and $N_2$ baryons of type 2 by the symbol \nucl{El}{N_{1}}{N_{2}}{A}.  Thus chemical oxygen in a $d$-$s$ world would be \nucl{O}{4}{4}{8}.  This oxygen has the nuclear binding of \nucl{Be}{4}{4}{8} in our world, plus some additional Coulomb energy.  Since \nucl{Be}{4}{4}{8} is unbound in our world, and the additional Coulomb energy in a $d$-$s$ world only unbinds it further relative to its decay products (\nucl{Be}{2}{2}{4}, the ($\Sigma^{-}$, $\Xi^{-}$) analog of the tightly bound $\alpha$ particle), we conclude that chemical oxygen is not stable in $s$-$d$ worlds, \emph{regardless of the $s$-$d$ mass difference}. 

In worlds that differ from ours only in the values of the $u$-$d$ quark mass difference, the analog method works somewhat differently.  The binding energies of nuclei are the same as in our world, however the systematics of weak decays change.  It is straightforward to determine the limits of nuclear stability.  The deuteron provides a simple example.  It is bound by $B(\nucl{H}{1}{1}{2}) = 2.22$ MeV in our world.  Changing the mass of the proton and neutron does not alter this.  However when $M_{p}-M_{n}>B(\nucl{H}{1}{1}{2})-m_{e}= 1.71$ MeV, the deuteron can decay to two neutrons by electron capture ($^2$H$\, e^{-}\to nn\nu_{e}$).  Alternatively, when $M_{n}-M_{p}> B(\nucl{H}{1}{1}{2})+m_{e}= 2.73$ MeV, the deuteron $\beta$-decays to two protons ($^2$H$\to ppe^{-}\bar \nu_{e}$).  These are the simplest examples of \emph{weak nucleon emission}, which figures importantly in determining the limits of congenial worlds.

\subsection{Generalizing the Weizs\"acker semi-empirical mass formula}
\label{SEMF}
Weizs\"acker's semi-empirical mass formula (SEMF) provides an excellent effective Hamiltonian for the nuclear ground state, when the number of each participating baryon species is conserved.  It is usually written as a formula for the nuclear binding energy, but with a change in sign and addition of the nucleon rest masses, it describes the total energy, which is better suited to our purposes.  In our world,
\beq
H(\{N_{i}\}) = \sum_{i} N_{i}M_{i}  -\epsilon^\oplus_{\rm v} A + \epsilon^\oplus_{\rm s} A^{2/3} + \epsilon^\oplus_{\rm c} \frac{Z(Z-1)}{A^{1/3}} + \epsilon^\oplus_{\rm a} \frac{I_{3}^{2}}{A} +\eta\frac{\Delta^\oplus}{A^{1/2}} 
\label{4.1}
\eeq
where $A=\sum N_{i}=Z+N$, $I_{3}=(Z-N)/2$, and as before, the superscript $\oplus$ denotes parameter values in our world.  The volume, surface, Coulomb, and asymmetry terms are self-explanatory.  The last term is the pairing energy.  $\eta$ is $+$ for odd-odd, $-$ for even-even, and $0$ for odd-even and even-odd nuclei.  We take the values of the coefficients, $\epsilon^\oplus_{j}$, from \cite{Lilley}.  They are listed in Table~\ref{t4.1}.
\begin{table}

\begin{center}
\begin{tabular}{|c|c|}
\hline
Parameter & Value (MeV) \\
\hline
$\epsilon^\oplus_{\rm v}$& 15.56\\
$\epsilon^\oplus_{\rm s}$& 17.23\\
$\epsilon^\oplus_{\rm c}$&0.7\\
$\epsilon^\oplus_{\rm a}$&93.12\\
$\Delta^\oplus$ & 12\\
\hline
$\epsilon^\oplus_{0}$ & 32.0 \\
$\bar\epsilon^{\,\oplus}_{\rm v}\equiv\epsilon^\oplus_{\rm v}+\frac{\epsilon_{0}^\oplus}{2^{2/5}}$& 39.8\\
$ \bar\epsilon^{\,\oplus}_{\rm a}  \equiv \epsilon^\oplus_{\rm a}-\frac{2^{8/5}\epsilon_{0}^\oplus}{3}$ & 60.8 \\
\hline
\end{tabular}
\end{center}
\caption{\footnotesize \small Parameters in the SEMF, as quoted in \cite{Lilley}.  The first five are the standard parameters as defined in \Eq{4.1}.  The last three are the parameters that enter the Fermi-gas modified SEMF developed in Sec.~\ref{s4.22}.  Note that the superscript $\oplus$ is a reminder that these parameters are fit to nuclear masses in our world. }
\label{t4.1}
\end{table}

To consider different nuclear constituents, we must generalize the SEMF in two ways.  First we must allow for different symmetry groups.  As we shall see the SEMF \emph{assumes isospin symmetry}, and if we are to consider worlds with three light quarks we must generalize to $SU(3)$ flavor symmetry.  Second, although isospin or $SU(3)$ may be an excellent symmetry of the strong interactions if quark masses are small enough, the symmetry may be dramatically broken by kinematic effects when a baryon becomes so massive as to drop out of nuclear dynamics.  This effect requires some explanation.  Consider a sequence of universes where the $u$ and $d$ quark masses are fixed at the values in our world, and the $s$ quark mass is slowly increased from a few MeV.  For the smallest $s$ quark mass, all eight octet baryons would be expected to participate in nuclear physics and $SU(3)$-flavor symmetry rules.  When the $s$ quark mass increases to the point that the lightest strange baryon mass exceeds the proton and neutron Fermi energy in the nucleus, strange baryons disappear from nuclear dynamics.  This occurs when the hyperon masses are only a few tens of MeV heavier than the nucleon mass, in a regime where the $SU(3)$ symmetry of the strong interactions is still quite a good symmetry.  We believe that this essentially kinematic, but massive, violation of flavor symmetry can be incorporated into the SEMF by modeling part of the volume and asymmetry energies by a degenerate Fermi gas.

\subsubsection{Restoring flavor symmetry to the SEMF}
\label{s4.21}
Imagine that the proton and neutron masses were exactly the same, so the strong interactions were exactly $SU(2)$ invariant.  Two terms in our world's SEMF, the Coulomb and asymmetry terms, appear to violate $SU(2)$ flavor symmetry. The Coulomb term violates $SU(2)$ because it originates in  electromagnetic interactions which, of course, are not $SU(2)$ invariant.  However, what are we to make of the appearance of the explicitly $SU(2)$ symmetry \emph{violating} operator $\propto I_{3}^{2}$ in the asymmetry term of \Eq{4.1}?  The asymmetry energy is a strong interaction effect and must depend only on $SU(2)$ flavor invariant operators.  

The solution to this puzzle is that $I_{3}^{2}$ should be replaced by the quadratic Casimir operator of $SU(2)$, $I(I+1)$, and that electromagnetic violations of $SU(2)$ allow $I$ to adjust to its minimum value, which is $I=I_{3}$.  In the exact $SU(2)$ limit and without electromagnetism, nuclei with a given $A$ would form degenerate multiplets labeled by the total isospin $I$.  The lightest multiplet would have the lowest value of $I$.  Since $I\ge I_{3}$ and $I_{3}$ is conserved in the absence of the weak interactions, this occurs when $I=I_{3}$.

Thus, electromagnetism allows the states with $I>I_{3}$ to decay quickly to the ground state, so for all intents and purposes the $SU(2)$-symmetric asymmetry term, $I(I+1)|_{I=I_{\rm min}}$ can be replaced by $I_{3}(I_{3}+1)$.   Switching between this and the usual form, $I_{3}^{2}$, produces a negligible shift in the computed values of the binding energy $B(A)$ and the charge $Z(A)$ of the lightest nucleus with given $A$.  That is, the two forms of the SEMF have nearly the same {\it valley of stability}.  Therefore, a more illuminating way to write the SEMF close to the $SU(2)$-flavor symmetry limit, is
\beq
H^{(2)}(N_{1},N_{2}) = N_{1}M_{1}+N_{2}M_{2}  -\epsilon^\oplus_{\rm v} A + \epsilon^\oplus_{\rm s} A^{2/3} + \epsilon^\oplus_{\rm c} \frac{Z(Z-1)}{A^{1/3}} + \epsilon^\oplus_{\rm a} \left.\frac{I(I+1)}{A}\right|_{I=I_{3}} +\eta\frac{\Delta^\oplus}{A^{1/2}} 
\label{4.2}
\eeq
where the superscript on $H^{(2)}$ reminds us that this effective Hamiltonian applies near the $SU(2)$ limit (regardless of which two quark species are light).

The same analysis must be performed near the $SU(3)$ symmetry limit.  The parameters in the SEMF are no longer the same as in our world (so we replace $\epsilon_{i}^\oplus$ and $\Delta^\oplus$ by unknown $\epsilon_{i}$ and $\Delta$), the variable $\eta$ must be redefined to count the number of unpaired baryons, and the operator that appears in the asymmetry term must be \emph{the quadratic Casimir operator of} $SU(3)$, which we denote by $G^{2}$.  The value of $G^{2}$ depends on the representation of $SU(3)$ to which the nucleus belongs.  $SU(3)$ representations can be labeled by two parameters, $\lambda=n_{1}-n_{2}$ and $\mu=n_{2}-n_{3}$, where $n_{1},n_{2},n_{3}$ are the number of boxes in the rows of the Young Tableau that defines the representation.  The expression for $G^{2}(\lambda,\mu)$ is
\begin{equation}
\label{4.3}
G^{2}(\lambda,\mu) = \frac{1}{3}\left(\lambda^{2}+\lambda\mu+\mu^{2}\right) +\lambda+\mu
\end{equation}
and the $SU(3)$-SEMF can be written as
\beq
H^{(3)}(\{N_{j}\}) = \sum_{j=1}^{8}N_{j}M_{j}  -\epsilon_{\rm v} A + \epsilon_{\rm s} A^{2/3} + \epsilon_{\rm c} \frac{Z(Z-1)}{A^{1/3}} + \epsilon_{\rm a} \left.\frac{G^{2}(\lambda,\mu)}{A}\right|_{\stackrel{\lambda_{\rm min}}{\mu_{\rm min}}} +\eta\frac{\Delta}{A^{1/2}} ~. 
\label{4.3a}
\eeq

In the $SU(3)$ case, the electromagnetic and strong interactions will quickly adjust the flavor content of the nucleus to minimize the asymmetry energy.  In contrast to the $SU(2)$ case, strong interactions play an essential role here.  In the two flavor case the two conserved quantities, $A$ and $I_{3}$, uniquely fix the number of both types of  baryons, which therefore remain fixed as $I$ adjusts to its minimum.  With three flavors there are as many as eight participating baryons, but only three conserved quantities, $A$, $I_{3}$, and the hypercharge, $Y$.  So the strong and electromagnetic interactions together will adjust the abundance of the eight species to minimize the ground state energy with fixed $A$, $I_{3}$ and $Y$\footnote{The reader will notice that the appearance of the baryon numbers $\{N_{i}\}$ in \Eq{4.3a} is problematic since they are dynamically determined (though constrained by the values of $A$, $I_{3}$ and $Y$).  Below we shall explain how they are fixed in a Fermi gas model.}.    

We need, therefore, to compute the minimum value of $G^{2}(\lambda,\mu)$ at fixed values of $I_{3}$ and $Y$.  Fixing $I_{3}$ and $Y$ fixes a point in the $SU(3)$ weight diagram (labeled as usual with $I_{3}$ along the horizontal and $Y$ along the vertical axis).  A representation is specified by a hexagon (or when $\lambda$ or $\mu$ vanishes, a triangle).  The values of $\lambda$ and $\mu$ give the lengths in $I_{3}$ of the topmost and bottommost horizontal lines in the diagram.  It is easy to see that the value of $G^{2}$ at fixed $I_{3}$ and $Y$ is minimized when the point $(I_{3},Y)$ appears on the boundary and at a corner of the polygon.   The $G^{2}(\lambda_{\rm min},\mu_{\rm min})$ resulting from this procedure can be explicitly computed:
\begin{equation}
\left. G^{2}(\lambda,\mu)\right|_{\stackrel{\lambda_{\rm min}}{\mu_{\rm min}}} = I_{3}^{2} + \frac{3}{4}Y^{2} + \frac{1}{2}\left(|N_u-N_d|+|N_d-N_s|+|N_s-N_u|\right)~.
\label{4.3b}
\end{equation}
Here $N_{u,d,s}$ are the coordinates of the point in question along axes that count the number of $u$, $d$, and $s$ quarks.  These terms can be written as piecewise continuous functions of $I_{3}$ and $Y$.  Whether or not they can be ignored depends on how close the system is to the symmetry point $I_{3}=Y=0$.  In Sec.~\ref{masslessuds}, where we use this formula, we keep these terms.  Although they make a significant quantitative difference, they do not affect the physical conclusions in that section.  \Eq{4.3a} and \Eq{4.3b} together provide the $SU(3)$ flavor generalization of the SEMF that we seek.

\subsubsection{Kinematic flavor symmetry violation in a Fermi gas model}
\label{s4.22}
All of the \emph{qualitative} features of the SEMF except for the pairing energy can be accounted for by treating the nucleus as a non-interacting Fermi gas of baryons under constant pressure.  Furthermore the Fermi gas model includes, in a physically intuitive way, the symmetry breaking effects that are essential for our analysis of worlds with different quark masses.  In particular, if the mass of a given baryon species exceeds the Fermi energy, then the species disappears from the nucleus.  This wholesale flavor symmetry violation must be taken into account even when the \emph{dynamical} breaking of flavor symmetries is small.

The non-interacting, degenerate Fermi gas under constant pressure gives a \emph{one parameter} model for nuclear energies.  If the pressure is chosen to reproduce the measured Fermi energy in heavy nuclei, then the surface, Coulomb and asymmetry energies are predicted.  The surface energy arises from the depletion of the density of states near the surface of the nucleus.  The Coulomb energy is that of a uniformly charged sphere whose volume is fixed by the pressure and the atomic mass. The asymmetry energy arises because the kinetic energy is at a minimum when the proton and neutron Fermi momenta are equal and grows when they differ.  However, the degenerate Fermi gas model accounts for only a fraction of the actual effects as parameterized in the SEMF.  This is not surprising:  some of the nuclear surface energy comes from the different interactions of nucleons on the surface relative to the interior of the nucleus, and some of the asymmetry energy comes from the dependence of the nuclear forces on isospin (the simplest example being the difference in the $I=0$ (deuteron) and $I=1$ channels of proton-neutron scattering).
  
In order to model flavor symmetry violation in worlds where two baryons participate in nuclear physics we construct a degenerate Fermi gas model with the following ingredients.  (We address the three-flavor case later.)
\begin{itemize}
\item  We seek an effective Hamiltonian as a function of the masses and numbers of each of the baryon species.  When we wish, we can find the most stable species at each $A$ by minimizing over $N_{i}$ at fixed $A$.
\item  The kinetic energies of the baryons are given by a degenerate Fermi gas under constant pressure, with the pressure chosen to reproduce the Fermi momentum in heavy nuclei in our world.
\item  We take the surface and Coulomb energies directly from the SEMF, ignoring possible flavor symmetry violating effects in these terms because they are higher order.
\item  We compute the Fermi gas contribution to the asymmetry energy and subtract it from the phenomenological asymmetry energy term in the SEMF.  The remainder we ascribe to interactions, \emph{which we assume to be flavor independent} because we work in worlds where $SU(2)$ symmetry violation is small.
\item The important, dramatic flavor symmetry violation is thus contained in the Fermi gas kinetic energy.
\end{itemize}    

Let $N_{i}$ be the number of baryons of the $i^{\rm th}$ species in the nucleus and $k_{i}$ be its Fermi momentum.  $N_{i}$ and $k_{i}$ are related by
\beq 
\label{4.4}
k_{i} = \left( \frac{3 \pi^2 N_{i}}{V} \right)^{1/3}~,
\eeq
where the volume $V$ is yet to be determined.
The  total internal energy including the baryons' rest mass contribution, is
\beq 
\label{4.5}
U(\{N_{i}\}) =  \sum_{i}U_{i}(N_{i},V)= \sum_{i} \left(   \frac{k_{i}^{5}}{10M \pi^{2}}V + M_{i}N_{i}   \right)~, \eeq
The $i^{\rm th}$ chemical potential is
\beq   
\mu_{i} \equiv \left. \frac{ \partial U }{ \partial N_{i}} \right\vert_{V} = M_{i} + \frac{k_{i}^2 }{2M}~.
\label{4.6}
\eeq
and the pressure is
\beq
P \equiv - \left.\frac{\p U}{\p V}\right|_{N_{i}} = \sum_{i}\frac{k_{i}^{5}}{15M\pi^{2}}~.
\label{4.61}
\end{equation}
Note that we consistently \emph{keep} baryon mass differences in the rest energy, where they drive large symmetry violating effects, but \emph{ignore them} in the kinetic energy, since they enter as $\Delta M_{i}/M^{2}\sim 10^{-2} / M$, which is negligible.   We use $M$ to stand for the average mass of the (two) participating baryons. 

It is convenient to use the grand canonical ensemble to describe a gas under constant pressure. The grand potential is defined as a function of $\{\mu_{i}\}$ and $V$, 
\begin{equation}
\label{4.7}
\Omega(\{\mu_{i}\},V) = \sum_{i}\Omega_{i}(\mu_{i},V) = \sum_{i}U_{i}(N_{i},V) - \mu_{i} N_{i},
\end{equation}
and its intensive counterpart is
\beq 
\omega_{i}(\mu_{i}) \equiv \frac{\Omega_{i}(\mu_{i},V)} { V}= - \frac{1}{15 M \pi^2} \left[ 2M (\mu_{i} - M_{i}) \right]^{5/2}.
\label{4.8}
\eeq
The quantity $\Omega$ is constructed so that
\begin{equation}
\left.\frac{\p\Omega}{\p\mu_{i}}\right|_{V}=-N_{i}~.
\label{4.81}
\end{equation}
The total energy is the internal energy of the Fermi gas together with the work it does against the confining pressure $P_{0}$:
\beq E(\{N_{i}\},V) = U(\{N_{i}\},V)+P_{0}V = \sum_{i}\left( \omega_{i}(\mu_{i})V+ \mu_{i}N_{i}\right) + P_{0}V ~.
\label{4.9}
\eeq 
Dynamical equilibrium is achieved by minimizing $E$ with respect to $V$ at fixed $N_{i}$,
\begin{equation}
\label{4.10}
 \left. \frac{\partial E}{\partial V}\right|_{N_{i}} = \sum_{i}\left[\omega_{i}(\mu_{i}) + V\frac{d\omega_{i}}{d\mu_{i}}\left.\frac{\p \mu_{i}}{\p V}\right|_{N_{i}}+\left.\frac{\p \mu_{i}}{\p V}\right|_{N_{i}}N_{i}\right] +P_{0}\,=\,0~.
\end{equation}
But from \Eq{4.81},
\begin{equation}
V\frac{d\omega_{i}}{d\mu_{i}}=-N_{i}~,
\label{4.11}
\end{equation}
so
\begin{equation}
\label{4.12}
\sum_{i}\omega_{i}(\mu_{i}) +P_{0}=0~,
\end{equation}
and, substituting back into \Eq{4.9}
\begin{equation}
E(\{N_{i}\}) =\sum_{i}\mu_{i}N_{i}~.
\label{4.121}
\end{equation}

Equations~(\ref{4.12}) and (\ref{4.121}) are the fundamental dynamical principles for the system.
Using them it is possible to solve for all of the dynamical properties of the system (energy, volume, compressibility, {\it etc.\/}) in terms of $P_{0}$ and the baryon numbers, $\{N_{i}\}$.  Of course, we are interested in the total energy, which is
\begin{equation}
\label{4.13}
E(\{N_{i}\}) =  \sum_{i}N_{i}M_{i} + \epsilon_{0}\Big( \sum_{i}N_{i}^{5/3} \Big)^{3/5}~,
\end{equation}
with  
\beq  
\label{4.14}
\epsilon_{0} =\frac{1}{2M^{3/5}} \left( 15 \pi^{2}P_{0}   \right)^{2/5}  . 
\eeq

We fix the value of $P_{0}$ based on the Fermi momentum  of nucleons in heavy nuclei in our universe.  Quasi-elastic electron scattering on heavy nuclei yield values of $k_{P}^\oplus= k_{N}^\oplus= 245$ MeV \cite{Maieron:2001it}, which yield a pressure  $P^\oplus_{0} = 0.827$  MeV/fm$^{3}$ and $\epsilon^\oplus_{0} = 32.0$ {MeV}. 

\subsubsection{Modified SEMF in the two flavor case}
\label{s4.23}
In this section we consider the two flavor case with two participating baryons with $A=N_{1}+N_{2}$ and $I_{3} = (N_{1}-N_{2})/2$.  Equation (\ref{4.13}) contains the kinetic contribution to the asymmetry energy which we seek.  In the usual SEMF, however, this contribution is inseparable from the dynamic contribution which has been fit to the form $\epsilon_{\rm a}I_{3}^{2}/A $ for small asymmetries.  To determine the dynamical contribution, we must expand $E(N_{1},N_{2})$ around $N_{1}=N_{2}$, identify the term with the form of the asymmetry energy, and subtract it from the phenomenological fit, leaving the dynamical term in the asymmetry.  Expanding $E$ as a power series in $I_{3}$, we obtain
\beq 
E (N_{1},N_{2}) = N_{1}M_{1}+N_{2}M_{2}+\frac{1}{2^{2/5}}\epsilon_{0} A + \frac{2^{8/5}}{3}\epsilon_{0}\frac{I_{3}^{2}}{A}+ {\cal O}(I_{3}^{4}/A^{3})
\label{4.15}
\eeq
When we explicitly include the Fermi gas kinetic energy we must compensate by removing the term proportional to $A$ in \Eq{4.15} from the volume energy and the term proportional to $I_{3}^{2}/A\approx I_{3}(I_{3}+1)/A$ from the asymmetry energy.  Putting this together we obtain an SEMF
\begin{eqnarray}
H^{(2)}(N_{1},N_{2}) &=& N_{1}M_{1}+N_{2}M_{2}+ \epsilon_{0} \Big( N_{1}^{5/3}+N_{2}^{5/3} \Big)^{3/5}\nonumber\\
 &-& \left(\epsilon_{\rm v}+\frac{\epsilon_{0}}{2^{2/5}}\right) A + \epsilon_{\rm s} A^{2/3}  + \left(\epsilon_{\rm a}-\frac{2^{8/5}\epsilon_{0}}{3}\right) \frac{I_{3}(I_{3}+1)}{A}\nonumber\\
&+& \epsilon_{\rm c} \frac{Z(Z-1)}{A^{1/3}}+\eta\frac{\Delta}{A^{1/2}} ~,
\label{4.16}
\end{eqnarray}
which incorporates the flavor symmetry violating effects that come from the Fermi gas kinetic energy.  As long as the low-energy nuclear interactions are similar enough to those in our world, the $\epsilon_i$'s in Eq. (\ref{4.16}) can be replaced with $\epsilon_i^\oplus$'s and $\Delta$ can be replaced with $\Delta^\oplus$.

It is convenient to define modified coefficients of the volume and asymmetry terms,
\begin{eqnarray}
\bar\epsilon^{\,\oplus}_{\rm v}&=&\epsilon^\oplus_{\rm v}+\frac{\epsilon_{0}^\oplus}{2^{2/5}}~; \nonumber\\
\bar\epsilon^{\,\oplus}_{\rm a} &\equiv&\epsilon^\oplus_{\rm a}-\frac{2^{8/5}\epsilon_{0}^\oplus}{3}~.
\label{4.17}
\end{eqnarray}
These parameters---along with $\epsilon_{0}^\oplus$---are listed in Table~\ref{t4.1}.

In the context of the Fermi gas model, the barred parameters represent the \emph{dynamical} contributions to the volume and asymmetry energy in contrast to the kinematic contributions that are contained in the term proportional to $\epsilon_{0}^\oplus$.  In terms of these parameters the modified SEMF becomes,
\begin{eqnarray}
H^{(2)}(N_{1},N_{2}) &=& N_{1}M_{1}+N_{2}M_{2}+ \epsilon_{0}^\oplus\Big( N_{1}^{5/3}+N_{2}^{5/3} \Big)^{3/5} - \bar\epsilon^{\,\oplus}_{\rm v}A + \epsilon^\oplus_{\rm s} A^{2/3}\nonumber\\ &+& \bar\epsilon^{\,\oplus}_{\rm a}\frac{I_{3}(I_{3}+1)}{A}+ \epsilon^\oplus_{\rm c} \frac{Z(Z-1)}{A^{1/3}}+\eta\frac{\Delta^\oplus}{A^{1/2}} 
\label{4.18}
\end{eqnarray}

\subsubsection{Modified SEMF in the three flavor case} 
\label{s4.24}
The analysis of the previous section can be generalized to the three flavor case by replacing the operator in the asymmetry term by the operator we derived in Section~\ref{s4.21},
\begin{eqnarray}
H^{(3)}(\{N_{i}\}) &=& \sum_{i=1}^{8}N_{i}M_{i}+ \epsilon_{0}\Big(\sum_{i=1}^{8} N_{i}^{5/3}\Big)^{3/5} - \bar\epsilon_{\rm v}A + \epsilon_{\rm s} A^{2/3}\nonumber\\ &+& \bar\epsilon_{\rm a}\frac{I_{3}^{2}+\frac{3}{4}Y^{2}}{A}+ \epsilon_{\rm c} \frac{Z(Z-1)}{A^{1/3}}+\eta\frac{\Delta}{A^{1/2}} ~,
\label{4.19}
\end{eqnarray}
where, for simplicity we have dropped the small corrections to the quadratic Casimir operator that are linear in the quark number differences.  These terms can be included with little increase in complexity.  Note, also, that we have removed the superscript ``$\oplus$'' on the parameters as a reminder that we do not know the values of the SEMF parameters in the $SU(3)$ limit because the physics of nuclear binding is qualitatively different than in the two flavor case.  For the same reason, there is no point in trying to separate $\bar\epsilon_{\rm a}$ into the dynamical and Fermi-gas kinetic pieces since neither is separately known.  

\Eq{4.19} neglects an important piece of the dynamics that appears when we try to generalize the SEMF from two to three flavors.  In the two flavor case the two quantum numbers conserved by the strong and electromagnetic interactions, $A$ and $I_{3}$, are sufficient to fix the numbers of both nuclear species.  In the three flavor case, fixing $A$, $I_{3}$, and $Y$ does not fix the number of all eight baryon species.  Strong interaction processes like 
\begin{equation}
\label{4.20}
\Sigma^{0}+\Lambda\leftrightarrow p + \Xi^{-}
\end{equation}
will establish ``chemical'' equilibrium among the various baryon species in order to minimize the energy of the system.  This is equivalent to minimizing $H^{(3)}(\{N_{i}\})$ with respect to the $N_{i}$ at fixed $A$, $I_{3}$, and $Y$.  In general this minimum cannot be found analytically even in the simple Fermi gas model.  However in Section~\ref{congenial} we will argue that the case when the baryon mass differences are small compared to the scales of nuclear physics ($\epsilon_{0}$, $\epsilon_{\rm v}$, {\it etc.\/}) is of particular interest.  In this case the minimization problem can be solved analytically  to first order in the mass differences, giving an effective Hamiltonian (still ignoring the linear terms in the $SU(3)$ Casimir operator), 
\begin{eqnarray}
H^{(3)}(A,I_{3},Y) &=& \left.H^{(3)}(\{N_{i}\})\right|_{N_{i}=N_{i}^{0}(A,I_{3},Y)}\nonumber\\
 &=& \overline M A+ \frac{\epsilon_{0}}{8^{2/5}}A +\frac{4\epsilon_{0}}{9\sqrt[5]{2}}
 \frac{I_{3}^{2}+\frac{3}{4}Y^{2}}{A}+\frac{1}{3}I_{3}\Delta M_{I}+\frac{1}{4}Y\Delta M_{Y} 
  - \bar\epsilon_{\rm v}A + \epsilon_{\rm s} A^{2/3}\nonumber\\ &+& \bar\epsilon_{\rm a}\frac{I_{3}^{2}+\frac{3}{4}Y^{2}}{A}+ \epsilon_{\rm c} \frac{Z(Z-1)}{A^{1/3}}+\eta\frac{\Delta}{A^{1/2}}~.
\label{4.20.2}
\end{eqnarray}
Here $\overline M$ is the average (over the octet) baryon mass, and $\Delta M_{I}$ and $\Delta M_{Y}$ are the isospin and hypercharge weighted baryon mass differences, 
\begin{equation}
\label{4.201}
\Delta M_{I}= \sum_{i=1}^{8} M_{i}I_{3i}\quad \mbox{and}\quad
\Delta M_{Y}= \sum_{i=1}^{8} M_{i}Y_{i} ~.
\end{equation}

As expected, if the baryon mass differences vanish, the  minimum in $H^{(3)}(A, I_{3},Y)$ is at $I_{3}=Y=0$ and the expansion of the Fermi gas kinetic energy to second order in $I_{3}$ and $Y$ yields terms of the same form as the volume and asymmetry terms.  Collecting terms with the same functional dependence, effectively reversing the logic after \Eq{4.16}, we can write the SEMF expanded to leading order in baryon mass differences about the $SU(3)$ limit,
\begin{eqnarray}
H^{(3)}(A,I_{3},Y)  
 &=& \bar M A+  \frac{1}{3}I_{3}\Delta M_{I}+\frac{1}{4}Y\Delta M_{Y} 
  -  \epsilon_{\rm v}A + \epsilon_{\rm s} A^{2/3}\nonumber\\ &+&  \epsilon_{\rm a}\frac{I_{3}^{2}+\frac{3}{4}Y^{2}}{A}+ \epsilon_{\rm c} \frac{Z(Z-1)}{A^{1/3}}+\eta\frac{\Delta}{A^{1/2}} ~,
\label{4.21}
\end{eqnarray}
where $\epsilon_{\rm v}$ and $\epsilon_{\rm a}$ are defined in terms of $\bar\epsilon_{\rm v}$, $\bar\epsilon_{\rm a}$, and $\epsilon_{0}$ by relations of the form of \Eq{4.17}.  Equation (\ref{4.21}) will allow us to explore the charge versus mass relationship of nuclei in worlds with very small quark mass differences.

\subsubsection{Nuclear stability}
\label{weakemission}

For completeness we review the conditions for the various decay mechanisms that figure into our studies of congeniality.  In our world weak interactions---various forms of $\beta$-decay---act to minimize the energy at fixed baryon number and strong interactions, notably $\alpha$ emission, lead to baryon number changing decays.  In the worlds we consider these processes are augmented by weak baryon emission.

The criteria for fixed baryon number weak decays are elementary.  For $\beta^{\pm}$ decay, $M_i > M_{f}+m_{e}$, and for electron capture $M_{i}+m_{e}>M_{f}$, where $i$ is the initial state and $f$ the final one.  If nuclei are made of baryons of equal charge, {\it e.g.\/} the ($\Sigma^{-}$, $\Xi^{-}$) world mentioned in Sec. \ref{s4.1}, weak decays are somewhat suppressed because they would correspond to strangeness changing neutral currents.  This suppression will be discussed further in Sec~\ref{dsworld} when it arises.

The criterion for strong particle emission ({\it e.g.\/} $\alpha$ decay) is that the final state be more bound than the initial state.  We need only consider the most stable nucleus, denoted $[A]$, for a given value of $A$ since we are only interested in absolute stability.  Let $b_{\rm max}(A)$ be the binding energy \emph{per baryon} of this nucleus, and let $[a]$ denote the nuclear fragment ({\it e.g.\/} the $\alpha$ particle) that it might emit and $b(a)$ its binding energy per baryon.  When the baryon number is large enough that the binding energy can be regarded as a smooth function of $A$, the decay $[A]\to [A-a]+[a]$ is energetically allowed if
\begin{equation}
\label{4.22}
b_{\rm max}(A)+A\frac{db_{\rm max}(A)}{dA} < b(a)~.
\end{equation}

The other decay process important for determining boundaries of stability is \emph{weak nucleon emission}, which does not occur naturally in our world, but becomes important at large baryon mass differences.  In weak nucleon emission, a heavier baryon weakly decays into a lighter baryon while it is emitted (with an accompanying lepton pair as appropriate). The nucleus gains the energy from the mass difference between the baryons. Exactly such a process is responsible for the semi-leptonic decays of hypernuclei in our universe.

Next, we compute the threshold for weak nucleon emission.  For concreteness we imagine that the relevant baryons are the proton and neutron.  There are two possibilities:  First, if $M_{p}>M_{n}$ then the proton in the nucleus can capture an atomic electron to give a neutron, which is emitted from the nucleus, and a neutrino,
\begin{eqnarray}
\mbox{\sc weak neutron emission:}\qquad ^{[A]}[Z] + e^{-} &\to& ^{[A-1]}[Z-1] + n+\nu_{e}~,\nonumber\\
^{[A]}[Z] &\to& ^{[A-1]}[Z-1] + n + e^{+}+\nu_{e}~.
\label{4.22.1}
\end{eqnarray}
Positron emission is also possible, as described in the second line of \Eq{4.22.1}, but the threshold is higher, so we limit our attention to electron capture.  Second, if $M_{n}>M_{p}$ then the neutron in the nucleus decays to $pe^{-}\bar\nu_{e}$ and the proton is ejected,
\begin{eqnarray}
\mbox{\sc weak proton emission:}\qquad
^{[A]}[Z]&\to& ^{[A-1]}[Z] + p + e^{-}+\bar\nu_{e}~.
\label{4.22.2}
\end{eqnarray}
For definiteness we consider the first case, weak neutron emission, and quote the result for weak proton emission at the end.  Weak neutron emission is energetically allowed when
\beq
M(N,Z)+m_{e} > M(N,Z-1) + M_{n}~.
\label{4.23}
\eeq

Let $B(A,I_{3})$ be the nuclear binding energy in terms of $A=N+Z$ and $I_{3}=\frac{1}{2}(Z-N)$.  Let $\Delta M = M_{n}-M_{p}$.
\Eq{4.23} can be rewritten in terms of binding energies,
\beq
B(A,I_{3}) - B(A-1, I_{3} - \half) <  m_{e} - \Delta M~.
\label{4.24}
\eeq
For very light nuclei this condition would have to be considered on a case by case basis.  For larger $A$ we can approximate the discrete difference as a derivative, 
\begin{equation}
\left.\frac{\partial B }{\partial A}\right|_{I_{3}} +\half \left.\frac{\partial B }{\partial I_{3}}\right|_{A} < m_{e} - \Delta M.
\label{4.25}
\end{equation}
We are interested in nuclei already at the minimum of the valley of stability set by the weak interactions, so $I_{3} = I_{3}(A)$ is set by 
\beq 
\left.\frac{\partial M(A,I_{3}) }{\partial I_{3}}\right|_{A} = 0~.
\label{4.26} 
\eeq
Since $M(A,I_{3})=  \frac{1}{2}A(M_{n}+M_{p}) - I_{3} \Delta M -B(A,I_{3})$, a nucleus at the weak interaction valley's minimum satisfies
\beq 
\left.\frac{\partial B }{\partial I_{3}}\right|_{A} = -\Delta M~. 
\label{4.27}
\eeq 
Substituting this back into \Eq{4.25}, we find that the nucleus at the bottom of the valley of stability is unstable to weak neutron emission when
\begin{equation}
\left.\frac{\partial B }{\partial A}\right|_{I_{3}}  <  m_{e} - \frac{1}{2}{\Delta M}~.
\label{4.28}
\end{equation}

Finally we relate $\partial B/\partial A|_{I_{3}}$ to the derivative of the binding energy along the valley of stability, which is defined by 
\begin{equation}
B_{\rm max}(A)=B(A,I_{3}(A))
\label{4.29}
\end{equation}
where $I_{3}(A)$ is determined by \Eq{4.26}.  Differentiating $B_{\rm max}(A)$, we obtain, 
\begin{eqnarray}
 \frac{d B_{\rm max}(A) }{d A} &=& \left.\frac{\partial B }{\partial A}\right|_{I_{3}} + \left.\frac{\partial B }{\p I_{3}}\right|_{A}\frac{d  I_{3}(A) }{d A}\nonumber\\
&=&\left.\frac{\partial B }{\partial A}\right|_{I_{3}}  - \Delta M\frac{d I_{3}(A) }{d A}~.
\label{4.30}
\end{eqnarray}
from which we can substitute into \Eq{4.28}, to obtain a condition for instability in terms of the the familiar function $B_{\rm max}(A)$, 
\begin{equation}
\mbox{\sc weak neutron emission:}\quad\frac{d B_{\rm max}(A) }{d A}  <  m_{e} - \Delta M \left(\frac{1}{2} +\frac{d I_{3}(A) }{d A} \right) ~.
\label{4.31}
\end{equation}
The nuclear asymmetry energy keeps $d I_{3}/dA$ small throughout the regime of stable nuclei. In our universe it is zero for nuclei at the bottom of the valley of stability up to $A\sim 40$, and increases only slowly thereafter.  It is of order $0.1$ for heavy nuclei.  Therefore to a good approximation (ignoring also the small mass of the electron) the criterion for instability by weak neutron emission reduces to
\begin{equation}
\label{4.32}
\mbox{\sc weak neutron emission:}\quad\frac{dB_{\rm max}(A)}{dA}\lesssim - \frac{1}{2}\Delta M~.
\end{equation}

The instability for weak proton emission, which occurs when the neutron is considerably heavier than the proton, are
\begin{align}
\mbox{\sc weak proton emission:} &\quad\frac{d B_{\rm max}(A) }{d A}  <  - m_{e} + \Delta M \left(\frac{1}{2} -\frac{d I_{3}(A) }{d A} \right)\nonumber\\
 &\quad\frac{dB_{\rm max}(A)}{dA}\lesssim   \frac{1}{2}\Delta M ~.
\label{4.33}
\end{align}
In worlds with two light quarks and two baryons participating in nuclear physics, the binding energy per nucleon in medium to heavy nuclei is approximately 8 MeV.  Therefore weak nucleon emission instability sets in throughout a broad region of $A$  when $|\Delta M|\gtrsim 16$ MeV.

Our result in Eq. (\ref{4.33}) differs by approximately a factor of two from the result claimed in Sec. 4.3 of \cite{Hall:2007ja}.  There the authors compute the constraint on the proton-neutron mass difference from the condition that complex nuclei be stable against $\beta$-decay.  Their Eq. (27), which reads \teq{m_{n}-m_{p}-m_{e}\lesssim E_{\rm bin}}, should read \teq{m_{n}-m_{p}-m_{e} \leq -\left. \p B / \p I_{3} \right|_{A}}, since the process they are considering takes place at fixed $A$.  In any event, this is not a condition for the stability of complex nuclei, but rather the condition that sets the isospin (or, equivalently, the $N/Z$ ratio) at the bottom of the valley of stability, as a function of $\Delta M$.  The process that limits the stability of complex nuclei is weak nucleon emission, and it is a coincidence that our own result differs from the result of \cite{Hall:2007ja} by a factor of $\sim 2$.

%%%%%%%%%%
%%% CONGENIAL AND UNCONGENIAL WORLDS
%%%%%%%%%%
\section{Congenial and uncongenial worlds}
\label{congenial}

Using the tools developed in Secs. \ref{baryonmasses} and \ref{nuclearmasses}, we now proceed to investigate which values of the quark masses (along our choice of a slice through the parameter space of the SM as defined in Sec. \ref{rules}) would correspond to congenial worlds.  We shall consider in turn the case of 
\ben
\item[(a)]one light quark leading to a single light baryon,
\item[(b)]two light quarks with equal electric charge, 
\item[(c)]two light quarks with different charge,
\item[(d)]one light quark leading to two light baryons,
\item[(e)]three light quarks.
\een
A complete characterization of (e) is left for future investigation.

\subsection{One light quark leading to a single light baryon}
\label{onebaryon}

A universe where only one baryon is light enough to have a chance of participating in nuclear physics would be uncongenial.  For instance, if the $d$ and $s$ were so much heavier than the $u$ that the decuplet $uuu$ state ({\it i.e.\/}, the $\Delta^{++}$) were lighter than all the octet states, then no complex nuclei would exist \cite{vev-selection}.  A possible bound state of two $\Delta^{++}$'s would have as its analog the dineutron, which is unbound due to its high asymmetry energy penalty.  The electromagnetic repulsion would further obstruct binding between $\Delta^{++}$'s.  The only chemical element would be helium (if the light quark has charge $2/3$) or hydrogen (if it has charge $-1/3$).

For the lightest decuplet baryon to be lighter than all the octet baryons somewhere in the space of quark masses corresponding to a fixed $m_T \equiv m_u + m_d + m_s$, the value of $m_T$ would have to exceed some lower bound, which can be estimated as follows:  Consider the extreme case $m_u =0$, $m_{d,s} = m_T/2$, where the possibility of a light $\Delta^{++}$ is greatest.  The lightest octet baryons would be the proton and the $\Sigma^+$, both with mass fixed to $940$ MeV on our slice through SM parameter space.  From \Eq{e.an.12} and Table \ref{t.an.2} (ignoring the electromagnetic correction) we then have:
\beq
M_p = 940 \hbox{~MeV} = (783 \pm 101  \hbox{~MeV})\,\,\frac{\Lambda_{\rm QCD}}{ \Lambda_{\rm QCD}^\oplus}  + (2.71\hbox{~MeV})\,\,x_{T}\,, 
\label{onequark}
\eeq
where the factor 2.71 MeV comes from evaluating the coefficients in \Eq{e.an.12} when $x_{8}=-x_T/6$ and $x_{3}=-x_T/ 2 \sqrt{3}$, which follow when $m_{u}=0$ and $m_{s}=m_{d}$. 

To accurately estimate the mass of the $\Delta^{++}$ in a world with $m_{u}=0$ and $m_{d}=m_{s}$, we would need the decuplet analog of the $\sigma$-term \cite{bernard}.  For our purposes, however, it will suffice to make the quark model motivated assumption that the mass of the $\Delta^{++}$ is independent of $m_{d}$ and $m_{s}$ and that its dependence on $m_{u}$ is the same as the proton's (which, incidentally, agrees with the estimate of \cite{bernard}).  This gives
\beq
\langle \Delta^{++}|m_{u}\bar uu|\Delta^{++}\rangle \approx \frac{m_{u}^{\oplus}}{\hat m^{\oplus}}\times 35\pm 5\,\mbox{MeV}\, \approx 25\pm 4 \, \mbox{MeV}~,
\eeq
so we estimate the mass of the $\Delta^{++}$ to be $\approx 1210$ MeV in a world with $\Lambda_{\rm QCD}$ equal to its value in our world, but with the $u$ quark mass set to zero.  In other words,
\beq
M_\Delta =1,210\,\mbox{MeV}\,\,\frac{ \Lambda_{\rm QCD}}{\Lambda_{\rm QCD}^{\oplus}}~.
\eeq
Then the maximum value of $x_T$ such that the $\Delta^{++}$ remains heavier than the proton for $m_u = 0$ is  
\beq
x_T^{\rm max} = \frac{940}{2.71} \left( 1 - \frac{783 \pm 101 }{1,210} \right) = 122 \pm 29~.
\label{xTmax}
\eeq

Let us summarize the situation at fixed $x_{T}$:  For worlds with $x_{T}$ well below $x_{T}^{\rm max}$, octet baryons are the lightest stable baryons throughout the mass triangle, and decuplet states can be ignored.  As $x_{T}$ grows beyond $x_{T}^{\rm max}=122\pm 29$ MeV, a decuplet state is less massive than the lightest octet baryon state in three regions at the perimeter of the triangle near the midpoint of each side ($P$ = 1/6, 1/2, and 5/6).  These regions are uncongenial.  This is an important effect because these regions (type (ii) in the classification of Fig.~\ref{regions}) may otherwise have been congenial.  We return to this point in Subsection~\ref{neutronsigmaworld} below.

\subsection{Two light quarks of the same charge}
\label{dsworld}

Consider first the case of two light quarks with charge $-1/3$. We shall see that these worlds are uncongenial because it is impossible to make stable nuclei with charge greater than 4.  To establish this it is sufficient to consider the case in which the light quarks are degenerate in mass.  Otherwise, the nuclei at the bottom of the valley of stability (defined by the condition $\partial M/ \partial N_2 |_{A=N_1+N_2}= 0$, where $N_{1,2}$ are the numbers of each of the participating baryons) would be less bound than in the degenerate case, due to the asymmetry energy penalty.

In these worlds the participating baryons are the $\Sigma^{-}$ and the $\Xi^{-}$.  Because both baryons have charge $-1$, the charge of a nucleus, $Z = -A$. With no mass or charge difference between the baryons, the valley of stability lies on the line $N_{\Sigma^{-}} = N_{\Xi^{-}}$.\footnote{Since $d$ and $s$ have the same charge, one might worry that weak transitions between the two participating baryons are flavor-changing neutral processes, and therefore suppressed by the Glashow-Iliopoulos-Maiani (GIM) mechanism.  In our world, however, the heavier $\Xi^-$ decays into a $\Sigma^-$ and a photon with a partial lifetime $\sim 10^{-14}$ s \cite{PDG}, primarily via a ``long-distance penguin'' diagram.  This can be (very roughly) thought of as the tree-level process \teq{\Xi^- \to \Lambda + \pi^-}, followed by the strong recombination of the $\pi^-$ and $\Lambda$, to form a $\Sigma^-$ (with a photon emitted).  Here we will assume that, when energetically favorable, the conversion between $\Xi^-$ and $\Sigma^-$ can always occur over time scales short compared to those needed for the evolution of intelligent life.}

Our universe does have a small $M_n - M_p$ quark mass difference, but it is much smaller than the Fermi energy and thus is negligible in determining the valley of stability. Hence in our universe (before the Coulomb energy becomes significant) we can find analog nuclei that lie on the corresponding line $Z=A/2$.

Returning to the world with two light -1/3 charge quarks, we begin by considering very light nuclei for which the SEMF cannot be used.  Using the notation introduced in Sec.~\ref{s4.1} we denote the first nucleus, with $A=2$ and $Z=-2$, as \nucl{He}{1}{1}{2}.  Its analog in our world is deuterium. We may estimate its Coulomb energy by modeling the deuteron nuclear potential as a finite spherical well with depth $V_0$ and radius $a$. To determine $V_0$ and $a$ we impose the condition that the nuclear binding energy match its measured value for the deuteron, $B=2.22$ MeV, and the r.m.s. radius correspond to the observed deuteron diameter,$~4.28$~fm. The potential giving the correct wavefunction has $V_{0} = -19.54$ MeV and $a=2.96$~fm. Treating the Coulomb interaction as a first order perturbation about that finite spherical well interaction, we estimate the Coulomb energy to be $ \left\langle e^2 /r \right\rangle = 0.57$ MeV.

The binding energies we find for the nuclei with the chemical properties of helium (\nucl{He}{1}{1}{2}), beryllium (\nucl{Be}{2}{2}{4}), and carbon (\nucl{C}{3}{3}{6}) by considering the analog nuclei and correcting for the Coulomb term are, respectively: 1.7 MeV, 23.9 MeV, and 22.7 MeV.  Thus, in the universe of light $d$ and $s$ quarks, the decay of carbon by fission into beryllium and helium is exothermic by 2.9 MeV.  Since \nucl{Be}{2}{2}{4} has the same nuclear shell structure as the $\alpha$ particle in our universe (\nucl{He}{2}{2}{4}), this fission reaction is the equivalent of $\alpha$ decay, which now occurs for nuclei as light as $A=6$.  The underlying physical reason for this instability can be understood in our world:  the relatively weak nuclear binding energy of \nucl{Li}{3}{3}{6} compared with \nucl{He}{2}{2}{4} is almost enough to make it unstable to decay into deuterium and an $\alpha$ particle.  The additional Coulomb energy in the ($-1/3, -1/3$) world is enough to generate instability.

Through the SEMF and consideration of some specific analog nuclei, we find that heavier nuclei also undergo $\alpha$ decay, a process which is exothermic by about 8 MeV. In our universe $\alpha$ decay may be impeded for heavy nuclei by the Coulomb barrier through which the $\alpha$ particle must tunnel.  In this case, however, the Gamow model of $\alpha$ decay~\cite{gamow} shows that chemical carbon would have a very short lifetime---of the order of $10^{-18}$ seconds.

Universes in which the two light quarks both have charge $-1/3$ do not have stable carbon (or, for that matter, any nuclei with charge greater than four) and are therefore uncongenial.  For universes with two light quarks of charge 2/3, such as the $u$ and $c$ quarks, even more instabilities occur. Both participating baryons have a charge of 2, so the charge goes as $2A$.  The two  single baryons now correspond  to isotopes of chemical helium (\nucl{He}{1}{0}{1} and \nucl{He}{0}{1}{1}), and the stable $\alpha$ particle is chemical oxygen.  The lack of hydrogen necessarily makes this an uncongenial world.  As for carbon, there are two isotopes, \nucl{C}{2}{1}{3} and \nucl{C}{1}{2}{3} (analog to $^3_2$He and $^3_1$H in our world), but the additional Coulomb energy in this world destabilizes them.

\subsection{Two light quarks of different charges}
\label{udworld}

\subsubsection{Calculations with analog nuclei}
\label{udanalog}

Our universe has two quarks that are light enough to participate in nuclear physics, $u$ and $d$, with charge $+2/3$ and $-1/3$ respectively.  For all such worlds we will revert back to the usual notation of $N$ as the number of neutrons and $Z$ as the number of protons in a nucleus, denoted by $^{A}$El.

Let us consider first the case of $M_{p} > M_{n}$:  Examination of the various isotopes of carbon in our world indicates that $^{14}$C is the last to remain stable as $M_{p}-M_{n}$ increases.  The $^{14}$C nucleus in our world has a binding energy of 105.3 MeV.  It is stable against $\alpha$ emission (which is endothermic by 8.35 MeV) and other fission processes. It would undergo electron capture to form $^{14}$B if $M_{p}-M_{n} > 19.34 $ MeV, and decay by weak neutron emission (see \Eq{4.22.1}) into $^{13}$B if $M_{p}-M_{n} > 20.32 $ MeV.

The cutoff for congeniality, however, is not set by the requirement of stable carbon, but rather by stable hydrogen: When $M_{n} < M_{p} + m_{e}$, free protons may decay into neutrons by electron capture.  Deuterium could serve as a hydrogen isotope, but it is only weakly bound (by 2.22 MeV) and easily undergoes weak neutron emission if $M_{p}-M_{n} > 1.71$ MeV.

In our universe, the $\beta$-decay of tritium ($^3$H) into $^3$He is exothermic by about 0.02 MeV because, even though tritium is bound more strongly than $^{3}$He by 0.76 MeV, the neutron is heavier than the proton by 1.29 MeV. With a proton-neutron mass difference large enough for free protons to decay, tritium would {\it not} undergo $\beta$-decay.  Instead, it could only decay through weak neutron emission and subsequent dissociation into three free neutrons and a neutrino,
\begin{equation}
^{3}\mbox{H} +e^{-}\to 3(n)+ \nu_{e}~.
\end{equation}
The congeniality cutoff set by requiring some stable hydrogen isotope is therefore
\beq
M_{p}-M_{n} < B(^{3}\text{H}) - m_{e} = 7.97 ~\text{MeV} ~ .
\eeq
This is the main result for the limit of congeniality for the case $M_{p}>M_{n}$.\footnote{{Even though we have chosen not to deal with issues of nucleosynthesis in this paper, it is worth repeating the observation made in Sec. \ref{introduction} that stellar burning might be very different in this world, since the two tritium nuclei can make $^{4}$He plus two free neutrons by a \emph{strong interaction}.}}

Let us now consider the case $M_{n} > M_{p}$.  In this case, congeniality is determined by the stability of chemical carbon against all possible decay pathways.  Whether or not $\alpha$ decay and fission would be exothermic depends only on the binding energies of the nuclei involved and not on quark mass differences.  In our world, $^{12}$C has a binding energy of 92 MeV and is stable even against fission into three $\alpha$ particles. $^{10}$C has a binding energy of 60.34 MeV, making it stable against fission into two $\alpha$ particles and two free protons by 3.7 MeV, and thus even more stable against simple $\alpha$ decay (which would leave an unstable $^{6}$Be behind). 

Weak proton emission changes a given carbon nucleus into a lighter isotope.  $^{12}C$ decays by weak proton emission into $^{11}C$ if $M_{n} - M_{p} > 19.24$ MeV; $^{11}C$ decays if  $M_{n} - M_{p} > 13.64$ MeV; $^{10}C$ decays if $M_{n} - M_{p} >  21.80 $ MeV, and $^{9}C$ decays if $M_{n} - M_{p} >14.77$ MeV.

Finally, $\beta$-decay occurs spontaneously if
\beq
M(Z,N) > M(Z+1,N-1) + m_{e}~.
\label{betacondition}
\eeq
The various carbon isotopes $\beta$-decay into nitrogen if $M_{n} - M_{p}$ is greater than the following cutoffs:  18.64 MeV for  $^{12}C$, 15.25 MeV for $^{11}C$, and 24.40 MeV for  $^{10}C$. Thus, $^{12}C$ is stable against all decays if  $M_{n} - M_{p} < 18.64$ MeV.   $^{10}C$ becomes stable against $\beta^{+}$-decay already if $M_{n} - M_{p}$ is greater than a few MeV, and remains stable until it weak-proton-emits into $^{9}C$ for $M_{n} - M_{p} \geq 21.80$ MeV. For such large neutron-proton mass difference, the $^{9}C$ nucleus also weakly-emits into $^{8}C$, which then would $\beta$-decay away. Thus, the congeniality cutoff for the case of $M_{n}>M_{p}$ is set by the stability of  $^{10}C$ against weak proton emission:
\beq M_{n}-M_{p} <  21.80 ~\text{MeV} ~ .\eeq

Even though we have not included the requirement of stable oxygen in our definition of congeniality, as mentioned in Sec. \ref{introduction} most worlds with stable carbon and hydrogen contain also stable oxygen, and therefore the worlds we identify as congenial would usually allow the formation of water.  This world is no exception.

\subsubsection{Estimation of $dB/dA$ from the SEMF}
\label{SEMFerror}

\begin{figure*}[tb]
\centering
\subfigure[]{\includegraphics[width=0.3\textwidth]{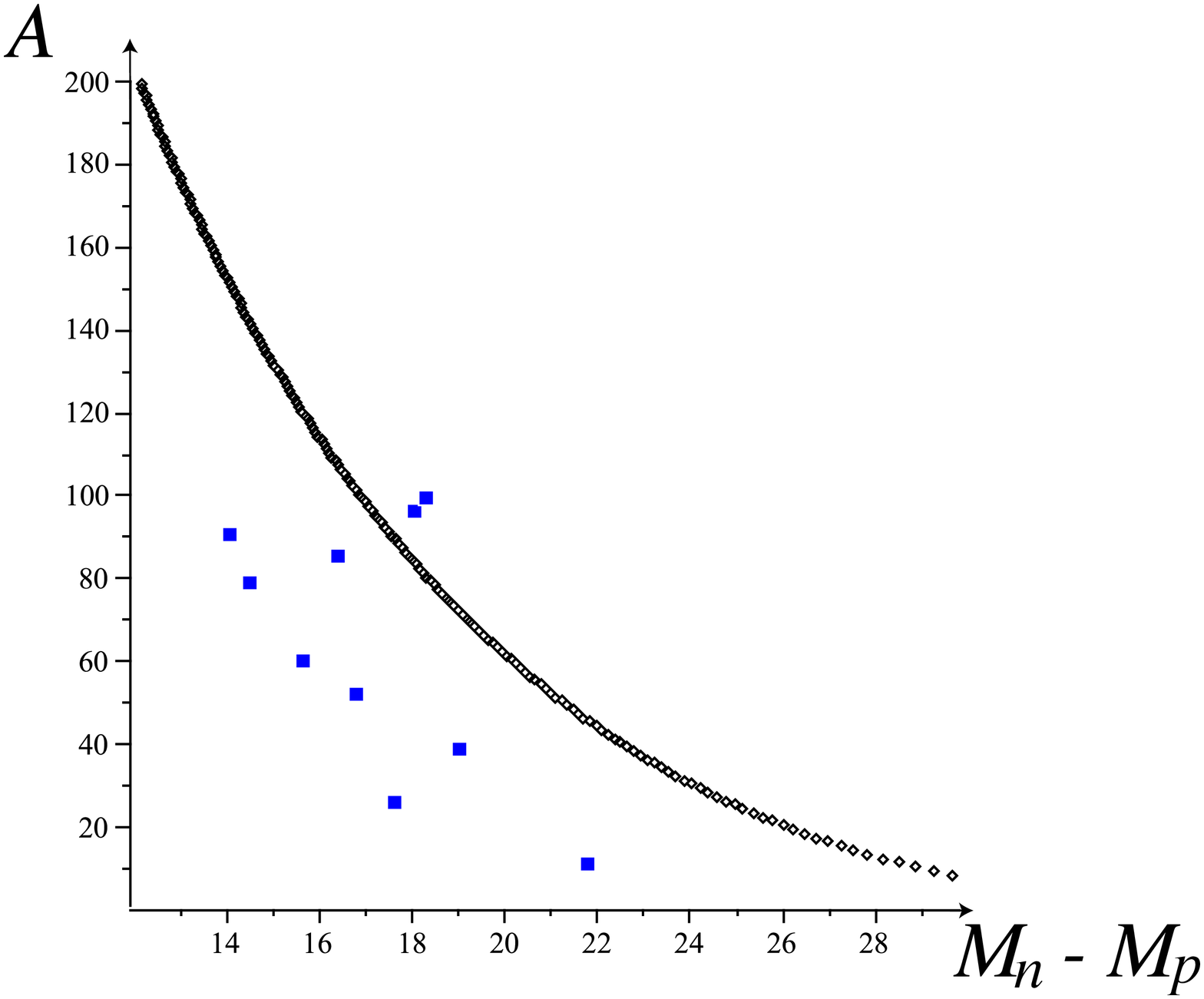}}\hspace*{1.5cm}
\subfigure[]{\includegraphics[width=0.3\textwidth]{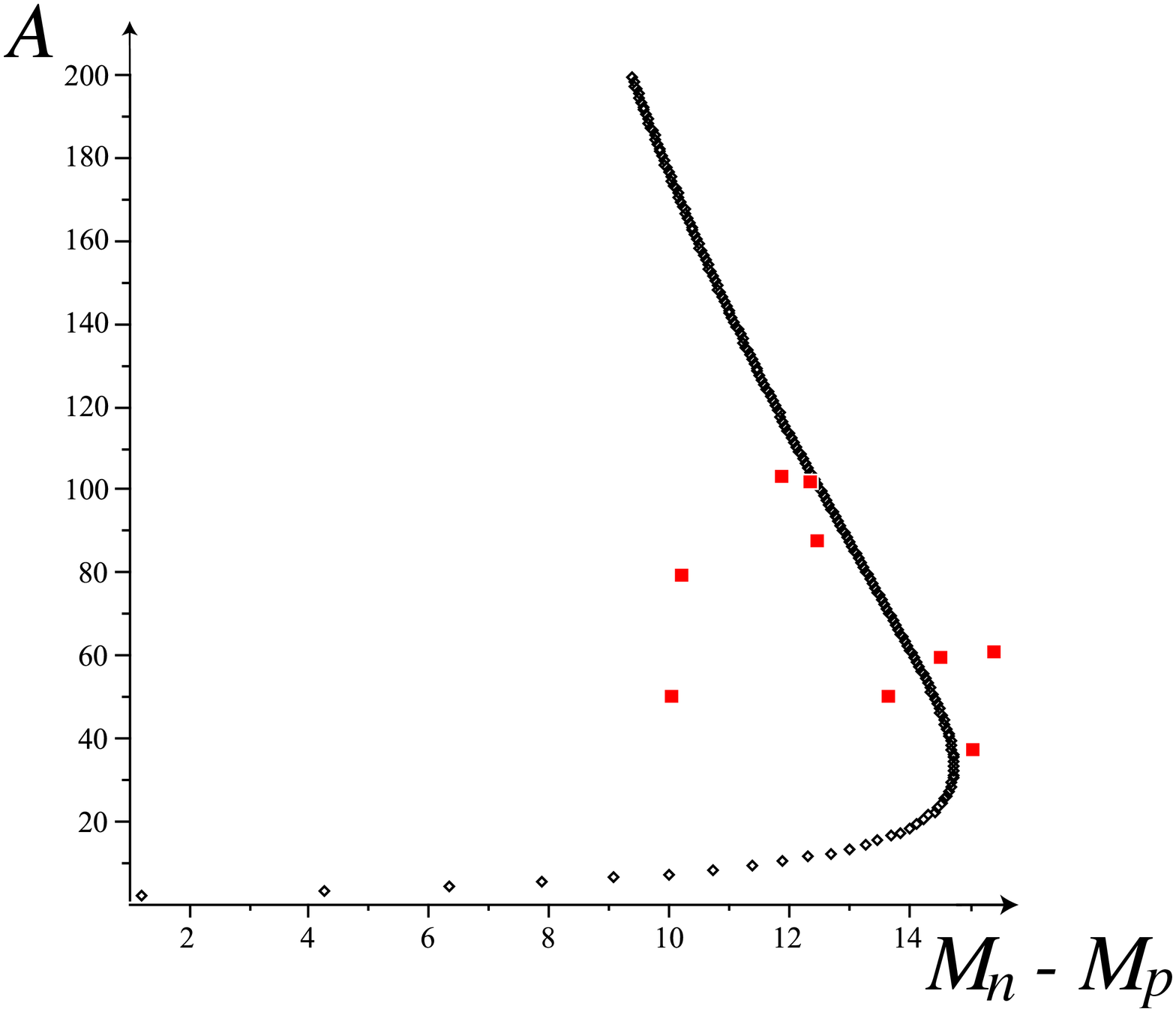}}
\caption{\footnotesize Upper limits on $M_n -M_p$ from stability against weak proton emission of nuclei at the bottom of the valley of stability, (a) for decay transitions involving a pairing term penalty, and (b) for decay transitions in which the nuclear binding energy gains a pairing bonus. The black curve is the prediction of the generalized SEMF, while the blue and red datapoints are obtained by considering analog nuclei.}
\label{Wvsanalog_comb}
\end{figure*}

In analyzing the boundaries of congeniality for worlds with two light baryons, we have been able to use the analog nucleus method to determine the stability of carbon, without ever referring to the generalized SEMF constructed in Sec. \ref{SEMF}. However, it is instructive to study the SEMF predictions about stability boundaries against weak proton emission for different nuclei. These are shown in Fig. \ref{Wvsanalog_comb}, together with the predictions of the analog nucleus analysis.  The bounds on $M_{n}-M_{p}$ are qualitatively similar to the bounds we obtained from study of carbon and oxygen by the analog method.  However, given the accuracy of the SEMF's predictions for $b_{\rm max}(A)$, it is at first surprising that the SEMF's predictions of $dB/dA \approx\delta M /2 $ are off by as much as $\sim 2$ MeV.

To understand why the SEMF does poorly in this case, one must bear in mind that the parameters of the SEMF---as quoted in Table \ref{t4.1}---are chosen so that the formula provides the best fit to the binding energy per nucleon $B/A$. We may write
\beq
\frac{B}{A} \equiv W(A) + W_{\rm error}(A)~.
\eeq
Note that the Weizs\"acker formula provides an excellent fit for $A\geq 10$, with  $ | W_{\text{error}}(A) | < 0.2$ MeV. However,
\beq
\frac{dB}{dA} = W(A) + A\frac{dW(A)}{dA} + W_{\rm error} (A) + A\frac{d}{dA}\left(W_{\rm error}(A)\right)~.
\label{dBdAerror}
\eeq
Since the magnitude of the variation in $W_{\rm error}$ from $A$ to $A+1$ is genericallly of order of $W_{\rm error}(A)$, the last term in \Eq{dBdAerror} can be quite large.  If we were to use the SEMF to estimate the boundaries of stability against weak neucleon emission, it might be necessary to refit the parameters to optimize the estimate of $dB/dA$, rather than $B/A$.

\subsubsection{Congeniality bounds on $u$, $d$ masses}
\label{udworld-quarks}

The condition for congeniality we found above for universes with light $u$, $d$ quarks is
\beq
-7.97 ~\text{MeV} \leq M_{n}-M_{p} \leq 21.80 ~\text{MeV} ~.
\label{udcong}
\eeq
To translate this into a condition on quark masses, we could use the results from $SU(3)$ perturbation theory presented in Sec. \ref{baryonmasses}.  However, note that the best fit to the parameters of the model gave us an error of about 2 MeV in the masses of the proton and the neutron.  For most of our purposes this is negligible, but in the case of worlds with only light $u$ and $d$ quarks we can improve on that accuracy by considering isospin symmetry.  Taking into account electromagnetic corrections, the $p$-$n$ mass difference is then 
\beq
M_p - M_n = (1.75 \, x_3 + 0.63 + 0.13)~ \hbox{MeV}~.
\eeq
Therefore Eq. (\ref{udcong}) translates to
\beq
-12.9 \leq x_{3} \leq 4.1 ~.
\label{udcongsu2}
\eeq

Referring back to \Eq{axes}, and taking $m_{T}^{\oplus}=100$ MeV, this translates into limits,
\begin{equation}
\label{udmasslimit}
-22.3\,\mbox{MeV} \le m_{u}-m_{d}\le 7.1\,\mbox{MeV}~.
\end{equation}
Our universe with $M_{p} - M_{n} = -1.29$ MeV is at $x_{3} = -1.17$, comfortably away from the edges of the congeniality band.

\subsubsection{Nuclei with more than two baryon species}
\label{3baryons}

If the $s$ quark becomes light enough, a third baryon, the $C_{\rm low}$ (which in our world is the $\Lambda$) will start to participate in nuclei. This marks the boundary along the $x_8$ direction of the congeniality region for two light $u$, $d$ quarks: while universes beyond this boundary may still be congenial, the analysis in Sec. \ref{udworld-quarks} no longer applies. 

Consider the case $x_3 = 0$, so that $M_n = M_p$.  Our universe is almost indistinguishable from this one in terms of nuclear composition. In the following analysis we will refer to the $C_{\rm low}$ simply as $\Lambda$, both for the sake of the reader's intuition and because if $x_{3}=0$, the Hamiltonian commutes with total isospin, so $C_{\rm low}$ = $\Lambda$.

Inside a nucleus composed of $Z$ protons and $N$ neutrons, a regular nucleon (which without loss of generality we will take to be a neutron) will spontaneously turn into a $\Lambda$ particle if 
\beq
M(Z,N, N_{\Lambda}=0)~ >~ M(Z,N-1,N_{\Lambda}=1)~.
\label{hn1}
\eeq
From now on when the mass $M$ or the binding energy $B$ is given as a function of three particle number variables, the last entry will be $N_{\Lambda}$ (the number of $\Lambda$'s).

Hypernuclei have been experimentally investigated in our world \cite{hypernuclei-review}. For a hypernucleus that contains only one $\Lambda$ particle, the binding energy of the $\Lambda$ as a function of $Z$ and $N$ can be expressed as:
\beq
B_{\Lambda}(Z,N) \equiv B(Z,N,1) - B(Z,N,0)~.
\eeq
Thus the mass of a hypernucleus is
\beq
M(Z, N-1, 1) = M_{\Lambda} + M(Z, N-1, 0) - B_{\Lambda}(Z,N-1) ~.
\label{hn2}
\eeq

Figure 2 of ~\cite{hypernuclei} provides experimental values of $B_{\Lambda}(Z,N)$ for various hypernuclei. For large nuclei around $A=120$, the $\Lambda$ binding energy roughly saturates to $B_{\Lambda} = 23$ MeV. For normal nuclei in our universe, $B/A$ saturates to around 8 MeV, so
\beq
M(Z, N-1, 0) = M(Z, N, 0) - M_{n} + 8 ~\hbox{MeV}~.
\eeq
Combining this equation with Eq.(\ref{hn1}) and Eq.(\ref{hn2}) we find that the $\Lambda$ participates in nuclear physics if
\beq
M_{\Lambda} - M_{n} \lesssim 15 ~\text{MeV} ~.
\label{14p5}
\eeq

A theoretical calculation using only the kinetic Fermi gas model energy yields a cutoff of $M_{\Lambda} - M_{n} < 2^{-2/5}\epsilon_{0} = 24 ~\text{MeV}$.  The discrepancy between this result and the value of 15 MeV deduced from real hypernuclei must arise from $\Lambda$-nucleon interactions, which are not included in the two flavor SEMF.  These interactions discourage the $\Lambda$ from participating in nuclei until the $\Lambda-N$ mass difference decreases to $\sim 15$ MeV.  Also note that performing the previous analysis for carbon, we find that the $\Lambda$ will not participate in carbon nuclei unless $M_{\Lambda} - M_{n} < 8$ MeV.  Therefore, we do expect universes to remain congenial a little bit beyond the boundary defined by \Eq{14p5}.

This analysis applies rigorously only around $x_{3} = 0$.  In that case the constraint on $x_8$ for the $\Lambda$ not to participate in nuclear physics is 
\beq
x_8 \leq -3.5~.
\label{2baryonlimit}
\eeq
For $|x_{3}| > 0$ , the boundary between the worlds with nuclei containing only two baryons and worlds with nuclei containing three or more baryons would lie at $x_8 < -3.5$.  The reason for this is that, due to the asymmetry energy penalty for nuclei composed of the two lightest baryons when these are not degenerate, it can be energetically favorable for a nucleus to capture a heavier $\Lambda$.

\begin{figure}[tb]
\includegraphics[width=.45\textwidth]{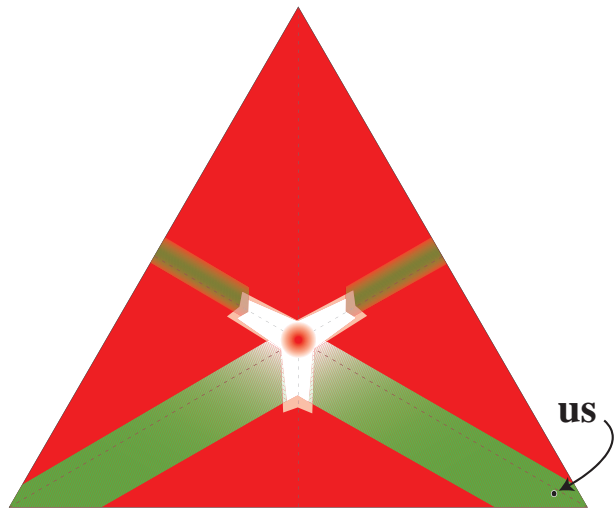}
\caption{\footnotesize Congeniality results  on the triangle slice of $x_T=x_T^\oplus=100$. Our universe is the point by the bottom right corner marked ``us.'' The green bands are congenial, the red background is uncongenial and the central white region corresponds to worlds with more than two baryons participating in nuclear physics, worlds which we cannot at the moment fully characterize as congenial or uncongenial. Fuzzy borders imply uncertainty in determining boundaries.}
\label{congtriangle}
\end{figure}

Figure \ref{congtriangle} shows the quark mass triangle for $x_T = 100$, with the areas corresponding to congenial worlds colored in green and the areas corresponding to uncongenial worlds colored in red.  Our world corresponds to the point in the lower right-hand corner labelled ``us.''  The central region---where more than two baryons participate in nuclear physics---is shown in white, with a red dot in the middle to indicate that worlds where all three quark masses are identical are very probably uncongenial, as we shall argue in Sec. \ref{masslessuds}.  The rationale for the two upper, narrow green bands shall be discussed in Sec. \ref{neutronsigmaworld}.

\subsection{One light quark with two participating baryons}
\label{neutronsigmaworld}

If one quark is much lighter than the other two, then the two lightest octet baryons may participate in nuclei, leading to congenial domains different in character from the one in which we live.  These are the type (ii) regions shown in Fig.~\ref{regions}.  They lie close to the midpoint of an edge of the quark mass triangle.  To be definite, we consider the case where $m_{d}\ll m_{s}\approx m_{u}$.  The other charge possibilities are considered later.  For $m_{d}$ small enough and $m_{u}$ and $m_{s}$ close enough, there is reason to expect nuclei composed of neutrons and $\Sigma^{-}$ to form and to have qualitatively similar systematics to nuclei in our world.  This region corresponds to $P\approx 1/6$ in the perimeter plots shown in Figs.~\ref{b_perimeter} and \ref{b_perimeter_adjusted}.  

On the other hand, there are important differences that distinguish this congenial region from type (iii) regions like the one in which our world lies.  Type (ii) regions are smaller in extent, and harder to characterize quantitatively.  First, as discussed in Sec.~\ref{onebaryon}, if $m_{u}$ and $m_{s}$ are too large, the decuplet state (in this case $\Delta^{-}$ or $ddd$) may undercut all octet baryons, leading to an uncongenial world.  If this happens for a given $x_{T}$, it excludes a region close to the boundary of the quark mass triangle near $P=1/6$.  For example, in the case treated at the beginning of this section ($m_{u}=0$) we found that the $\Delta^{++}$ became the lightest baryon if $x_{T}\ge 122\pm 29$ MeV.  Note that at the lower limit of the uncertainty this would occur at the perimeter of the quark mass triangle with $x_{T}=100$, in which our world sits (see Fig.~\ref{congtriangle}).

As the light quark mass increases (moving into the interior of the quark mass triangle), this problem diminishes, but another one appears.  Notice (see Fig.~\ref{b_perimeter_adjusted}) that for the same $x_{T}$, other baryons are much closer in mass to the lightest nearly degenerate pair in this region than in type (iii) regions.  In the case of a light $d$ quark near $P=1/6$, the proton, $\Xi^{-}$, and $C_{\rm low}$ have masses less than 75 MeV heavier than the neutron and $\Sigma^{-}$.  In contrast in the type (iii) congenial domain near $P=0$, the $\Lambda$ is more than 170 MeV heavier than the p-n doublet.  As a result, other baryons begin to participate in nuclear physics sooner (as $m_{d}$ and $m_{s}$ are increased) than in our world.  We make a rough guess of how far from the center of the mass triangle this congeniality band ends, by supposing that, as in the case of the light $u$ and $d$ worlds, the third lightest baryon must be less that $15$ MeV heavier than the two lightest baryons for it to participate in nuclear physics.  Translating this into quark masses we obtain the boundaries shown in Fig.~\ref{congtriangle}.

Any attempt make quantitative estimates of nuclear binding in type (ii) worlds is obstructed by the fact that the pseudoscalar meson spectrum is qualitatively different from our world.  This can be seen in Fig.~\ref{spectra}.  Near $P=1/6$ the hadron spectrum has an $SU(2)$ symmetry similar to isospin.  However the lightest pseudoscalar meson is an isosinglet similar to the $\eta$ in our world.  Not much heavier is a complex doublet similar to the kaons in our world.  We suspect that the dominant contribution to the nuclear force still comes from exchange of a \emph{scalar} isoscalar meson analogous to the $f_{0}(600)$, but we have no tools to analyze the problem quantitatively.  

It is possible to say something about the width of the congeniality band for type (ii) congenial regions.  The strength of nuclear binding sets a limit on the $n-\Sigma^{-}$ mass difference similar to the limit we derived from considerations of weak nucleon emission in the previous section.  This translates into a limit on the $s$-$u$ quark mass difference.  However the neutron-$\Sigma^{-}$ mass difference grows twice as fast with $m_{u}-m_{s}$ than the proton-neutron mass difference grows with $m_{u}-m_{d}$.  All else being equal, therefore, the congeniality band has half the width of the one described by \Eq{udcongsu2} in which our world sits.  This is shown in Fig.~\ref{congtriangle}.

Finally we mention the extension of these ideas to other quark charges.  First consider worlds where the three light quarks have charges $2/3, -1/3,$ and $-1/3$ ($uds$).  We have discussed the case where the $d$ quark is very light.  A light $s$ quark yields identical results.  In worlds where the $u$ quark is light, the proton ($uud$) and $\Sigma^{+}$ ($uus$) are the possible constituents of nuclei.  However since both have positive charge, we do not expect congeniality, for the same reason that we failed to find a congenial region when the $\Sigma^{-}$ and $\Xi^{-}$ were the light doublet.  Finally, if the light quarks have charges $2/3, 2/3$ and $-1/3$ ($cud$), the candidate worlds with one light quark have either two neutral baryons, when the $d$ quark is lightest, or baryons of charge +1 and +2, when the $c$ or $u$ quark is lightest.  Neither can be congenial.

\begin{figure*}[bt]
\centering
\subfigure[]{\includegraphics[width=0.1\textwidth]{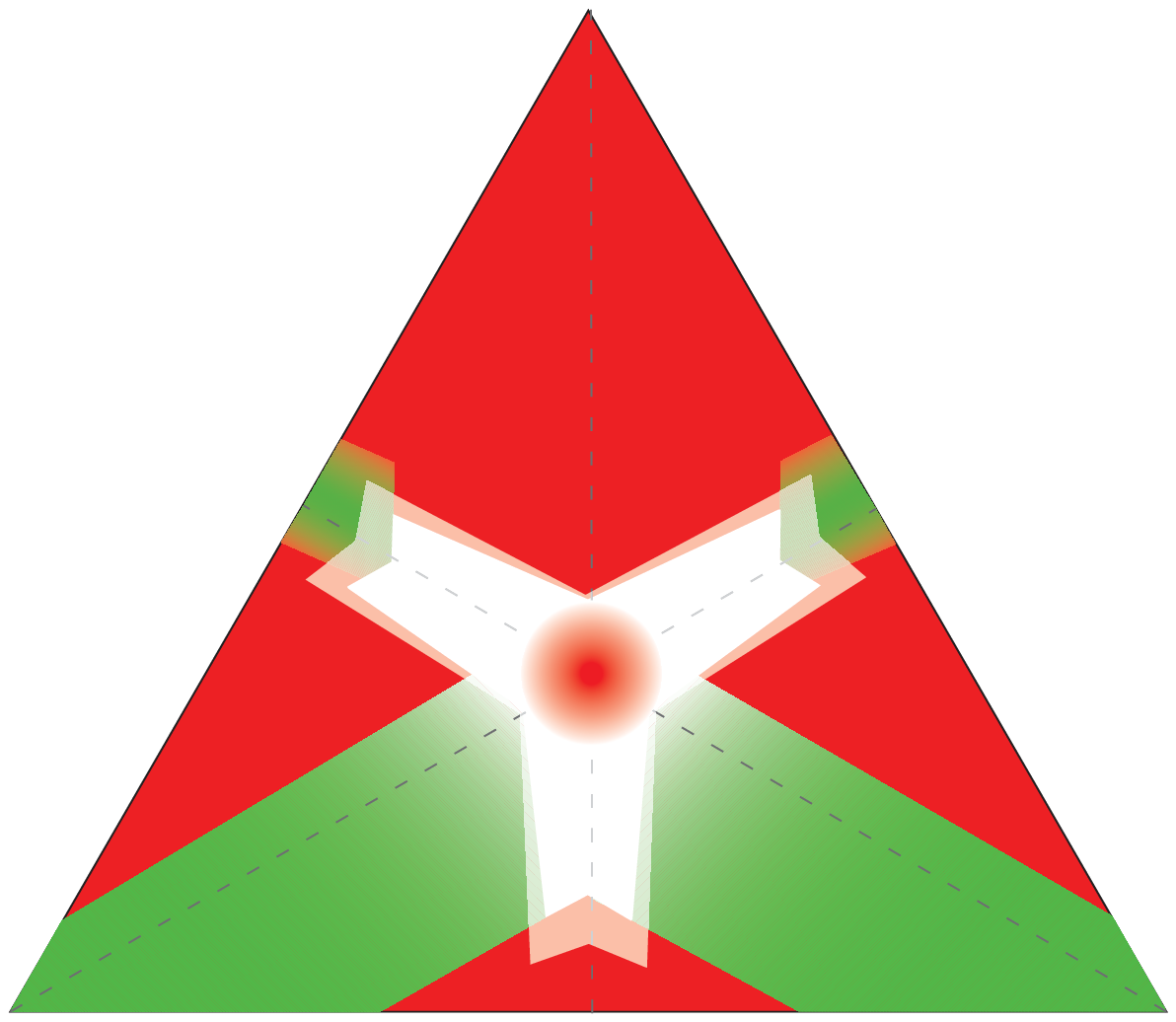}}\hspace*{1.5cm}
\subfigure[]{\includegraphics[width=0.4\textwidth]{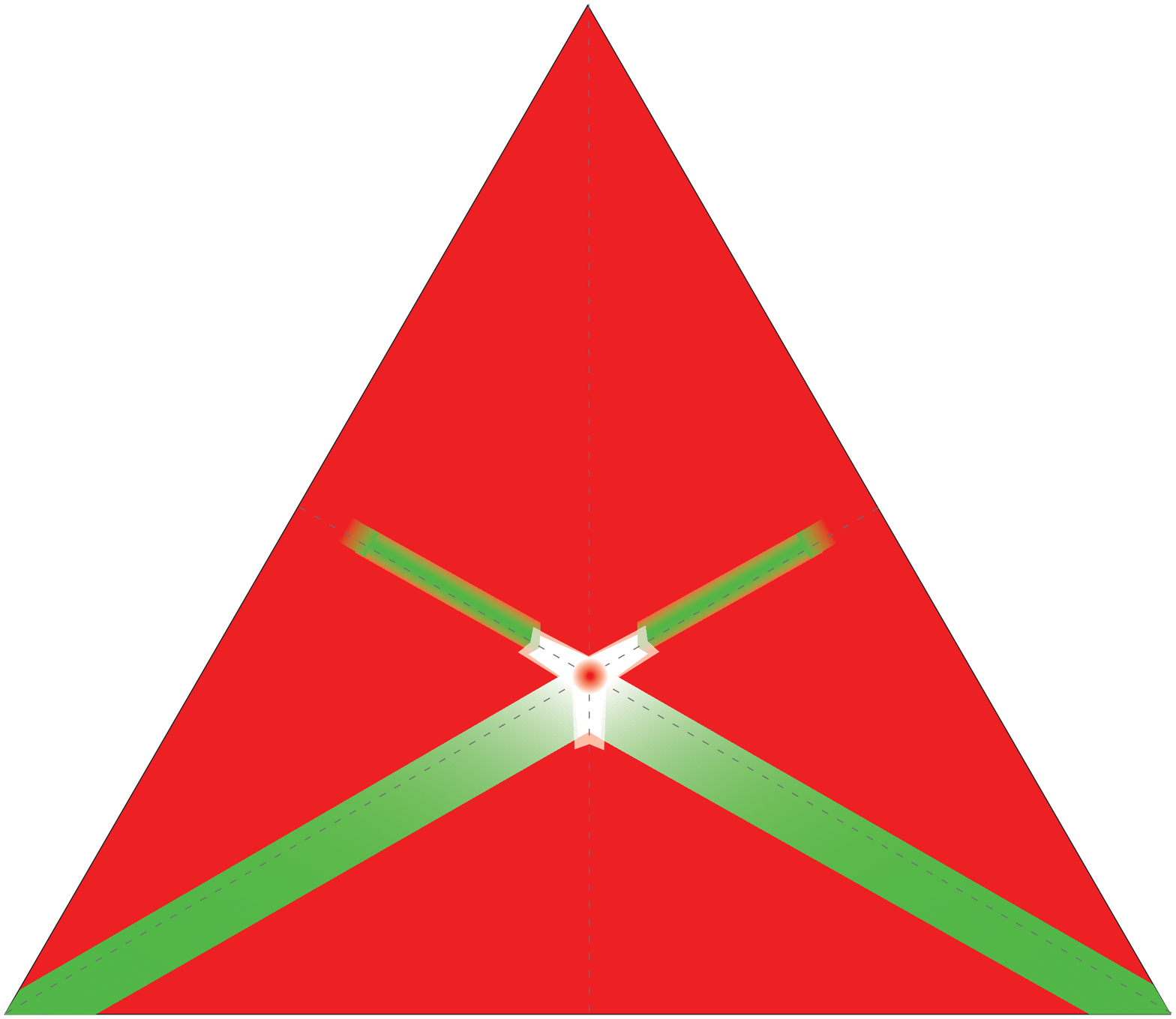}}
\caption{\footnotesize Congeniality results  on the triangle slice of (a) $x_T=50$ and (b) $x_T = 200$.  The coloring conventions are the same as for Fig. \ref{congtriangle}.  Both (a) and (b) are drawn on the same $x_{3,8}$ scale, but not on the same scale as the triangle in Fig. \ref{congtriangle}.}
\label{congtriangles-50-200}
\end{figure*}

For the $uds$ worlds, the regions of congeniality are shown in Fig. \ref{congtriangles-50-200} for the quark mass triangles with $x_T = 50$ and with $x_T = 200$.  Notice that the width of the bands of congeniality, and the size of the central region where more than two baryons participate in nuclear physics, are independent of $x_T$.  For large $x_T$, however, the narrow upper bands of congeniality (corresponding to neutron-$\Sigma^-$ worlds) end before reaching the perimeter, for the reasons discussed in Sec. \ref{onebaryon}.

\subsection{Three light quarks}
\label{threelightquarks}

There are two distinct ways in which more than two nucleons could participate in nuclear physics.  First, more than two quarks could have nearly degenerate masses.  For three light quarks, this would correspond to area very near the center of the quark mass triangle, for any value of $x_{T}$.  However, like the accidental degeneracy discussed in Sec.~\ref{lightandheavy}, this would be quite unlikely if the {\it a priori\/} quark mass distributions are independent of one another.  The second and more relevant case is that in which three (or more) quarks are light even though their mass differences might be of the same order as the masses themselves.  Both cases can be treated simultaneously because the essential issue is flavor $SU(3)$ breaking.  

If quark masses are logarithmically distributed (as we have argued they might well be) then the landscape would be dominated by worlds in which many of the quark masses are very light compared to $\Lambda_{\rm QCD}$.  If we add leptons to the discussion, we already have evidence from the neutrino sector of the possibility of very small masses.  Unfortunately, the analysis of worlds with more than two baryons participating in nuclear physics requires better methods than those currently available.

It is possible, nonetheless, to make some qualitative statements about worlds with more than two very light quarks.  To simplify matters we consider the case where the quark masses are negligible and the only contributions to the baryon mass differences are electromagnetic.  Since we are limited to considering at most three light quarks, there are four charge assignments of interest:  (a) $2/3, -1/3, -1/3$ (which we label $uds$ for convenience); (b) $2/3, 2/3, -1/3$ ($cud$); (c) $-1/3, -1/3, -1/3$ ($dsb$); and (d) $2/3, 2/3, 2/3$ ($uct$).

\subsubsection{A possible dihyperon instability}
\label{dihyp}

The first relevant question about the domain of three light quarks is whether octet baryons are stable.  It is possible that in this limit the dihyperon, $H$---composed of two quarks of each flavor---could be less massive than two octet baryons \cite{dihyperon}.  If this were the case, then regular nuclei would decay into a state made of $A/2$ dihyperons.  Since the $H$ is a boson and therefore not subject to the Pauli exclusion principle, this would lead to worlds radically different from ours.  Because no state made of dihyperons has unit charge, organic chemistry as we know it would not be possible.

The various unsuccessful searches for a stable $H$ suggest that, in our world, $H$ is close to or heavier than twice the mass of $\Lambda$, 2,231 MeV, which is the threshold for its strong decay.  For $M_H \leq 2 M_\Lambda$,  the dihyperon would still not be perfectly stable, since it could decay slowly by strangeness-changing weak interactions, unless it were also lighter than twice the nucleon mass ($\approx$ 1,880 MeV).  Thus $2 M_N$ sets the absolute threshold for the stability of $H$.

The quark content of the $H$ ($uuddss$) matches that of two $\Lambda$'s, two $\Sigma$'s, or $N \Xi$.  Flavor $SU(3)$ violation lowers the mass of two $\Lambda$s relative to $\Sigma\Sigma$ or $N\Xi$.  The $H$, on the other hand, is an $SU(3)$ singlet and its mass is independent of $SU(3)$ breaking.  Since $SU(3)$ becomes exact when all three quarks are massless, a measure of the stability of the $H$ in that limit can be obtained by comparing its mass in our world with twice $\overline M$ (the average octet baryon mass) or $2 \times (1,151) = 2,302$ MeV.  Thus, in the absence of $SU(3)$ violation, the threshold for $H$ stability with respect to strong decay would rise from $2 M_{\Lambda} = 2,231$ MeV to $2 \overline M= 2,302$ MeV.  The threshold for absolute stability is even more dramatically affected:  It is $\approx$ 1,880 MeV in our world, but would be 2,302 MeV without $SU(3)$ symmetry violation.

This argument is far from rigorous, but it does suggest that the $H$ might be more stable in a world of massless quarks than it is in our world.  This is an important point on which a lattice calculation might shed light.  Determining the stability of $H$ in a world of massless quarks would be interesting from an environmental perspective, especially if the qualitative features of nuclei built of dihyperons could be established.  One might speculate, for example, that an instability to forming dihyperons puts an environmental lower bound on the number of very light quarks in a congenial world, in which case we might owe our existence to (among other things) the relatively large mass of the strange quark in our world.  For the rest of this section, we assume that the dihyperon \emph{is not} stable when three quarks are very light.

\subsubsection{Worlds with massless $u$, $d$, and $s$ quarks}
\label{masslessuds}

Of the four charge assignments listed at the beginning of this section, the first, $2/3, -1/3, -1/3$ is special because the sum of the charges is zero.  It is also the one of most immediate interest to us, because it is closest to our world and occupies the apex of the quark mass prism we have been exploring.

If all octet baryon masses were exactly equal, then the minimum of the valley of stability would occur at electric charge neutrality, {\it i.e.\/}, at equal numbers of each quark species.  At low $A$, charged nuclei would quickly $\beta$-decay or electron-capture until they reached the stable, electrically neutral nucleus with baryon number $A$.  The only possibility of a stable, charged nucleus would develop at very large $A$, where the asymmetry term in the SEMF becomes less important.  Thus, even small electromagnetic mass differences make a big difference in worlds where $u$, $d$, and $s$ quarks are all massless. To estimate the effect, we used the electromagnetic mass shifts given in Table~\ref{t.an.2}, together with the three-flavor SEMF of \Eq{4.21} (in which baryon species abundances have already been adjusted to maximize binding at fixed $Y$ and $I_{3}$).  The parameter that drives the appearance of stable charged nuclei for zero quark masses is $\Delta M_{I} = -0.15$ MeV.  We then compute the lowest value of $A$ at which carbon ($Z=6$) is a stable against $\beta$-decay and/or electron capture.  First we minimize the nuclear mass with respect to $Y$ since weak non-leptonic interactions will effectively adjust the $d$ and $s$ quark abundances to maximize binding.\footnote{Weak scattering processes like $\Lambda n\leftrightarrow nn$ are mediated by $us\leftrightarrow du$ at the quark level.}  This yields $Y_{\rm min}$ as a function of $A$, $I_{3}$, and the SEMF parameters. Then we re-express the nuclear mass in terms of $Z=I_{3}+\fract{1}{2}Y$.   We focus on $Z=6$.  For small values of $A$, the state with $Z=6$ decays to lower charge by electron capture.  The criterion for it to be stable is therefore
\beq
M(Z=6,A) + m_e > M(Z=5,A) ~.
\eeq
The value of $A$ at which this is satisfied depends most sensitively on the strength of the asymmetry term in the SEMF, $\epsilon_{a}$.\footnote{It also depends on the Coulomb interaction, but we do not see any reason to change $\epsilon_{c}$ from its value in our world.}  For $\epsilon_{a}=\epsilon_{a}^{\oplus}$ we find that carbon becomes stable only for $A\gtrsim 2\times 10^{5}$.  Even if such nuclei were stable against particle emission, the likelihood of these gargantuan atoms coming together to form life forms seems very remote.

The existence of very light, strongly bound nuclei analogous to the $\alpha$ particle in our world, would very likely prevent the formation of large $A$ nuclei in worlds with massless $u$, $d$, and $s$ quarks.  Shell effects and the pairing term both favor the binding of nuclei made of two baryons of each species, starting at $A=4$ (in direct analogy to the $\alpha$ particle) and continuing up to $A=16$, a particle we call the ``\emph{arkon}'', after its Biblical analog which carried every species two-by-two.  These $s$-state nuclei would have their own complicated stability pattern, but there is every reason to expect one or more of these species to be very tightly bound relative to generic heavy nuclei (as is the $\alpha$ particle in our world), as well as electrically neutral.  Heavy nuclei like the $A\approx 10^{5}$ carbon which we have just considered might then very well be unstable against emitting such light, neutral objects.

We conclude that there is good reason to believe that a world with very light $u$, $d$, and $s$ quarks would be uncongenial.  Quantitative bounds on the quark masses, however, await better theoretical tools.

\subsubsection{Other light quark charge assignments}

The three other charge assignments listed at the beginning of this section:  $cud$, $dsb$, and $uct$, are intriguing.  In the last case, $uct$, all octet baryons have charge $Z=+2$, so there is no hydrogen and these worlds are uncongenial.  The minimum of the nuclear asymmetry energy occurs at $Z=+A$ for $cud$ worlds and $Z=-A$ for $dsb$ worlds.  This implies a greater mismatch between the Coulomb and asymmetry energy minima than in our world (where the asymmetry energy minimum occurs at $Z=A/2$).  Since the conflict between asymmetry and Coulomb energies destabilizes large $A$ nuclei, we expect that, given similar SEMF parameters, the nuclear mass table in these worlds would likely terminate at smaller values of $A$ than in our world.  On the other hand, the very light nuclei that are important to us, hydrogen, carbon, and oxygen, might well be stable.  For example, in the $cud$ world the baryons with the quark content $ccu$ and $uuc$ have $Z=+2$, so a baryon number four nucleus with the structure of the $\alpha$ particle would have the chemistry of oxygen.  Intriguing though these worlds may be, we leave them for future study, hopefully with more powerful analytic tools.

\subsection{Final congeniality results in a light $u$, $d$, $s$ world}

We can now complete the description of the congeniality triangle introduced at the beginning of this section.  The results are summarized in Figs.~\ref{congtriangle} and \ref{congtriangles-50-200} .  The green congeniality bands beginning in the corners at $P=0$ and $P=2/3$ are the type (iii) congeniality domains discussed extensively in Sec.~\ref{udworld}.  The bands beginning at the centers of edges, at $P=1/6$ and $P=1/2$ are the type (ii) congeniality bands discussed in Section~\ref{neutronsigmaworld}.  The center of the triangle is red because the neutral worlds discussed in this subsection are very likely to be uncongenial.  The domains shown in white are regions where three or more baryons participate in nucleus building, but the quark mass differences are too large to apply the symmetry arguments of Sec. \ref{masslessuds}.  Note that the type (iii) congeniality domains extend further into the triangle than the type (ii) domains because the next lightest baryon is split further from the lightest doublet in that case.  

%%%%%%%%%%
%%% ACKNOWLEDGMENTS
%%%%%%%%%%
\begin{acknowledgments}

The authors thank T. William Donnelly, John Donoghue, Yasunori Nomura,  Jos\'e Pel\'aez, Michael Salem, Matthew Schwartz, Iain Stewart, and Mark Wise for conversations, correspondence, and suggestions regarding this work.  RLJ thanks Leonard Susskind for encouragement and AJ thanks Michael Salem for advice on the manuscript.  This work was supported in part by the U.S. Department of Energy under contract DE-FG03-92ER40701.

\end{acknowledgments}

%%%%%%%%%%
%%% APPENDIX:
%%% FIRST ORDER PERTURBATION THEORY FOR BARYON MASS DIFFERENCES
%%%%%%%%%%
\appendix
\section{First order perturbation theory for baryon mass differences}
\label{a.an.1}

Even though it is a textbook exercise, for the sake of completeness we present here the first order perturbation analysis relating octet baryon mass differences to quark mass differences.  We label states in the baryon octet by $|8;I_{3},Y,\nu\rangle\equiv |8;\mu\rangle$, where $\nu$ is the label that distinguished states of the same $I_{3}$ and $Y$ that mix due to $SU(3)$ flavor violation.  The mass of an octet baryon identified by $\mu$ is $\braket{8;\mu }{H}{ 8;\mu}$. Using the Wigner-Eckart Theorem, we may express the expectation value of the last two terms in $H_{\rm flavor}$ in Eq. (\ref{Hoctet}) as:
\beq
\braket{8;\mu }{ \Theta_{\alpha}^{8}}{ 8;\mu} = \sum_{\gamma = 1}^{2} 
\left( \begin{array}{ccc}
8 & 8 &  8_{(\gamma)} \\
\mu & \alpha  &\mu  \end{array} \right)
\left \langle 8 || \Theta^{8} | |8  \right\rangle_{\gamma} ~.
\label{wigner-eckart}
\eeq
The double-barred term is the Wigner reduced invariant matrix element (WRME) and the term in parenthesis is the $SU(3)$ Clebsch-Gordan coefficient.  The sum over the index $\gamma$ reflects the fact that there are two irreducible 8-dimensional representations in the decomposition of $8 \otimes 8 = 27 \oplus 10 \oplus \overline{10} \oplus 8_{1} \oplus 8_{2} \oplus 1$.  The $SU(3)$ Clebsch-Gordan coefficient can be decomposed as
\beq
\left( \begin{array}{ccc}
8 & 8 &  8_{(\gamma)} \\
\mu & \alpha  &\mu  \end{array} \right) = 
C^{I_{\mu} I_{\alpha} I_{\mu}}_{I_{\mu 3} I_{\alpha 3} I_{\mu 3}} 
\left( \begin{array}{ccc}
8 & 8 &  8_{(\gamma)} \\
I_{\mu} Y_{\mu} &  I_{\alpha} Y_{\alpha} & I_{\mu}Y_{\mu} 
\end{array} \right) ~,
\label{clebsch-gordan}
\eeq
where the $C$ symbol is the $SU(2)$ coefficient and the term in parenthesis on the right-hand side of Eq. (\ref{clebsch-gordan}) is the isoscalar coefficient.  Their values are quoted, for instance, in \cite{deswart} (whose notation we have adopted).

We may therefore express the masses of the octet baryons in terms of five parameters: a term $A_{0} \equiv H_{\rm  QCD} + m_{0}\braketnoline{\Theta_{0}^{1}}$ common to all the baryons, the two WRME's, $m_{3}$, and $m_{8}$.  For example, the mass of the proton can be expressed as
\beqa
M_p = A_{0} &+&  \frac{m_{3}}{\sqrt{3}} \left[\frac{3\sqrt{5}}{10} \left\langle 8||\Theta^{8}||8 \right\rangle_1  + \frac{1}{2} \left\langle 8||\Theta^{8}||8 \right\rangle_2  \right] \nonumber \\
&+& m_{8} \left[-\frac{\sqrt{5}}{10} \left\langle 8||\Theta^{8}||8 \right\rangle_1 + \frac{1}{2} \left\langle 8||\Theta^{8}||8 \right\rangle_2  \right]~.
\label{proton}
\eeqa
where the $SU(2)$ Clebsch-Gordan coefficients are shown in front of the terms in square brackets.

Following the notation of \cite{CG}, we define two new parameters proportional to the WRME's,
\beq
D \equiv \frac{\sqrt{15}}{10} \left\langle 8||\Theta^{8}||8 \right\rangle_1~, ~~ F \equiv \frac{\sqrt{3}} 6 \left\langle 8||\Theta^{8}||8 \right\rangle_2~,
\label{DF}
\eeq
so that Eq. (\ref{proton}) becomes
\beq
M_p = A_{0} + \left(\frac{3F-D}{\sqrt{3}}\right)m_{8} + (F+D)m_{3} ~.
\eeq
Analogous results for the other baryons on the periphery of the octet weight diagram are summarized in the first six entries in Table~\ref{t.an.1}.

The Hamiltonian of Eq. (\ref{baryonH}) is not   diagonal in the ($I_{3},Y$) basis.  In other words, not all of the basis states in the octet representation are mass eigenstates.  If we label the eigenstates of $I^{2}$ the $\Lambda$ ($I=0$) and the $\Sigma^{0}$ ($I=1$), then in that basis $\left( \Lambda, \Sigma^{0} \right)$ and in the $I_{3}=Y=0$ subspace,
 
\beq 
H  = A_{0} +  \frac{2D}{\sqrt{3}}
\left( \begin{array}{cc}
m_{8} & m_{3}  \\
m_{3} & -m_{8} \end{array} \right) + H_{\rm EM}~.
\eeq
Defining $\Delta \equiv \sqrt{m_{8}^{2} + m_{3}^{2}}$, the mass eigenstates in the $(\Lambda,\Sigma^{0})$ basis are,
\beq
C_{\rm high} = \frac{1}{\sqrt{2\Delta}} \left( \begin{array}{c}
\sqrt{\Delta + m_{8}}   \\
\sqrt{\Delta - m_{8}}  \end{array} \right)~
\text{and} ~~ 
 C_{\rm low}  = \frac{1}{\sqrt{2\Delta}} \left( \begin{array}{c}
-\sqrt{\Delta - m_{8}}   \\
\sqrt{\Delta + m_{8}}  \end{array} \right)~.
\eeq
This mixing is negligible (for most purposes) in our world, since $m_3 / m_8 = 0.02$, and $\Delta = 1.0002 \, m_8 \approx m_8$.  Therefore to a few percent, $C_{\rm low} \approx \Lambda$, and $C_{\rm high} \approx \Sigma^{0}$. However, this mixing is significant in worlds where $m_{3}$ and $m_{8}$ are comparable.  In those cases total isospin no longer approximately commutes with the Hamiltonian, and the states $C_{\rm high}$ or $ C_{\rm low} $ cannot be identified with total isospin eigenstates.

%%%%%%%%%%
%%% REFERENCES
%%%%%%%%%%

\bibliographystyle{aipprocl}   % if natbib is available
%\bibliographystyle{aipprocl} % if natbib is missing

%%%%%%%%%%%%%%%%%%%%%%%%%%%%%%%%%%%%%%%%%
% You probably want to use your own bibtex database here
%%%%%%%%%%%%%%%%%%%%%%%%%%%%%%%%%%%%%%%%%
%\newpage

\end{document}